\pdfoutput=1
\documentclass [a4paper, 12pt]{article}
\usepackage[inner=1in,outer=1in,tmargin=1in,bmargin=3cm]{geometry}
\usepackage{mathtools}
\usepackage{amssymb}
\usepackage[utf8]{inputenc}
\usepackage[T1]{fontenc}
\usepackage[english]{babel}
\usepackage{parskip}
\usepackage{tensor}
\usepackage{hyperref}
\usepackage{amsmath}
\usepackage{afterpage}
\usepackage{graphicx}
\usepackage{tikz}
\usepackage{tikz-3dplot}   %TikZ is required for this to work.  Make sure this exists before the next line
\usepackage{ae,aecompl}
\usepackage[capposition=bottom]{floatrow}
\usepackage{empheq}
\usepackage{amsthm}
\usepackage{cite}
\usepackage{cases}
\usepackage{hyperref}
\usepackage{pgfplots}
\usepackage{comment}
\usepackage[autostyle, english = american]{csquotes}
\MakeOuterQuote{"}

\hypersetup{
	colorlinks,
	linkcolor={red!80!black},
	citecolor={blue!80!black},
	urlcolor={blue!80!black},
	linktocpage=page
}

\usetikzlibrary{decorations.pathmorphing,calc}

\DeclareMathOperator{\Tr}{Tr}

\providecommand{\abs}[1]{\lvert#1\rvert}

\newcommand{\mup}{^{\mu}}
\newcommand{\mud}{_{\mu}}
\newcommand{\nup}{^{\nu}}
\newcommand{\nud}{_{\nu}}
\newcommand{\au}{^\alpha}
\newcommand{\ad}{_\alpha}
\newcommand{\bu}{^\beta}

\newcommand{\munuu}{^{\mu\nu}}
\newcommand{\munud}{_{\mu\nu}}
\newcommand{\nn}{\nonumber}

\newcommand{\pt}{\partial}

\newcommand{\be}{\begin{equation}}
	\newcommand{\ee}{\end{equation}}
\newcommand{\ba}{\begin{eqnarray}}
	\newcommand{\ea}{\end{eqnarray}}
\newcommand{\baa}{\begin{array}}
	\newcommand{\eaa}{\end{array}}

\newcommand{\na}{\nabla}
\newcommand{\ijd}{_{ij}}
\newcommand{\iju}{^{ij}}

\newcommand{\puf}{^{(1)}}

\newcommand{\pus}{^{(2)}}

\newcommand{\hb}{\bar{h}}
\newcommand{\abd}{_{\alpha\beta}}
\newcommand{\abu}{^{\alpha\beta}}
\newcommand{\ttu}{^\rmi{TT}}
\newcommand{\ttd}{_\rmi{TT}}
\newcommand{\lam}{$\Lambda$}
\newcommand{\cdt}{\cdot\cdot\cdot}

\newcommand{\bfx}{{\bf x}}
\newcommand{\emax}{\eta'_\rmi{max}}

\newcommand{\bh}{\bar{h}}
\newcommand{\bg}{\bar{g}}
\newcommand{\br}{\bar{R}}
\newcommand{\nab}{\bar{\nabla}}

\newcommand{\nr}[1]{(\ref{#1})}  %puts parentheses to eq numbers
\newcommand{\rmi}[1]{{\mbox{\scriptsize #1}}}  %practical for sub- or superscripts in roman
\newcommand{\fra}[2]{{\textstyle{\frac{#1}{#2}\,}}}  %%Smaller size fractions!
\newcommand{\OL}{\Omega_\Lambda}
\newcommand{\Om}{\Omega_m}

\newcommand{\fr}[2]{{\frac{#1}{#2}\,}}

\newcommand{\GN}{G}

\newcommand{\etco}{{\eta_\rmi{coal} } }
\newcommand{\coal}{_\rmi{coal}}

\newcommand{\tailup}{^\rmi{tail}}
\newcommand{\lconeup}{^\rmi{LC}}

\numberwithin{equation}{section}

\begin{document}
	
	\begin{flushright}
		HIP-2022-7/TH\\
	\end{flushright}
	\vspace{1cm}
	
\begin{center}
	
	\centerline{\Large {\bf Gravitational wave memory and its tail in cosmology}}
	
	\vspace{8mm}

	\renewcommand\thefootnote{\mbox{$\fnsymbol{footnote}$}}
	Niko Jokela,${}^{1,2}$\footnote{niko.jokela@helsinki.fi}
	K. Kajantie,${}^2$\footnote{keijo.kajantie@helsinki.fi} and
	Miika Sarkkinen${}^{1}$\footnote{miika.sarkkinen@helsinki.fi}
	
	\vspace{4mm}
	${}^1${\small \sl Department of Physics} and ${}^2${\small \sl Helsinki Institute of Physics} \\
	{\small \sl P.O.Box 64} \\
	{\small \sl FIN-00014 University of Helsinki, Finland} 
	
	\vspace{0.8cm}
\end{center}

\vspace{0.8cm}

\setcounter{footnote}{0}
\renewcommand\thefootnote{\mbox{\arabic{footnote}}}

\begin{abstract}
	\noindent 
	%Gravitational wave memory effect is a permanent relative displacement that two freely falling test masses experience after a passage of a gravitational wave burst. The effect is well understood in flat spacetime background where it has been studied for decades now. However, at the largest possible scales the universe is not flat but curved due to cosmic expansion. To take into account this fact we studied the memory effect in the FLRW cosmological model with matter and the cosmological constant $\Lambda$. We focused on the nonlinear Christodoulou memory effect coming from second order gravitational radiation sourced by first order gravitational radiation. Towards this purpose, we solved the Green's equation for the wave operator in an expanding universe. Green's function so obtained is known to have two parts: the familiar d'Alembertian Green's function and the tail two-point function. With the tail two-point function we solved numerically the tail part of the memory signal sourced by radiation from an equal mass quasicircular binary system at various cosmic distances. We did this for redshifts 1,...,30, binary black hole lifetimes 10^3,...,10^8 yr and masses 10,...,10^6 M_sun. We found that typically the tail memory signal is extremely small compared to the corresponding light cone signal, but for binaries with very long lifetimes it might be measurable in the future gravitational wave detectors.
	
	We study gravitational wave memory effect in the FRW cosmological model with matter and cosmological constant. Since the background is curved, gravitational radiation develops a tail part arriving after the main signal that travels along the past light cone of the observer. First we discuss first order gravitational wave sourced by a binary system, and find that the tail only gives a negligible memory, in accord with previous results. Then we study the nonlinear memory effect coming from induced gravitational radiation sourced by first order gravitational radiation propagating over cosmological distances. In the light cone part of the induced gravitational wave we find a novel term missed in previous studies of the cosmological memory effect. Furthermore, we show that the induced gravitational wave has a tail part that slowly accumulates after the light cone part has passed and grows to a sizeable magnitude over a cosmological timescale. This tail part of the memory effect will be a new component in the stochastic gravitational wave background. %propagating to an observer operating at the present, early $\Lambda$-dominated era.

\end{abstract}

\newpage
	
\tableofcontents
\newpage
%%%%%%%%%%%%%%%%%%%%%%%%%%%
\section{Introduction}
%%%%%%%%%%%%%%%%%%%%%%%%%%%

In flat spacetime gravitational radiation, as also the electromagnetic one, propagates along the light cone. In curved spacetime, like the  expanding Friedmann-(Lema\^{\i}tre-)Robertson-Walker (FRW) metric of cosmology, radiation develops another component moving subluminally inside the source's future light cone, the tail \cite{DeWitt:1960fc,Friedlander:2010eqa,Ford:1977dj,Faraoni:1991xe,Caldwell:1993xw,Wiseman:1993aj,Iliopoulos:1998wq,deVega:1998ia,Balek:2007hh,Poisson:2011nh,Harte:2013dba,Ashtekar:2015lxa,Chu:2015yua,Chu:2016ngc,Chu:2016qxp,Chu:2021apx,Kehagias:2016zry,Yagdjian:2021pab,Copi:2022ire}. The main purpose of this paper is to quantitatively analyse the properties of the tail of radiation from a black hole binary propagating in the background of the $\Lambda$CDM model, also known as the concordance model. In particular, we wish to study how the tail affects the memory effect \cite{Zeldovich:1974gvh,Braginsky:1986ia,Christodoulou:1991cr,Thorne:1992sdb,Wiseman:1991ss,Will:1996zj,Favata:2008ti,Favata:2008yd,Favata:2009ii,Favata:2010zu,Bieri:2013ada,Bieri:2015jwa,Bieri:2017vni,Garfinkle:2016nhe,Garfinkle:2022dnm,Chakraborty:2022qvv}. One usually expects a very small tail signal, but we show that there is a sizeable long time tail \cite{Chu:2015yua,Chu:2016ngc}, subluminally propagating excitations arrive at the detector long after the merger signal. This, therefore is an observable and potentially distinguishable from the standard light cone memory strain.

\begin{figure}
		\begin{center}
			\includegraphics[width=150mm]{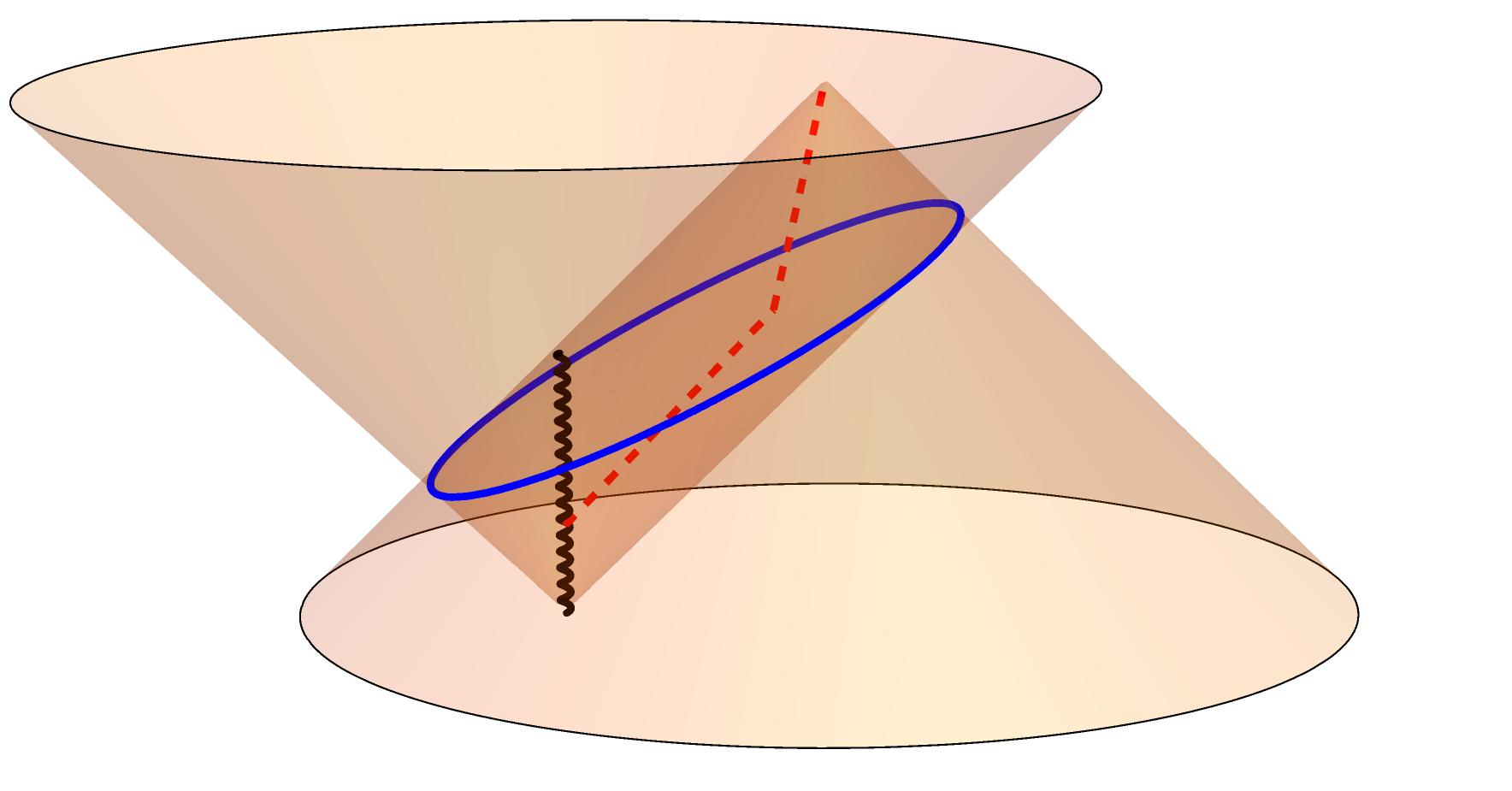}
		\end{center}
	\caption{\small An illustration of the integration region. The black wiggly line represents the spacetime trajectory of the black hole binary, located at the spatial origin; direction of time is upwards. The red dashed line is a GW signal first emitted from the binary and propagating at the speed of light, then this signal sources a tail GW at a point in the bulk, and the tail signal travels to the observer at the tip of the cone that forms the future boundary of the region. The blue elliptical curve tracks the intersection of the cones. The third spatial coordinate is suppressed. For more details, see Fig. \ref{kuvaintregion}.
	}\label{intersectingcones}
\end{figure}

The physical set-up is as follows, see Fig. \ref{intersectingcones}. We have an equal-mass compact binary with component masses $M$ in a quasi-circular orbit at some large redshift $z$. The binary emits gravitational radiation and coalesces with a certain lifetime $\tau$ at a certain conformal time $\etco$. This radiation, of first order in metric perturbations, is observed in gravitational wave detectors and its evolving light cone memory signal is well known \cite{Favata:2010zu,Bieri:2013ada,Garfinkle:2016nhe}. We find that the first order tail actually effectively also propagates along the light cone, there is no long time tail. The energy-momentum tensor of this first order radiation sources second order radiation which is known to lead to a null or non-linear memory \cite{Christodoulou:1991cr,Thorne:1992sdb,Wiseman:1991ss,Bieri:2013ada}. We compute the evolution of the light cone memory signal and indentify a new expansion dependent contribution. Further, the tail arriving \emph{before} the merger signal is very small, but the total effect of subluminal excitations arriving \emph{after} the merger leads to a sizable evolving memory signal. The quantitative computation of this tail memory, originating from one merger, possibly long ago, is the main result of this paper. Computation of the total effect from all earlier mergers, with various lifetimes, redshifts, and mass distributions, remains a problem for future study.

In the technical side, the computation basically boils down to evaluating the metric tensor perturbation from the integral
\begin{equation}
	h_{ij}(\eta, {\bf x}) = 4 G \int d^4 x' \frac{\delta(u)}{\abs{{\bf x-x'}}} \frac{a(\eta')}{a(\eta)}t_{ij}(\eta',{\bf x'}) + 4 G \int d^4 x' \theta(u) B(x,x') \frac{a(\eta')}{a(\eta)}t_{ij}(\eta',{\bf x'}) \ ,
\end{equation}
where $u = \eta - \eta' - \abs{{\bf x-x'}}$ is retarded time, $\eta$ conformal time, $a$ is the scale factor, $t_{ij}$ the source stress-energy and $B$  the tail two-point function. The first term is the signal which travels on the light cone and the second one is the tail term. $B$ is determined from an equation with homogeneous part $B''(\eta)-(\vec{\na}^2-V)B$,  $V(\eta)=a''(\eta)/a=\fra16 a^2\mathcal{R}$ where $\mathcal{R}$ is the Ricci scalar. In the concordance model $V$ exists within the range $0<\eta<\eta_\rmi{max}=4.4457/H_0$, between early Big Bang and the comoving visibility limit. We find that $V$  is surprisingly symmetric under reflection  around the middle of its range of existence, $ \eta_\rmi{max}/2$. Special numerical techniques are developed to compute the tail propagator $B(x,x')$.
%and to perform the two TT projections required for second order radiation.

The paper is organized as follows. In Sec. 2, we outline the derivation of nonlinear memory effect in Minkowski spacetime. In Sec. 3, we discuss how the tail arises in a curved cosmological background, compute numerically the tail propagator for the \lam CDM model, summarize first order gravitational radiation, and discuss the tail associated with it. In Sec. 4, we proceed to the main task: computing the  nonlinear tail memory induced by radiation from a binary black hole source. Appendices discuss the specifics of the concordance model and the tail propagator therein, metric perturbation theory, and angular integration techniques.

We employ the mostly plus signature convention $- + + +$.
Greek letters are used to denote the spacetime indices $0,1,2,3$ and Latin letters $i,j,k,..$ to denote 3d spatial indices, either Cartesian or spherical. 3d vectors are written in boldface.
Overdot $\dot{ }$ denotes the derivative with respect to $t$  and prime $'$ the derivative with respect to  $\eta$. The speed of light $c = 1$, mostly. 
\section{Memory effect in flat spacetime revisited}\label{flatspacemem}
%%%%%%%%%%%%%%%%%%%%%%%%%%%
The memory effect in flat spacetime is theoretically a well understood phenomenon. For the sake of completeness, here we outline its derivation to spare the reader from a detour into the literature, since the main computation of our paper parallels it closely. A reader already familiar with the topic may well skip this section.

\subsection{Memory effect as a displacement}
In flat spacetime, the perturbed metric is
\begin{equation}
	g\munud = \eta\munud + h\munud \ , \quad \abs{h\munud} \ll 1 \ ,
\end{equation}
where $\eta\munud$ is the Minkowski metric and $h\munud$ is a small perturbation. We define the trace-reversed metric perturbation as
\begin{equation}
	\bar{h}\munud = h\munud - \frac{1}{2}\eta\munud h\ , \quad h \equiv \eta\munuu h\munud \ ,
\end{equation}
and impose the Lorenz gauge condition $\pt\mup \bar{h}\munud=0$. Then the linearized Einstein's equation takes the form of a wave equation with source
\begin{equation}\label{hbareinstein}
	\Box \bar{h}\munud = -16\pi \GN T\munud \ ,
\end{equation}
where $\Box$ is the d'Alembert operator $\pt\mud \pt\mup = -\pt_t^2 + \vec{\na}^2$, and $T\ijd$ is the source stress-energy tensor. Away from all sources, we are allowed to choose the transverse-traceless (TT) gauge defined by
\begin{equation}
	h_{0\mu}\ttu = \pt\mup h\munud\ttu = h = 0\ , \quad h\equiv \eta\munuu h\munud\ttu \ .
\end{equation}
Note that in the TT gauge $h\munud = \bar{h}\munud$.
The TT tensor perturbation then satisfies
\begin{equation}
	\Box \bar{h}\ijd\ttu = 0 \ ,
\end{equation}
which describes a physical gravitational wave (GW) traveling in vacuum. A GW incited by a source is then obtained by inverting Eq. \nr{hbareinstein} and projecting into the TT gauge. We use the algebraic projection method \cite{Ashtekar:2017ydh}, the use of which has recently been clarified in \cite{Garfinkle:2022dnm}.

It is often convenient to express the TT tensor perturbation in terms of polarization modes. If ${\bf n}$ is the direction of propagation of the GW, we choose an orthonormal pair of vectors ${\bf u}, {\bf v}$ such that $u^i n_i = v^i n_i = 0$. In this basis the polarization tensors read
\begin{equation}\label{polar}
	e^+\ijd = \frac{1}{\sqrt{2}}(u_i u_j - v_i v_j)\ , \quad e^\times \ijd =  \frac{1}{\sqrt{2}}(u_i v_j + u_j v_i) \ .
\end{equation}
These tensors satisfy
\begin{equation}
	n^i e^A\ijd = 0\ , \quad \Tr e^A = 0\ , \quad e^A\ijd e_B\iju = \delta^A_B\ , \quad A,B = +, \times \ .
\end{equation}
It is convenient to choose the orthonormal vectors ${\bf u}, {\bf v}$ to be the normalized coordinate basis vectors ${\bf e_\theta}, {\bf e_\phi}$ of the spherical coordinate system. In spherical coordinates, the polarization tensors are then given by the matrices
\begin{equation}\label{polarization}
	e^+\ijd = \frac{1}{\sqrt{2}}
	\begin{pmatrix}
		0 && 0 && 0 \\
		0 && 1 && 0 \\
		0 && 0 && -1
	\end{pmatrix}\ , \quad
	e^\times \ijd = \frac{1}{\sqrt{2}}
	\begin{pmatrix}
		0 && 0 && 0 \\
		0 && 0 && 1 \\
		0 && 1 && 0
	\end{pmatrix} \ .
\end{equation}
TT gravitational radiation can now be decomposed into its plus and cross modes as
\begin{equation}
	h\ttu\ijd = h_A e^A\ijd = h_+ e^+ \ijd + h_\times e^\times \ijd\ .
\end{equation}
%TÄSSÄ VOISI EHKÄ SELITTÄÄ RELAKSOIDUSTA EINSTEININ YHTÄLÖSTÄ JA LANDAU-LIFSCHITZ PSEUDOTENSORISTA

\begin{figure}[t!]
	\begin{center}
		\includegraphics[width=8cm]{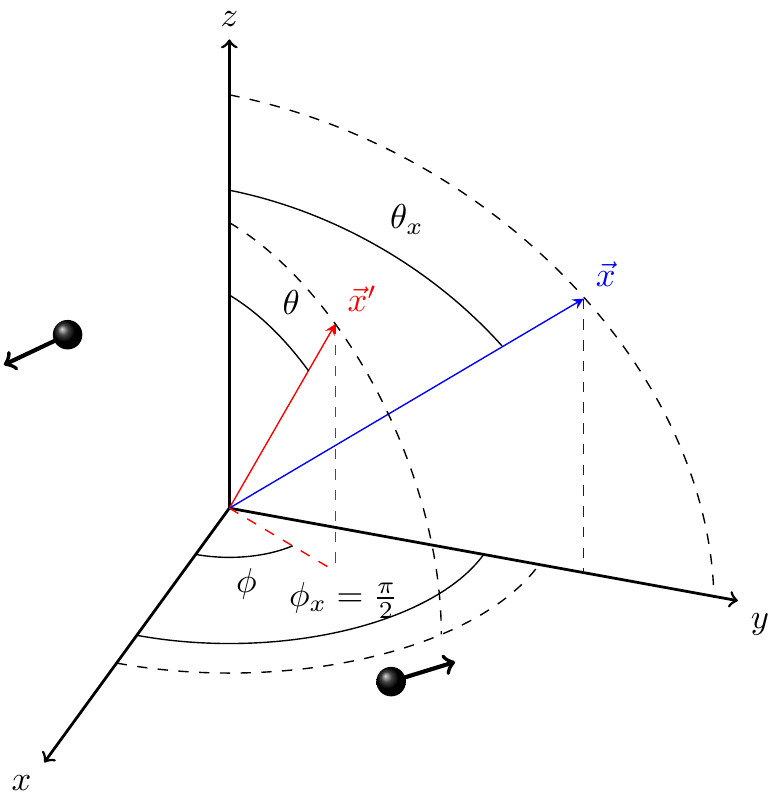}
	\end{center}
		\caption{\small Coordinate system used. The equal mass binary rotates in the $x-y$ plane around the origin of the
			figure. Its gravitational radiation at the point $ {\bf x'}=r'(\sin\theta\cos\phi,\sin\theta\sin\phi,\cos\theta)$ induces further gravitational radiation that observer receives at	${\bf x}=r(0,\sin\theta_x,\cos\theta_x)$ at time $u=t-r$ according to (\ref{inspmem}) and (\ref{ringmem}). 
		}\label{coordsyst}
		
\end{figure}

A system of test particles in free fall experience a tidal force induced by a passing GW. The memory effect manifests itself as a permanent change of distance between two freely falling test masses after the burst of gravitational radiation has passed. Tidal effects between a pair of observers moving geodesically juxtaposed to each other is described by the deodesic deviation equation
\begin{equation}\label{deviation}
	u^\rho \na_\rho (u^\sigma \na_\sigma \xi\mup) =  \tensor{R}{^\mu_\nu_\rho_\sigma}u^\nu u^\rho \xi^\sigma \ ,
\end{equation}
where $\tensor{R}{^\mu_\nu_\rho_\sigma}$ is the Riemann curvature tensor,\footnote{We use the convention $[\na\mud\, , \na\nud] V_\rho = \tensor{R}{_\mu_\nu_\rho^\sigma}V_\sigma$ for the Riemann tensor. Covariant derivative $\na$ acting on a $(k,l)$ tensor is given by
	\begin{align}
		\na_\alpha \tensor{T}{^{\mu_1}^{...}^{\mu_k}_{\nu_1}_{...}_{\nu_l}} &= \pt_\alpha \tensor{T}{^{\mu_1}^{...}^{\mu_k}_{\nu_1}_{...}_{\nu_l}} + \Gamma_{\alpha\beta}^{\mu_1}\tensor{T}{^\beta^{...}^{\mu_k}_{\nu_1}_{...}_{\nu_l}} + ... + \Gamma_{\alpha\beta}^{\mu_k}\tensor{T}{^{\mu_1}^{...}^{\beta}_{\nu_1}_{...}_{\nu_l}} \nn \\ 
		&\quad - \Gamma_{\alpha\nu_1}^{\beta}\tensor{T}{^{\mu_1}^{...}^{\mu_k}_{\beta}_{...}_{\nu_l}} - ... - \Gamma_{\alpha\nu_l}^{\beta}\tensor{T}{^{\mu_1}^{...}^{\mu_k}_{\nu_1}_{...}_{\beta}} \ , \nn
	\end{align}
where $\Gamma$ is the Levi-Civita connection.} $\xi^\mu$ is the spacelike deviation vector pointing from one geodesic to the adjacent one, and $u\mup$ the four-velocity of the observer. In the rest frame of one of the observers, written in terms of the metric perturbation this is
\begin{equation}
	\frac{d^2 \xi_i}{dt^2} = \frac{1}{2}\ddot{h}\ttu\ijd \xi^j \ .
\end{equation}
Integrating over time $t$ twice and assuming that the changes in $\xi^i$ are tiny compared to $\xi^i$ itself, we get
\begin{equation}\label{memoryeq}
	\Delta \xi_i \approx \frac{1}{2}\Delta h\ttu\ijd \xi^j \ .
\end{equation}
Thus, given a process that generates a GW strain with different values in the asymptotic past and future, the distance between test masses will be permanently shifted, i.e., there is a memory effect. 

\subsection{Black hole binary system as a GW source with memory}
A well-known source for GWs with memory is a binary black hole (BH) system \cite{Wiseman:1991ss,Favata:2009ii}. We shall in this section summarize equations for second order radiation, GW induced by radiation from the binary and the Minimal Waveform Model (MWM), designed to characterize the final merger of the binary.

For simplicity, consider an equal-mass BH binary with component mass $M$, in a quasicircular orbit with radius $R(t)$. We choose a coordinate system where the center of mass of the binary is located at the origin and the binary rotates in the $x-y$ plane, as in Fig. \ref{coordsyst}. The binary loses energy in the form of gravitational radiation, which makes the orbit shrink gradually. The smaller the orbital radius, the greater the power of radiation is; hence, the system is in a runaway process where the component masses are on a collision course. 
%the radiation luminosity peaks at the time of coalescence. 
Taking the energy of radiation to be a further source of GWs, the associated stress-energy is described by the Isaacson formula \cite{Isaacson:1968zza}
\begin{equation}\label{isaacsonflat}
	t\ijd = \frac{1}{32 \pi \GN} \big\langle \pt_i h^{TT}_{kl} \pt_j h_{TT}^{kl} \big\rangle \ ,
\end{equation}
where the angular brackets denotes average over several cycles. The first order GW inside the brackets is given in the quadrupole approximation by the quadrupole formula \cite{Misner:1974qy}
\begin{equation}\label{quadformula}
	h\ijd\ttu(t,{\bf r}) \approx \frac{2 G}{r}\ddot{Q}\ijd\ttu(t-r) \ ,
\end{equation}
where $Q\ijd$ is the quadrupole moment of the source evaluated at retarded time $t-r$, defined as the trace-free part of the mass moment $M\iju$:
\begin{equation}\label{quaddef}
	M\iju = \int d^3 x\,x^i x^j T^{00}, \quad Q\iju = M\iju - \frac{1}{3}\delta\iju \Tr M\iju \ .
\end{equation}
Concretely, the mass moment for the binary is:
\begin{equation}\label{quad}
	\left(M\iju\right) =
	\left(
	\begin{array}{ccc}
		2 M R^2 \cos ^2(\Omega t) & M R^2 \sin (2 \Omega t) & 0 \\
		M R^2 \sin (2 \Omega t) & 2 M R^2 \sin ^2(\Omega t) & 0 \\
		0 & 0 & 0 \\
	\end{array}
	\right) \ ,
\end{equation}
where $R$ is the physical orbital radius (n.b. not the separation between the bodies), and $\Omega$ is the angular frequency.
The TT projection in \nr{quadformula} is done with respect to the direction ${\bf r}$ with the Lambda tensor, defined by
\begin{equation}\label{ttdef}
	X\ttu\ijd = \Lambda_{ij,}^{\,\,\,\, kl}({\bf n})X_{kl}, \quad \tensor{\Lambda}{_i_j_,^k^l}({\bf n}) = P_{ik}P_{jl}- \frac{1}{2}P\ijd P_{kl}, \quad P\ijd = \delta\ijd - n_i n_j,\quad {\bf n} = {\bf r}/\abs{{\bf r}} \ ,
\end{equation}
where $X$ is any symmetric tensor. We may then change the spatial derivatives in \nr{isaacsonflat} to time derivatives using the retarded time dependence in \nr{quadformula}, discard terms that are higher order in $1/r$, and write
\begin{equation}\label{tijflat}
	t\ijd \approx \frac{n_i n_j}{r^2} \frac{dL}{d\Omega} \ ,
\end{equation}
where $dL/d\Omega$ is the directional luminosity of radiation: 
%and $n_i$ the unit radial vector pointing in the direction where $t\ijd$ is evaluated. 
%For a quasi-circular equal-mass binary system, the angle-dependent luminosity is
\begin{equation}\label{isaacson}
	\frac{dL}{d\Omega}(t,r,\Omega) = \frac{r^2}{32 \pi \GN} \big\langle \pt_t h^{TT}\ijd \pt_t h_{TT}\iju \big\rangle (t,r,\Omega) \approx \frac{\GN}{8\pi}\big\langle \dddot{Q}\ijd^{TT} \dddot{Q}\iju_{TT} \big\rangle (t-r,\Omega) \ ,
\end{equation}
Gravitational waves generated by the source \nr{tijflat} are then given by
\begin{equation}
	h^{TT}\ijd = 4 \GN \int d^4 x' \frac{\delta (t-t'-\abs{{\bf x}-{\bf x'}})}{\abs{{\bf x}-{\bf x'}}} \frac{(n_i' n_j')^{TT}}{r'^2} \frac{dL}{d\Omega'} \ ,
\end{equation}
where the TT projection is now performed with the Lambda tensor with respect to the position vector of the observer;\footnote{As pointed out in \cite{Maggiore:2007ulw}, strictly speaking the TT projection can only be performed in the absence of stress-energy, which is of course possible when the source is compact. In our case, however, the source is not compact; radiation spreads out all over the 3d space. Nevertheless, the radiation energy density falls off like $r^{-2}$ so a vacuum approximation is reasonable at large distances from the binary source. The TT gauge choice is hence viable in the wave zone of the first order source.} notice that the above formula involves two TT projections: one depending on the integration angle, the other in the direction of the observer. The first step is to write the integral in spherical coordinates and introduce a retarded time coordinate $u' = t' - r'$ by inserting a delta function inside the integral by $1 = \int du'\delta(u' - \eta' + r')$, and then perform the $t'$-integral:
\begin{equation}
	h^{TT}\ijd = 4 \GN \int d\Omega' du' d r' \frac{\delta(t-u'-r'-\abs{{\bf x}-{\bf x'}})}{\abs{{\bf x}-{\bf x'}}} (n_i' n_j')^{TT} \frac{dL}{d\Omega'}(u',\Omega') \ .
\end{equation}
We can now get rid of the final delta function by doing the $r'$-integral. This rids us of the delta functions while constraining the values of $u'$ and $r'$ to be such that
\begin{equation}
	r' =  \frac{(t - u')^2 - r^2}{2(t - u' - r \cos \theta_{xx'})} \equiv r'_0, \quad t' = u' + r'_0 \ ,
\end{equation}
where $\theta_{xx'}$ is the angle between ${\bf x}$ and ${\bf x'}$, and $r'_0$ is the positive root of $t-t'-\abs{{\bf x - x'}}=0$.
Furthermore, because of delta function identities the inverse distance factor gets replaced by
\begin{equation}
	\frac{1}{\abs{{\bf x - x'}}} \rightarrow \frac{\theta(t-r - u')}{t - u' - r \cos \theta_{xx'}} \ ,
\end{equation}
where the factor $\theta(t - r - u')$ comes from the fact that $r'$-integral was defined on the interval $(0,\infty)$. The upper limit for the $u'$ integral therefore is $t - r$ and lower limit can be taken to $-\infty$. We extract a $1/r$ factor from the integrand, and get
\begin{equation}
	h^{TT}\ijd = \frac{4 \GN}{r} \int d\Omega' \int^{t-r}_{-\infty} du' \frac{(n_i' n_j')^{TT}}{\frac{u - u'}{r} + 1 - \cos \theta_{xx'}}  \frac{dL}{d\Omega'}(u',\Omega') \ ,
\end{equation}
where $u = t - r$ is retarded time. If we assume that the GW burst duration is very small compared to the astronomical distance $r$ that the burst traverses, we may approximate $(u-u')/r \approx 0$, and the integral simply becomes
\begin{equation}\label{basiceq}
	h^{TT}\ijd \approx \frac{4 \GN}{r} \int d\Omega' \int^{t-r}_{-\infty}du' \frac{(n_i' n_j')^{TT}}{1 - \cos \theta_{xx'}}   \frac{dL}{d\Omega'}(u',\Omega') \ .
\end{equation}
The angle-dependent luminosity in the above integral is given by \nr{isaacson} where we have a contraction of TT projected quadrupole moment tensors. Given the direction of GW propagation ${\bf n}$, the contraction can be written as
\begin{equation}\label{quadrutt}
	\dddot{Q}\ijd^{TT} \dddot{Q}\iju_{TT} = \Lambda_{ij,kl}\dddot{Q}\ijd \dddot{Q}_{kl} = \dddot{Q}\ijd \dddot{Q}\iju - 2 \dddot{Q}\ijd \dddot{Q}^{ik}n^j n_k + \frac{1}{2}\dddot{Q}\ijd \dddot{Q}_{kl} n^i n^j n^k n^l \ .
\end{equation}
The standard Keplerian relation
\begin{equation}
	\Omega = \left({\GN M \over 4 R^3}\right)^{1/2}
\end{equation}
between the angular frequency, orbital radius, and the mass can be employed to write the directional luminosity as
\begin{equation}\label{inspL}
	\frac{dL}{d\Omega}= \frac{\GN}{8\pi} \big\langle\dddot{Q}\ijd \dddot{Q}\iju - 2 \dddot{Q}\ijd \dddot{Q}^{ik}n^j n_k + \frac{1}{2}\dddot{Q}\ijd \dddot{Q}_{kl} n^i n^j n^k n^l\big\rangle = \frac{1}{32\pi \GN} \left(\frac{r_S}{2 R}\right)^5 \left(1 + 6 \cos^2 \theta + \cos^4 \theta \right)
\end{equation}
where $r_S = 2 G M$ is the Schwarzschild radius of the black hole and $\theta$ is the angle between the normal vector to the orbital plane and the observer's position vector. The angular structure will appear several times in later computations so define the shorthand notation
\begin{equation}\label{Fdef}
	\mathcal{F}(\cos\theta) = \big\langle\dddot{Q}\ijd \dddot{Q}\iju - 2 \dddot{Q}\ijd \dddot{Q}^{ik}n^j n_k + \frac{1}{2}\dddot{Q}\ijd \dddot{Q}_{kl} n^i n^j n^k n^l\big\rangle /\left(\frac{G^3 M^5}{2 R^5}\right) = \frac{1}{2}\left(1 + 6 \cos^2 \theta + \cos^4 \theta \right) \ .
\end{equation}
The total luminosity $L$ is obtained by integrating \nr{inspL} over all angles:
\begin{equation}
	L = \int d\Omega \frac{dL}{d\Omega} = \frac{2 c^5}{5 G}\left(\frac{r_S}{2 R}\right)^5 \ ,
\end{equation}
where we have reinstated the speed of light $c$ for a moment. Note that $c^5 /\GN$ is the Planck luminosity \cite{Cardoso:2018nkg}.
Equating the energy loss of the binary and the total GW luminosity as $dE/dt = - L$, where energy $E$ is the sum of kinetic and potential energy, yields a differential equation for the orbital radius $R(t)$ \cite{Peters:1964zz}:
\begin{equation}
	\dot{R} = - \frac{1}{5}\left(\frac{r_S}{R}\right)^3,
\end{equation}
which has the solution
\begin{equation}\label{radius}
	R(t) = r_S \left[ \frac{4}{5} \frac{t_\rmi{coal} - t}{r_S}\right]^{1/4} = R_i \left[1-\frac{t-t_i}{\tau}\right]^{1/4}, \quad R_i = r_S \left(\frac{4}{5}\frac{\tau}{r_S}\right)^{1/4} \ ,
\end{equation}
where $t_\rmi{coal}$ is the moment of coalescence, $t_i$ is the initial moment of the binary and $R_i$ its initial radius, and $\tau$ is the binary lifetime. 

Coming back to the integral (\ref{basiceq}), we see that it can be computed analytically for a binary inspiral. The time-dependent and angle-dependent factors can be neatly separated, which allows us to factorize the integral into the time integral and the $S^2$ integral. Substituting \nr{quadrutt} in the $S^2$ integral, we get terms with 2, 4, and 6 $n_i$'s. The $S^2$ integral can thus be decomposed into terms of the form
\begin{equation}
	\int d\Omega \frac{n_{i_1}...n_{i_k}}{1 - {\bf\hat{x}}\cdot {\bf n}} \ ,
\end{equation}
where $k$ takes the values 2, 4, and 6. Integrals of this type can be worked out by the method described in Appendix \ref{angularintegrals}. We will run across with this kind of integrals also later in Sec. \ref{radbyrad}. The result of angular integration in \nr{basiceq} is, after implementing the integration techniques of Appendix \ref{angularintegrals},
\begin{equation}
	4 G \int d\Omega' (n_i' n_j')\ttu \frac{dL}{d\Omega} = \frac{G^5 M^5}{R^5} \frac{1}{60} (17 + \cos^2\theta_x)\sin^2\theta_x \, \sqrt{2}e^+\ijd \ .
\end{equation}
Then in \nr{basiceq} there is just left the time integral, which gets contributions from the entire history of the binary system. Using the equation for the orbital radius (\ref{radius}), we may compute the inspiral part of the time integral:
\begin{equation}
	\int_{-\infty}^{u} du' \frac{G^5 M^5}{R^5} = \frac{5}{8}\frac{G^2 M^2}{R} \ .
\end{equation}
Combining these results, the formula for the memory strain cumulated during the inspiral boils down to
\begin{equation}\label{inspmem}
	h_{+}^\rmi{insp} = \sqrt{2}\frac{r_S^2}{2 r R} \frac{1}{192} (17 + \cos^2 \theta_x ) \sin^2 \theta_x \ ,
\end{equation}
where $\theta_x$ is the inclination angle of the observer position vector, see Fig. \ref{coordsyst}. Remarkably, the dominant factors are just the same as those in the first order metric perturbation. The numerical factor reduces the effect by about one order of magnitude.

\begin{figure}[!t]
	\begin{center}
		\includegraphics[width=0.7\textwidth]{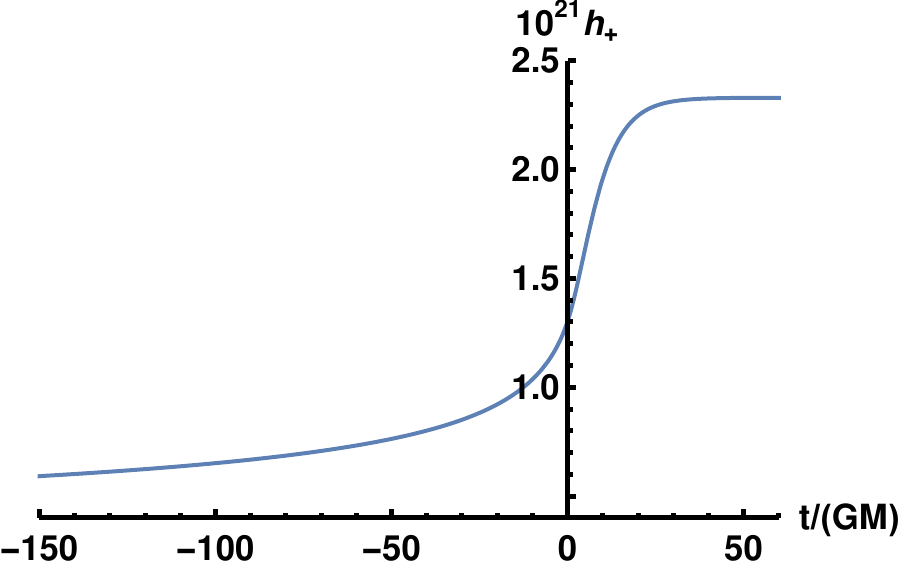}
	\end{center}
	\caption{\small Accumulation of the memory strain during the late inspiral and the ringdown in MWM for an equal-mass face-off ($\theta_x = \pi/2$) binary system at distance $r=10^8$ lightyears, with single BH mass $M =10 M_\odot$. On the $x$-axis the time is given in the units of $G M$. The origin of the time coordinate is set to the matching time between different phases.
	}\label{flatmemplot}
\end{figure}

The above result for the memory is divergent when $t\rightarrow t_\rmi{coal}$. To get a more realistic description of the memory waveform close to $t\coal$, one needs to introduce a cutoff radius and a corresponding matching time $t_\rmi{m}$ where one switches to a different approximation. A simple analytic model that does the job is the Minimal Waveform Model (MWM) \cite{Favata:2009ii} where one glues the inspiral waveform to the ringdown waveform given by a quasi-normal mode (QNM) expansion. The expansion is done for the complex strain built from the GW polarizations, typically written in terms of spin-weighted spherical harmonics as
\begin{equation}\label{complexstrain}
	h_+ - i h_\times = \frac{G M_f}{r} \sum_{l=2}^{\infty}\sum_{m=-l}^{l} h_{lm} \, {}_{-2} Y_{lm} \ .
\end{equation}
Here $M_f$ is the final BH mass and the GW mode functions $h_{lm}$ are
\begin{equation}
	h_{lm} = \sum_{n} A_{lmn} e^{-\sigma_{lmn}(t-t_\rmi{m})/(G M_f)} \ ,
\end{equation}
where $\sigma_{lmn}=i\omega_{lmn} + \tau^{-1}_{lmn}$ are the QNMs, $\omega_{lmn}$ being the QNM frequency and $\tau^{-1}_{lmn}$ the damping time, and the sum is over $n$ ($n=0$ corresponding to the fundamental mode and $n>0$ to overtones).
%Gravitational radiation is dominated by the $l=\abs{m}=2$ mode \cite{Cotesta:2022pci} \ack{note about m=-2: only m=2 if plus polarized radiation}
Similarly to \cite{Favata:2009ii}, we approximate the ringdown waveform by
\begin{equation}\label{QNMappro}
	h_{22} = \sum_{n = 0}^{n_\rmi{max}}  A_{22n} e^{-\sigma_{22n}(t - t_\rmi{m})/(G M_f) } \ .
\end{equation}
For the systems we have in mind the final Kerr BH mass $M_f$ is roughly $M_f \approx 0.95 M_\rmi{tot} = 1.9 M$ and the final spin $a_f \approx 0.7$ \cite{Baker:2008mj}. We choose $n_\rmi{max} = 1$ and perform the matching so that the time profile of the GW luminosity and its first and second time derivatives are continuous at $t_\rmi{m}$, which fixes the unknown factors $A_{22n}$ up to a phase. 
%(Here we deviate slightly from the treatment in \cite{Favata:2009ii} where matcing was done for the modes $h_{2\pm2}$.) 
The QNM frequencies can be looked up from, e.g., \cite{Berti:2005ys}. For an extensive review on the theory and applications of QNM's, see \cite{Berti:2009kk}. A reasonable choice for the matching radius is the radius of the Innermost Stable Circular Orbit (ISCO), which equals the Schwarzschild photon sphere radius of the component masses: $R_\rmi{m} = \frac{3}{2} r_S = 3 G M$. The corresponding matching time $t_\rmi{m}$ satisfies
\begin{equation}
	t_\rmi{coal} - t_\rmi{m} = \frac{405}{32} \GN M = \frac{405}{64} r_S \ .
\end{equation}
From \nr{isaacson} and \nr{QNMappro} we get that the ringdown luminosity is
\begin{equation}\label{Lring}
	\left(\frac{dL}{d\Omega}\right)_\rmi{ring} \approx \frac{1}{48\pi G}\Lambda_{ij,kl} \mathcal{Y}^{22}\ijd \mathcal{Y}^{22*}_{kl}\sum_{n,n'}  A_{22n}A^*_{22n'}\sigma_{22n}\sigma^*_{22n'} e^{-(\sigma_{22n}+\sigma^*_{22n'})(t-t_\rmi{m})/(G M_f)},
\end{equation}
where the symmetric trace-free basis tensors $\mathcal{Y}^{2m}\ijd$ and their relation to spin-2 spherical harmonics are given in Appendix \ref{angularintegrals}. The factor in front of the sum
actually gives the same angular dependence as in \nr{inspL}.
We require that
\begin{equation}\label{matchingcond}
	\left(\frac{dL}{d\Omega}\right)^{(p)}_\rmi{insp}(t_\rmi{m}) = \left(\frac{dL}{d\Omega}\right)^{(p)}_\rmi{ring}(t_\rmi{m}) \ ,
	%\frac{M}{M_f}\sum_{p=0}^{n_\rmi{max}} \frac{2}{3^{p+1}(\sqrt{3} r_S)^{p}} (V^{-1})_{p,n} (-1)^p\frac{d^p}{dx^p}\cos x \big\rvert_{x=\Phi(t_m)},
\end{equation}
for $p=0,1,2$, $(p)$ denotes the $p$th time derivative, and $(dL/d\Omega)_\rmi{insp}$ is given by \nr{inspL} and \nr{radius}. The complex matching constants are then solved from this numerically. We write
\begin{equation}
	A_{22n} = \sqrt{\frac{96\pi}{15}}\mathcal{A}_{22n}e^{i \phi_{22n}},\quad \Delta\phi = \phi_{220}-\phi_{221} \ ,
\end{equation}
where all the constants on the rhs are real and the numerical factors make the overall factor in \nr{Lring} similar to the one in (\ref{inspL}) by canceling numerical constants in $\mathcal{Y}^{22}\ijd$'s. We may then determine the moduli $\mathcal{A}_{22n}$ and the phase difference $\Delta\phi$ from the three conditions in \nr{matchingcond}. 
%For the matching radius given by the photon sphere radius and the ratio $M_f/M = 1.9$, the numeric values are
%\begin{equation}
%	\mathcal{A}_{220} = 1.269, \quad \mathcal{A}_{221} = 0.863, \quad \Delta\phi = 3.157 
%\end{equation}
%\ack{mod 2pi?}.
%\ack{Make sure these are correct}
%where the Vandermonde type matrix $V_{p,n}$ is defined by
%\begin{equation}
%	V_{p,n} = (\sigma_{22n})^p, \quad p,n = 0,1,2.
%\end{equation}
The ringdown contribution to the memory signal becomes
\begin{align}\label{ringmem}
	h_{+}^\rmi{ring} &= h_{+}^\rmi{insp}(t_\rmi{m}) + \frac{\GN M_f}{60r}  \sum_{n,m=0}^{n_\rmi{max}} \frac{\sigma_{22n}\sigma_{22m}^* \mathcal{A}_{22n} \mathcal{A}^*_{22m}}{\sigma_{22n}+\sigma_{22m}^*}e^{i(\phi_{22n}-\phi_{22m})}\left(1 - e^{-(\sigma_{22n}+\sigma_{22m}^*)\frac{t-t_\rmi{m}}{G M_f}}\right) \zeta(\theta_x) \nn \\
	&= \frac{G M}{r}\left[\frac{1}{288} + \frac{1}{60} \frac{M_f}{M}\sum_{n,m=0}^{n_\rmi{max}} \frac{\sigma_{22n}\sigma_{22m}^* \mathcal{A}_{22n} \mathcal{A}^*_{22m}}{\sigma_{22n}+\sigma_{22m}^*}e^{i(\phi_{22n}-\phi_{22m})}\left(1 - e^{-(\sigma_{22n}+\sigma_{22m}^*)\frac{t-t_\rmi{m}}{G M_f}}\right) \right] \zeta(\theta_x) \ ,
\end{align}
where
\begin{equation}
	\zeta(\theta_x) = \sqrt{2}(17+\cos^2\theta_x)\sin^2\theta_x \ .
\end{equation}
Also here only the plus polarized mode is present in the signal. 
In Fig. \ref{flatmemplot} we have the memory strain computed from MWM with black holes of mass $10 M_\odot$ and distance to the observer $r = 10^8$ lightyears. As we can see, the strain asymptotes to a non-vanishing constant value; by Eq. \nr{memoryeq}, we therefore have a displacement that remains after the GW burst, a memory effect.

%%%%%%%%%%%%%%%%%%%%%%%%%%%
\section{Gravitational waves in cosmological background}\label{perttheory}
%%%%%%%%%%%%%%%%%%%%%%%%%%%
We shall now proceed to extend the previous discussion of gravitational radiation in flat spacetime to the curved FRW metric appropriate for the concordance model, also known as the $\Lambda$CDM model. For our purposes, studying tail memory, it is a good approximation of reality. Treating the matter- and the cosmological constant -dominated cases separately is not new, but we observe that a joint discussion will lead to a somewhat surprising approximate symmetry in the equations of motion. The fact that we are in a curved spacetime will also lead to the appearance of a new component of the solution: the tail. One expects that its magnitude will be very small, but it is there and it is useful to know how small it is.

\subsection{Linearized equations of motion}
%%%%%%%%%%%%%%%%%%%%%%%%%%%
The line element of the FRW universe in conformal coordinates is
\begin{equation}
	ds^2 = a^2(\eta)\left(-d\eta^2 + \delta\ijd dx^i dx^j\right) \ ,
\end{equation}
where the spatial part is in Cartesian coordinates. The time coordinate here is conformal time, defined in terms of cosmological time by
\begin{equation}
	d\eta = \frac{dt}{a(t)} \ .
\end{equation}
We write the perturbed FRW metric as
\begin{equation}
	ds^2 = a(\eta)^2 \left(\eta\munud + h\munud \right)dx\mup dx\nup \ .
\end{equation}
The perturbed Einstein equation is
\begin{equation}
	\delta G\munud = 8\pi G \delta T\munud \ .
\end{equation}
This involves equations for the $h_{\eta\eta}, h_{\eta i},$ and $h\ijd$ components of the metric perturbation. We are only interested here in GW solutions so we focus on the tensor sector and restrict our attention to the gauge-invariant TT perturbation $h\ttu\ijd$.
The linearized equation of motion for the tensor perturbation is (see Appendix \ref{perturb})
\begin{equation}\label{linearized}
	\left(
	\Box -2 \frac{a'}{a} \pt_\eta
	\right) h\ijd\ttu = -16\pi G T\ijd\ttu \ ,
\end{equation}
where $\Box$ is the Minkowski space d'Alembert operator $\eta\munuu\pt\mud\pt\nud$ and where we simply write $T\ijd\ttu$ for the TT part of the stress-energy source. $T\ijd$ could be any form of stress-energy perturbation but we are here only interested in stress-energy of a compact binary system and, in particular, stress-energy of GWs emitted from such a system. The first order term in \nr{linearized} is responsible for the tail effects that we analyze below.
Note that for each tensor component, \nr{linearized} takes the form of a scalar inhomogeneous wave equation in the FRW universe, which we can solve for each component separately. The recipe for solving this equation when cosmic expansion is given by a power-law $a(\eta) = C \eta^\alpha, \alpha = -1, 2$ (i.e., vacuum-dominated and matter-dominated universes, respectively) is given in \cite{Burko:2002ge,Poisson:2011nh}, see also \cite{deVega:1998ia}. We set out to do this in \lam CDM. With a field redefinition $\psi\ijd = a h\ijd\ttu$, the first order term is eliminated and the equation becomes
\begin{equation}
	\left(
	\Box + \frac{a''}{a}
	\right)\psi\ijd = -16\pi G a T\ijd\ttu \ .
\end{equation}
%The Green's function for the covariant wave operator is the solution to the equation
%\begin{equation}\label{greeneq1}
%	\na_\alpha \na^\alpha G(x,x') = -4\pi \frac{\delta^{(4)}(x-x')}{\sqrt{-\det (g\munud)}}.
%\end{equation}
%The metric determinant in the denominator is needed to turn the scalar density Dirac delta into a proper coordinate invariant quantity. Analogously with the above field redefinition, we write
%\begin{equation}
%	g(x,x') := a(\eta)a(\eta') G(x,x').
%\end{equation}
%With the new Green's function, the Green's equation now becomes
We need to find Green's function $g(x,x')$ for the differential operator appering on the lhs, Green's equation for which reads
\begin{equation}\label{greeneq2}
	\left(
	\Box + V
	\right)g(x,x') = -4\pi \delta^{(4)}(x-x'), \quad V \equiv \frac{a''}{a} \ .
\end{equation}
%Note the lack of $\sqrt{-\det (g\munud)}$ above; the Green's equation is now for an operator and the corresponding Green's function only defined in a specific coordinate system. 
It is a well-established mathematical fact that a partial differential equation of type \nr{greeneq2} is solved by a Hadamard Ansatz\cite{Friedlander:2010eqa}:
\begin{equation}\label{hadamard}
	g(x,x') = \frac{\delta(u)}{\abs{{\bf x - x'}}} + B(x,x')\theta(u), \quad u = \eta - \eta' - \abs{{\bf x - x'}} \ .
\end{equation}
This Green's function consists of two pieces: the familiar flat-spacetime d'Alembert operator's Green's function involving a delta function that forces the signal to propagate along light cone from $x'$ to $x$, and a second piece that basically tells that there is a signal present also in case that $x$ lies inside the future light cone of $x'$; this is the so-called tail. The two-point function $B(x,x')$ determines the strength of the tail signal, and the step function dictates that the signal travel over timelike separation only.

Feeding the ansatz \nr{hadamard} into the Green's equation \nr{greeneq2} yields two equations for $B$, one of which is the main evolution equation and the other one a boundary condition on the null cone:
\begin{align}
	&\left(
	\Box + V
	\right) B = 0, &\eta - \eta' > \abs{{\bf x - x'}}, \label{Beq}\\
	&\left(x-x'\right)\mup \pt\mud B + B - \frac{1}{2} V = 0,  &\eta - \eta' = \abs{{\bf x - x'}} \ . \label{bdryeq}
\end{align}
The boundary condition can be integrated to give
\begin{equation}\label{bdrysol}
	B(x,x')\Big\lvert_{\abs{{\bf x - x'}}=\eta-\eta'} = \frac{1}{2(\eta-\eta')} \int_{\eta'}^{\eta} V(\zeta)d\zeta \ ,
\end{equation}
meaning that the general solution of \nr{Beq} must reduce to \nr{bdrysol} when points $x' = (\eta',{\bf x'})$ and $x = (\eta,{\bf x})$ become null separated. Note that since the potential $V = a''/a$ does not depend on spatial position, \nr{Beq} can be written in the form of a vanishing divergence:
\begin{equation}\label{conserv}
	\pt\mup \left(a\,  \pt\mud B - B \pt\mud a\right) = 0 \ .
\end{equation}
Interestingly, the above equation looks like a conservation law, but it is not clear what would be the associated Lagrangian and the symmetry thereof. More details on this equation can be found in Appendix \ref{Bapp}.

In a homogeneous and isotropic universe we can restrict the form of the position-dependence of two-point function $B(x,x')$ when we specialize to any Gaussian normal coordinate system adapted to the homogeneous and isotropic Cauchy slices, in our case the conformal coordinates $(\eta,{\bf x})$. The value of $B(x,x')$ cannot depend on the particular spatial coordinates of the two points; therefore $B(x,x')$ can only depend on the length of the relative position vector ${\bf x - x'}$:
\begin{equation}
	B(x,x') = B(\eta,\eta',\abs{{\bf x - x'}}) \ .
\end{equation}

\subsection{Solving for the tail part of the Green's function}
So far the only information we have fed in has been that the background is a spatially flat FRW universe. We will specialize to a \lam CDM universe and solve the tail part of the Green's function determined by that background. To numerically solve for the tail two-point function $B$ from the Green's equation we first define the Fourier modes by
\begin{equation}
	g(x,x') =\int  \frac{d^3 k}{(2\pi)^3} \, \tilde{g}(\eta,\eta',{\bf k}) e^{i {\bf k \cdot (x-x')}}=
	\frac{1}{2\pi^2 \rho}\int_0^\infty  \,dk\,k \sin k \rho \, \tilde{g}(\eta,\eta',k) \ ,\quad \rho\equiv \abs{{\bf x-x'}} \ ,
\end{equation}
where again, due to homogeneity and isotropy of the background, $g$ can only depend on ${\bf x}$ and ${\bf x'}$ through the modulus $\abs{{\bf x - x'}}$; hence $\tilde{g}$ only depends on the wave vector length $k$. % This allows us to do the angular integration, whereby we are only left with the integral over $k$:
In terms of these spatial Fourier modes,  the Hadamard Ansatz
\be
g(x,x')=g(\eta,\eta',\rho)={\delta(\eta-\eta'-\rho)\over \rho}+B(\eta,\eta',\rho)\theta(\eta-\eta'-\rho) \ , 
\ee
becomes
\be
\tilde g(\eta,\eta',k)=\int d^3x \, e^{-i{\bf k}\cdot\bfx}g(\eta,\eta',\rho)\equiv  \tilde g_\rmi{LC}(\eta,\eta',k)+\tilde g_\rmi{tail}(\eta,\eta',k) \ ,
\ee
where the light cone (LC) and tail parts are
\be\label{fourmodes}
\tilde g_\rmi{LC}(\eta,\eta',k)={4\pi\over k}\sin[(\eta-\eta')k]\theta(\eta-\eta') \ ,\quad  \tilde g_\rmi{tail}(\eta,\eta',k)={4\pi\over k}\int_0^{\eta-\eta'} d\rho \,\rho\,\sin(k\rho)B(\eta,\eta',\rho) \ .
\ee
In terms of Fourier modes the main equation $(\square+V)g(x,x')=-4\pi\delta^{(4)}(x-x')$  becomes
\be
\partial_\eta^2\tilde g(\eta,\eta',k)+(k^2-V(\eta))\tilde g(\eta,\eta',k)=4\pi\delta(\eta-\eta') \ .
\ee
Using the explicit form \nr{fourmodes} one computes (remembering that $f(x)\delta'(x)=f(0)\delta'(x)-f'(0)\delta(x)$)
\be
\partial_\eta^2\tilde g_\rmi{LC}(\eta,\eta',k)+k^2\tilde g_\rmi{LC}(\eta,\eta',k)=4\pi\delta(\eta-\eta')
\ee
so that the tail part should satify the inhomogeneous equation
\be
\partial_\eta^2\tilde g_\rmi{tail}(\eta,\eta',k)+(k^2-V(\eta))\tilde g_\rmi{tail}(\eta,\eta',k)= V(\eta)\tilde g_\rmi{LC}(\eta-\eta',k) \ .
\label{inhomeq}
\ee
The $4\pi\delta(\eta-\eta')$ term in the rhs of the main equation is entirely taken care of by the LC term. For numerical solution we need the initial values of the function and its derivative at some point. This point is naturally $\eta=\eta'$ and from above one can explicitly verify that
\be
\tilde g(\eta,k)=0 \ ,\quad \lim_{\eta \rightarrow \eta'+}\partial_\eta\tilde g(\eta,k)= \lim_{\eta \rightarrow \eta'+}\partial_\eta\tilde g_\rmi{LC}(\eta,k)=4\pi \ .
\ee
The first derivative comes entirely from $\tilde g_\rmi{LC}$. When solving Eq.\nr{inhomeq} numerically for $\tilde g_\rmi{tail}(\eta,\eta',k)$, we should then impose the initial conditions
\be
\tilde g_\rmi{tail}(\eta=\eta',k)=0 \ ,\quad \partial_\eta\tilde g_\rmi{tail}(\eta=\eta',k)=0
\label{initcond}
\ee
and the tail $B(\eta,\eta',\rho)$ is determined by inversion:
\be
B(\eta,\eta',\rho)={1\over 2\pi^2\rho}\int_0^\infty  \,dk\,k \sin k\rho \, \tilde g_\rmi{tail}(\eta,\eta',k) \ .
\label{modeinv}
\ee

The relation between the inhomogeneus equation for $\tilde g_\rmi{tail}$ and the homogeneous equation for $B$ can be made manifest by computing $\partial_\eta^2\,\tilde g_\rmi{tail}(\eta,\eta',k)$ from the representation \nr{fourmodes} and using the fact that $\partial_\eta^2\,B=(\vec{\nabla}^2+V)B$.  Contributions on $k=\eta-\eta'$ arise from derivatives with respect to the upper limit and from partial integrations when evaluating the effect of $\vec{\nabla}^2$.  One finds that
\be
\partial_\eta^2\,\tilde g_\rmi{tail}=(-k^2+V)\tilde g_\rmi{tail}+\tilde g_\rmi{LC}\,2(\rho\partial_\eta B+\rho\partial_\rho B+B)_{\rho=\eta-\eta'} \ .
\ee
Satisfying the inhomogeneous equation \nr{inhomeq} then implies that, on the LC,
\be
(\rho\partial_\eta B(\eta,\eta',\rho)+\rho\partial_\rho B(\eta,\eta',\rho)+B)_{\rho=\eta-\eta'}-{1\over2}V=0 \ .
\ee

\subsection{Numerically computed tail for the concordance model}
The above is for a general potential $V(\eta)$. We now specialize to the concordance model, analysed in detail in Appendix \ref{concordance}. Note first that for a purely matter-dominated expansion $a(\eta)\sim \eta^2$ we have $V(\eta)=a''(\eta)/a(\eta)=2/\eta^2$ and the full solution is
\be
g_m(\eta,\eta',\rho)={1\over \rho}\delta(\eta-\eta'-\rho)+{1\over\eta\eta'}\theta(\eta-\eta'-\rho) \ .
\ee
For purely exponential expansion $a(t)=e^{Ht}$, normalising conformal time by $\eta(t=0)=0$, we have $\eta=H^{-1}(1-e^{-Ht})$ and $a(\eta)=1/(1-H\eta)$ so that
\be
V_\Lambda(\eta)={2\over (\eta_\rmi{max}-\eta)^2} \ ,\quad \eta_\rmi{max}={1\over H} \ ,
\label{msol}
\ee
and the solution is
\be
g_\Lambda(\eta,\eta',\rho)={1\over \rho}\delta(\eta-\eta'-\rho)+{1\over(\eta_\rmi{max}-\eta)(\eta_\rmi{max}-\eta')}\theta(\eta-\eta'-\rho) \ .
\label{dSsol}
\ee

\begin{figure}
	\begin{center}
		\includegraphics[width=1.05\textwidth]{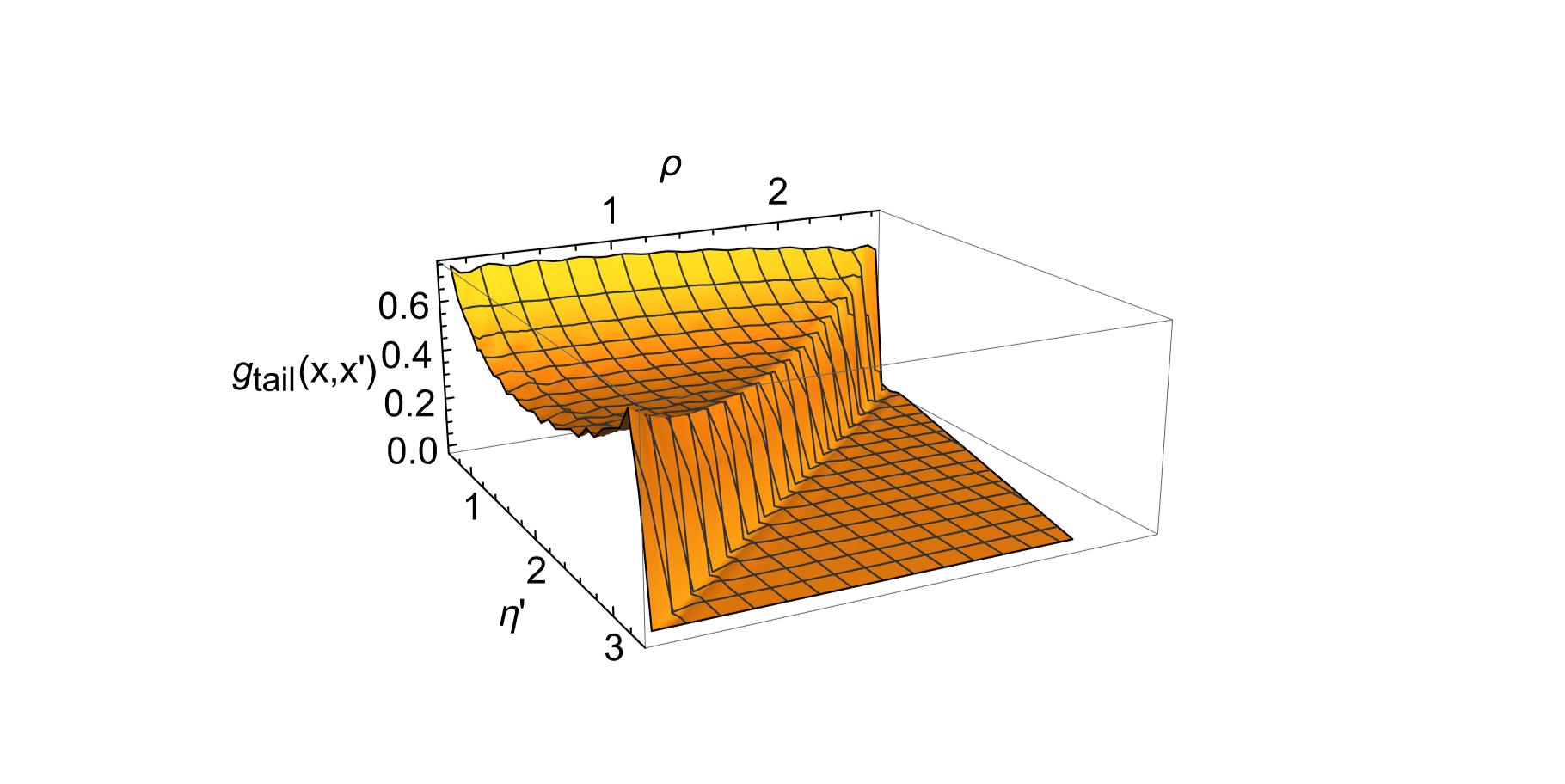}
	\end{center}
	\caption{\small The numerical solution to the part of the Green's function from which the delta function singularity is extracted away. The solution exhibits the step function structure as foreseen in the ansatz \nr{hadamard}: the tail part of the Green's function indeed is $B(\eta,\eta',\rho )\theta(\eta-\eta'-\rho)$, $\rho \equiv \abs{{\bf x - x'}}$, where $B(\eta,\eta',\rho)$ with $\eta = 3.3$ (present day) is responsible for the non-trivial behavior inside the light cone in the above figure. Minor bumps and slightly tilted walls in the plot are due to a finite cutoff of the numerical integral over the Fourier space radial coordinate.
	}\label{greenB}
\end{figure}

%In view of Eq. (\ref{hadamard}), we first separate the flat-space Green's function Fourier mode from the Fourier transform $\tilde{g}$: 
%\begin{equation}\label{extractdelta}
%	\tilde{g} = \left(\tilde{g} - \frac{4\pi}{k} \sin k (\eta-\eta') \right) + \frac{4\pi}{k} \sin k (\eta-\eta') \equiv \hat{g} + \frac{4\pi}{k} \sin k (\eta-\eta').
%\end{equation}
%With the delta function identity
%\begin{equation}
%	\frac{2}{\pi}\int_{0}^{\infty} dk \sin (k x) \sin (k x') = \delta (x - x') - \delta (x + x'),
%\end{equation}
%we see that the sine in (\ref{extractdelta}) gives, after performing the Fourier integral \nr{radialfourier}, the flat-space retarded Green's function:
%\begin{equation}
%	g(x,x') = \theta(\eta-\eta')\frac{\delta(\eta-\eta' - \abs{{\bf x - x'}})}{\abs{{\bf x - x'}}} + \frac{\theta(\eta-\eta')}{2\pi^2 \abs{{\bf x - x'}}} \int_0^\infty dk\,k \sin k\abs{{\bf x-x'}} \hat{g}(\eta,\eta',k).
%\end{equation}

For the concordance model the potential
\be
V(\eta)={a''\over a}=H_0^2(2\Omega_\Lambda a^2+{1\over2}\Omega_m a^{-1})
\ee
can be given analytically as a function of $t$ but requires numerics when given as a function of $\eta$, see Appendix \ref{concordance}. Not surprisingly, the potential shows the two-peak structure,
\be
V_{\Lambda\rmi{CDM}}(\eta)\approx {2\over\eta^2}+{2\over  (\eta_\rmi{max}-\eta)^2} \ ,\quad \eta_\rmi{max}=\eta(t\to\infty)=4.445744 \ ,
\label{2peak}
\ee
but what is surprising is how accurate this approximation is.

The tail is then computed by solving the Fourier mode $\tilde  g_\rmi{tail}(\eta,\eta',k)$ from \nr{inhomeq} with the initial conditions in \nr{initcond} and going back to configuration space by using \nr{modeinv}. The outcome of the computation is shown in Fig. \ref{greenB}, which depicts $B(\eta,\eta',\rho)\theta(\eta-\eta' - \rho)$ on the $\rho,\eta'$ plane for $\eta=3.31$.

%In detail, the computation of the tail part $B(x,x')$ then proceeds as follows
%\begin{itemize}
%	\item Solve numerically the function $\tilde g(\eta,\eta',k)$ from Eqs \nr{greeneq4} with the potential $a''/a=H_0^2(2\Omega_\Lambda a^2+{1\over2}\Omega_m a^{-1})$ (see Eq. (\ref{potential})).
%	\item With the solution so obtained, compute the integral
%	\begin{equation}\label{numint}
	%		{1\over 2\pi^2\,R}\int_0^{\infty} dk\,k\,\sin(k R)[\tilde g(\eta,\eta',k)-{4\pi\over k}\sin((\eta-\eta')k)].
	%	\end{equation}
%\end{itemize}

%The integral is formally over an infinite range but for numerical integration we need to introduce a cutoff $k_\rmi{max}$ for the upper limit. Values of $k_\rmi{max}$ up to 100 were used for $\eta=3.3$. Here the last term eliminates the light cone term and leaves the tail part $B(\eta,\eta',R)\theta(\eta-\eta' - R)$ of the Green's function.

The result has two notable features. For the first we see that the step function $\theta(\eta - \eta' - \abs{{\bf x - x'}})$ is built into the tail two-point function: the two-point function is only supported strictly inside the past light cone of an observer at $\eta$. Physically, this means that the observer only receives tail signals from points that are past timelike separated from the point where the observation takes place and that are also in the support of the source. For the second, in spite of the $1/\rho$ factor in Eq. \nr{modeinv}, the overall $\rho$ dependence is very mild. Both these numerical results are rather surprising if one just looks at the integrand. This near $\rho$ independence we observe is in agreement with that of the one-peak tail solutions
\begin{align}
	&B_\rmi{m}(\eta,\eta') = \frac{1}{\eta \eta'} \ ,
	&B_\rmi{\lam}(\eta,\eta') = \frac{1}{(\eta_{max}-\eta)(\eta_{max}-\eta')}
\end{align}
above. In fact, with an appropriate constant $C$ the sum
\begin{equation}\label{interpol}
	B_\rmi{\lam CDM}(\eta,\eta',R) \approx B_\rmi{m}(\eta,\eta') + B_\rmi{\lam}(\eta,\eta') - C
\end{equation}
is a good approximation to the exact numerical solution we found. The shift is determined by numerically minimizing the integrated difference between the analytic interpolation and the exact numerical result. The minimum is obtained by $C = 0.095 H_0^2$ with parameter values given in Appendix \ref{concordance}. In later computations we shall use the approximation  (\ref{interpol}) for the tail Green's function.

\subsection{Quadrupole radiation from compact sources}
As a first application of the formulas for the tail solution
we will consider a compact GW source consisting of point-masses and finally specialize to an equal-mass compact gravitationally bound binary system. We will discuss the GW waveform generated by such a system, the tail part of the waveform, in particular.

In general, the first-order GW solution $h\ijd$ is
\begin{align}\label{gwsol1}
	h\ijd(\eta,{\bf x}) &= 4 \GN \int d^4 x' g(x,x') \frac{a(\eta')}{a(\eta)} T\ijd(\eta',{\bf x'}) \nn \\
	&= 4 \GN \int d^4 x' \frac{\delta(u)}{\abs{{\bf x - x'}}} \frac{a(\eta')}{a(\eta)}T\ijd(\eta',{\bf x'}) + 4 \GN \int d^4 x' B(x,x')\theta(u) \frac{a(\eta')}{a(\eta)}T\ijd(\eta',{\bf x'}) \\
	&\equiv h\ijd\lconeup + h\ijd\tailup \ ,
\end{align}
where $u = \eta - \eta' - \abs{{\bf x - x'}}$, and  $T\ijd$ is the stress-energy of a compact source. 
%We denote the first term, the light cone part, by $h\lconeup\ijd$ and the second term, the tail part, by $h\tailup\ijd$. 
Generally, the stress-energy tensor of a collection $A$ of pointlike masses $m_A$ with spacetime trajectories $x_A\mup(\tau)$ is
\begin{equation}
	T\munuu(x) = \sum_A \frac{m_A}{\sqrt{-g (x))}} \int d\tau\, u\mup(x_A(\tau)) u\nup(x_A(\tau)) \delta^{(4)}(x - x_A(\tau)) \ .
\end{equation}
The stress-energy can be related to the mass moment of the source by
\begin{equation}\label{tijmij}
	\int d^3 x' \, T\iju \approx \frac{1}{2}\int d^3 x' \, x'^i x'^j \ddot{T}_{tt} = \frac{1}{2 a^5} \int d^3 x' \,a^3  \, (a x'^i)(a x'^j) \ddot{T}_{tt} \approx \frac{1}{2 a^5} \ddot{M}\iju \ ,
\end{equation}
where $M\iju$ is the mass moment, 3-dimensional spatial integral over $T^{00}(t,{\bf x})a(t) x^i a(t)x^j$. To get a physically meaningful quantity, we needed to have the proper volume element and proper distance inside the integral. Making the above approximations, we assumed that the compact source lives inside a gravitationally bound region, a galaxy for instance, so that the effect of Hubble expansion on the source dynamics can be ignored, in contrast to, e.g., \cite{Chu:2016qxp,Ashtekar:2015lxa} where the cosmic expansion generates additional terms in the stress-energy--quadrupole relations. Note that we cannot make anymore this assumption for the second order source studied in Sec. 4.

We derived the relation between 3d integral of stress-energy and mass moment for the contravariant mass moment tensor, which is a more fundamental quantity than the covariant one since the mass moment of the point masses should be given in terms of physical position vectors $a x^i$. In \nr{gwsol1} we have the stress-energy tensor with indices down so when relating this to the contravariant mass moment, we get a factor of $a^4$, which together with $a$ in the numerator in \nr{gwsol1} cancels the scale factors coming from \nr{tijmij}. Thus, the light cone part of the GW solution can be written as
\begin{equation}\label{linearlcone}
	h_{ij}^\rmi{LC,TT} \approx \frac{2G}{a(\eta)r} \ddot{Q}_{TT}\iju(t(\eta-r)) \ ,
\end{equation}
where the quadrupole moment $Q\iju$ was defined in \nr{quaddef}.
Note that the TT projection of the mass moment equals the one for the quadrupole moment so we were able to express the solution in terms of the latter. Above the 2nd derivative of the quadrupole moment should be evaluated at the retarded moment of time $t(\eta-r)$, which simply means that an observer at coordinate distance $r$ away, monitoring a process that in the source frame takes the time $\delta t$, measures the duration $\delta t/a(\eta-r) = (1+z)\delta t$ for the process. The tail part on the other hand is given by
\begin{equation}
	h_{ij}^{\rmi{tail,TT}} \approx \frac{2G}{a(\eta)a(\eta-r)} B(\eta,\eta-r,r) \left[\dot{Q}\ttd\iju(t(\eta-r))-\dot{Q}\ttd\iju(t_0)\right] \ .\label{lineartail}
\end{equation}
Note that the 2nd $a$ in the denominator comes from changing an $\eta'$ integral to a $t'$ integral. 

In the above discussion we approximated that $\abs{{\bf x - x'}} \approx r$ for a compact source. Also, we assumed that the tail two-point function and the scale factor do not vary much over the source lifetime, but the second time derivative of the quadrupole moment does. For the two-point function this is justified by the fact that it is a background-dependent quantity that changes at scales comparable to the radius of curvature of the background. Therefore the approximations we made only hold for sources with lifetimes small compared to the local Hubble parameter value. When this approximation no longer holds, we also get integrals of terms that involve derivatives of the scale factor and the tail two-point function. 
%\ack{unless we can apply the adiabatic approximation to the background quantities!}

Note that in the tail solution \nr{lineartail} we have a difference between the first time derivative of the source quadrupole moment evaluated at the retarded time, corresponding to the moment of measurement, and at the initial moment of the binary. For a system that at late times consists of widely separated masses moving at constant velocities, as measured in the rest frame of a distant observer, this yields a signal that evolves linearly in time. In contrast, the first order light cone signal in this case settles down to a constant value, giving the ordinary memory effect.

\subsection{A compact binary system}\label{binrad}

The above results for gravitational waves hold for any compact quadrupole sources with short lifetimes. Now we take our source to be an equal-mass quasicircular compact binary, with the coordinate system chosen such that the binary rotates in the $x-y$ plane and the center of mass coincides with the origin, see Fig. \ref{coordsyst}. The mass moment for the binary was given in Eq. \nr{quad}.
%\begin{equation}\label{quad}
%	\left(M\iju\right) =
%	\left(
%	\begin{array}{ccc}
%		2 M R^2 \cos ^2(\Omega t) & M R^2 \sin (2 \Omega t) & 0 \\
%		M R^2 \sin (2 \Omega t) & 2 M R^2 \sin ^2(\Omega t) & 0 \\
%		0 & 0 & 0 \\
%	\end{array}
%	\right) \ ,
%\end{equation}
%where $R$ is the physical orbital radius (n.b. not the separation between the bodies) and each of the BH's have mass $M$. 
We further fix the coordinates of the system so that the light cone merger signal arrives at $\eta_f=\eta_\rmi{today}=3.3051$ and we are, initially, interested in the waveform at $\eta<\eta_f$. We choose to specify the time of the merger by its redshift $z$, defined by
\begin{equation}
	1 + z = \frac{1}{a} \ ,
\end{equation}
so that the conformal time $\etco$ is determined from $a(\etco)=1/(1+z)$. The binary  lives at $r=0$ during the time $\eta_0 = \etco-\tau  <\eta' < \eta_\rmi{coal}$, $\tau$ = conformal lifetime of the binary. We give the lifetime in years of $t$ but convert it to $\eta$  using Eq. \nr{a(eta)}. At any observer time $\eta$, there arrives a light cone signal emitted from $\bfx'=0$ at $\eta'=\eta-r$ but also a tail signal emitted from $\eta_0 < \eta' <\eta-r$. The coordinate $r$ is determined by the merger signal: $r=\eta_f-\etco$. Note that in the observer frame the lifetime in the $\eta$ coordinate is the same, $\tau=\etco-\eta_0$ but in the $t$ coordinate it is redshifted by a factor $1+z$.\footnote{To concretize the numbers, choosing $z=20$, lifetime $= 10^8$a and $M=100M_\odot$ leads to $\etco=0.7968$, $\tau=0.1937$, $\eta_0=0.6031$, $R_i=1.2\times 10^{10}{\rm m}=40{\rm s}$ = $0.079{\rm au}$, $R_i/r_S=4\cdot 10^4$, $\Omega_i=4.5\cdot 10^{-5}/{\rm s}$, $T_i=1.40\cdot 10^5${\rm s}, $2\cdot 10^{10}$ rotations until merger.} As long as $t$ is close to $t_\rmi{coal}$, we can at $\eta_\rmi{coal}$  use the approximation ($a(\eta)$ is effectively constant during the binary lifetime)
\be
t_\rmi{coal}-t(\eta-r)\approx a(\eta_\rmi{coal})(\eta_\rmi{coal}-(\eta-r))= a(\eta_\rmi{coal})(\eta_f-\eta) \ .\label{deltdeleta}
\ee
But we can apply the same approximation at $\eta_f$:
\be
t_f-t\approx a(\eta_f)(\eta_f-\eta)={a(\eta_f)\over a(\eta_\rmi{coal})}(t_\rmi{coal}-t)=(1+z)(t_\rmi{coal}-t) \ . \label{redshiftdelt}
\ee
Redshift of cosmic time differences is thus built in the approximation (\ref{deltdeleta}).

We assume further that $\dot\Omega\ll\Omega^2$, so that $\dot R(t)$ can be neglected when taking time derivatives.  For the $h_+$ mode and for radiation in the $z$ direction (radiation with polar angle $\theta$ brings in a factor $\fra12(1+\cos^2\theta)$),   Eq.(\ref{gwsol1}) integrates to Eqs. (\ref{linearlcone}) and (\ref{lineartail}), which in explicit form are
\ba
h_+(\eta)&=&{r_S^2\over2a(\eta) rR(t)}\cos(\Phi(t))\\
&&-{B(\eta,\eta-r)\over a(\eta)a(\eta-r)}\sqrt{\fra12 r_S^3R(t)}\sin\Phi(t)
+{B(\eta,\eta_0)\over a(\eta)a(\eta_0)}\sqrt{\fra12 r_S^3R(t_0)}\sin\Phi_0 \ ,\label{tailterms}
\ea
where $r_S=2GM$ Schwarzschild radius, $r=\eta_f-\etco$, $t=t(\eta-r)$, $t_0=t(\etco-\tau)$,
\be\label{R(t)}
R(t)=r_S\left(\fr45{t_\rmi{coal}-t\over r_S}\right)^{1/4},\quad 2\Omega(t)=\sqrt{{r_S\over 2R^3(t)}}={5\over (20r_S)^{5/8}}{1\over (t_\rmi{coal}-t)^{3/8}}
\ee
and
\ba
\Phi(t)&=&\Phi_0+\int_{t_0}^t dt\,2\Omega(t)=\Phi_0+{8\over (20r_S)^{5/8}}[(t_{\rm coal}-t_0)^{5/8}-(t_\rmi{coal}-t)^{5/8}]\label{bigphi}
\\
&\approx&\Phi_0+1.75\cdot 10^{14}\left({M_\odot\over M}\right)^{5/8}[(H_0t_{\rm coal}-H_0t_0)^{5/8}-(H_0t_\rmi{coal}-H_0t)^{5/8}]
\ea
and
\be
B(\eta,\eta')={1\over \eta\eta'}+{1\over(\eta_\rmi{max}- \eta)(\eta_\rmi{max}-\eta')}-C,\quad \eta_\rmi{max}=4.4457,\quad  C=0.095 H_0^2 \ .
\ee
For more details on approximating the tail Green's function $B$, see below Eq. \nr{interpol}.

For the tail part one had to integrate over the lifetime ($t_0<t'<t(\eta-r)$) of the binary a quantity which is a second $t'$ derivative of $Q_{ij}$. This leaves only the values of the first derivative at the end points. The upper limit depends only on the retarded time $\eta-r$, the lower limit is a constant. 
%In this sense the tail signal in this leading approximation also propagates along the light cone. Just looking at the equation of motion suggests dispersive subluminal propagation \cite{balekpolak}.

\begin{figure}[!t]
	\begin{center}
		\includegraphics[width=0.6\textwidth]{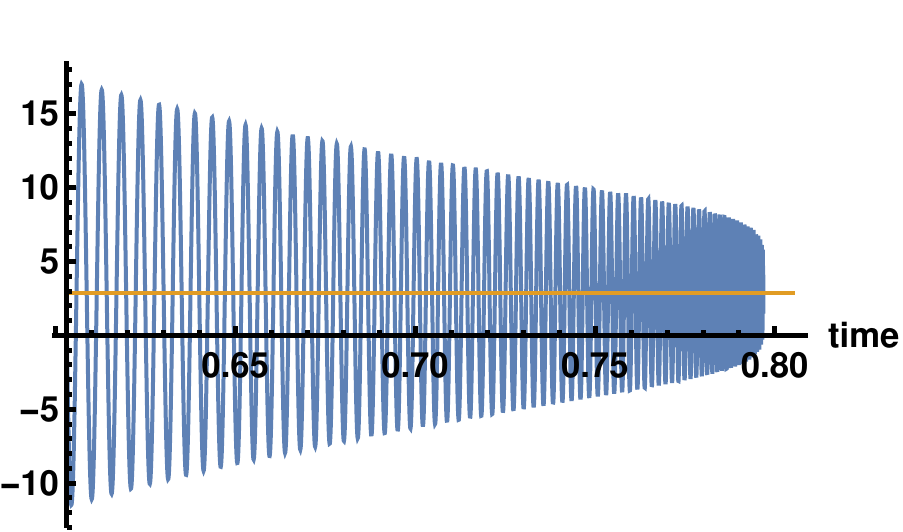}
	\end{center}
	\caption{\small A qualitative picture of the tail signal over the lifetime of the binary, plotted vs time in the source frame. The horizontal line is the initial stage dependent last term in (\ref{tailterms}). With the parameters in the footnote, the initial magnitude of the mode would be $1.2\cdot 10^{-39}$.
	}\label{tailpic}
\end{figure}

The first order light cone term amplitude is well-known (see e.g. \cite{Maggiore:2007ulw}):
%presented here so that the tail term can be compared with it. Normally it is presented by replacing $R$ withe the more directly observable frequency $f=\Omega/(2\pi)$. The key point is that $r$ and $rR$ are given in the binary frame. Redshifting both to the observer frame ($r$ is then replaced by the luminosity distance) produces a factor $(1+z)^2$, which can be combined with the $M^2$ in $r_S^2$. This is particularly economical  in the case of arbitrary masses when the chirp mass (= $M/2^{1/5}$ for equal masses) and frequency are used. In some more detail:
\be
{r_S^2\over 2a(\eta)rR(t)}={r_S\over 2a(\eta)r}\left(\fr54 {r_S\over t_\rmi{coal}-t}\right)^\fra14={ (H_0r_S)^\fra54\over 2a(\eta)H_0r}
\left(5\over 4a(\eta_\rmi{coal})H_0(\eta_f-\eta)\right)^\fra14 \label{ligcone} \ ,
\ee

%{1\over a(\eta)r}\left({GM\over 2^{1/5}}\right)^\fra54\left({5\over t_\rmi{coal}-t}\right)^\fra14
%={(H_0r_{S\odot})^\fra54\over \eta_f^\rmi{num}-\eta_\rmi{coal}^\rmi{num}}\left({(1+z)M\over M_\odot}\right)^\fra54\left({5\over 2^6H_0(t_\rmi{coal}-t)}\right)^\fra14,
%\ee
where $(H_0r_S)^{5/4}\approx 4.92\times10^{-29}(M/M_\odot)^{5/4}$. All quantities in the denominator are ${\cal O}(1)$ so the overall magnitude is basically determined by $H_0r_S$. The usual scaling of $M$ by $1+z$ is obtained from $1/a(\eta_\rmi{coal})=1+z$ and by changing $r$ to luminosity distance.

%$\eta^\rmi{num}\equiv H_0\eta$. In the first form $t$ is in the source frame, in the second in the observer frame.

%The first order result (\ref{ligcone}) can be compared with the second order light cone computation plotted in Fig. \ref{lconefig}. There the terms $H_0r$ and $a(\eta_\rmi{coal})$ in  (\ref{ligcone}) are scaled away. For $z=20$ in the middle of the range at $\eta=3.3$ the curve (\ref{ligcone}) is on the level of $6\cdot 10^{-27}$, the second order result in Fig. \ref{lconefig} is basically down by the factor $17/197$ in Eq.(\ref{inspmem}). The striking merger peak in Fig. \ref{lconefig}, of course, is only mildly reflected in the weak $1/(\eta_f-\eta)^{1/4}$ divergence in (\ref{ligcone}).

The tail in (\ref{tailterms}) consists of two terms, either an oscillating term for $t=t(\eta-r)$ from the upper limit of the time integral or a constant initial state term at $t=t_0$ from the lower limit. However, the value of the constant $\Phi_0$ is unknown. The last term reflects radiation from complex astrophysical phenomena associated with the formation of the binary, see for example \cite{vanSon:2021zpk}. The magnitude of the tail is made explicit by scaling out $H_0$ from \nr{tailterms}:
\ba
(H_0r_{S})^{\fra{15}{8}}\left({a(\eta_\rmi{coal})\over 20}\right)^\fra18
&&\hspace{-0.7cm}\left[-{B(H_0\eta,H_0(\eta-r)) \over a(\eta)a(\eta-r)}\,\left(H_0(\eta_f-\eta)\right)^\fra18
\sin\Phi(t)\nonumber\right.\\
&&\left.+{B(H_0\eta,H_0\eta_0) \over a(\eta)a(\eta_0)}\left(H_0(\eta_f-\eta_0)\right)^\fra18\sin\Phi_0\right] \ ,
\ea
where everywhere one has just the numerical values $H_0\eta$ of the etas and $\Phi(t)$ is as in Eq. (\ref{bigphi}). A qualitative picture of the waveform over the lifetime of the binary is plotted in Fig.\ref{tailpic}. The overall magnitude is given by $(H_0r_{S\odot})^{15/8}=3.45\cdot 10^{-43}$, somewhat compensated by $(M/M_\odot)^{15/8}$, the other terms are ${\cal O}(1)$. The suppression relative to the leading light cone term is  $(H_0r_{S})^{5/8}\approx 7.01\cdot 10^{-15}(M/M_\odot)^{5/8}$ so there clearly is no hope to observe the tail. But it is there! And most importantly, in the second order mechanism a new channel allowing for subluminal propagation opens up and leads to a  much larger tail contribution, discussed later in Sec. 4.

Authors of Ref. \cite{Tolish:2016ggo} proved that there is no memory effect associated with the tail. While the first order tail studied here tends to be minuscule, it exists nonetheless and gives a non-zero contribution even after the merger, which seems to contradict the theorem in \cite{Tolish:2016ggo}. However, this tension is accounted for by the fact that the memory effect they search for is a derivative of a delta function in spacetime curvature, which is not a restriction we make in our study. Therefore, the apparent discrepancy dissolves.

\subsection{Stress-energy of gravitational waves}
The first order gravitational radiation induced by a compact binary only causes the ordinary memory effect, which is small compared to the nonlinear or null memory \cite{Christodoulou:1991cr}. Therefore we are first and foremost interested here in the GW stress-energy that sources the nonlinear memory effect. For more on induced gravitational waves, see, e.g., \cite{Gong:2019mui,Ota:2021fdv,DeLuca:2019ufz}. The GW stress-energy tensor is calculated by taking the average over several wavelengths \cite{Isaacson:1968zza}:
\begin{equation}
	t\munud = \frac{1}{32\pi \GN}  \Big\langle\pt\mud \gamma^{TT}_{\rho\sigma} \pt\nud \gamma^{\rho\sigma}_{TT} \Big\rangle \ , \quad g\munud = \bg\munud + \gamma\munud = a^2(\eta\munud + h\ttu\munud) \ , \quad \gamma\munud = a^2 h\ttu\munud \ .
\end{equation}
Now, the first order gravitational wave entering the Isaacson formula above is a sum of two pieces, the light cone part and the tail one, as we saw earlier. Expanding the product $\pt\mud(\gamma\lconeup + \gamma\tailup)\pt\nud(\gamma\lconeup + \gamma\tailup)$ inside the average yields three kinds of terms: light cone--light cone, light cone--tail, and tail--tail terms. The last one is quadratic in the first order tail and hence utterly small, the second one includes $\dddot{Q}\ttu\ijd \ddot{Q}\ttd\iju$ that averages to zero. Thus we are only left with the light cone--light cone term that includes third time derivatives of the quadrupole moment.
The relevant component of the stress-energy tensor hence is
\begin{equation}
	t\ijd(\eta,r,\Omega) = n_i n_j\frac{\GN}{8\pi r^2} \left(\frac{a(\eta-r)}{a(\eta)}\right)^2  \Big\langle \dddot{Q}^{TT}_{kl}\dddot{Q}^{TT}_{kl}\Big\rangle(\eta-r,\Omega) \ ,
\end{equation}
where $n_i$ is the unit vector aligned with the propagation direction of the GW. Note that the scale factors coming from raising the indices are canceled when we use $H\munud = a^2 h\munud$; also note that terms involving derivatives of the scale factor are wiped out under the averaging operation. With the equal-mass quadrupole moment \nr{quad}, stress-energy becomes
\begin{equation}
	t\ijd(\eta,r,\Omega) = \frac{n_i n_j}{32\pi r^2}\frac{\GN^4 M^5}{R^5} \left(\frac{a(\eta-r)}{a(\eta)}\right)^2  (1+6 \cos ^2 \theta +\cos ^4 \theta ) \ ,
\end{equation}
where we also used \nr{quadrutt}. Here $\theta$ is the angle between ${\bf \hat{z}}$ and ${\bf n}$; in the chosen orientation for the coordinate system the result does not depend on the azimuthal angle $\phi$. Imposing the quasi-circular approximation, whereupon the orbital radius becomes a time-dependent function according to \nr{radius}, we finally get 
\begin{align}\label{tijcosmo}
	t\ijd(\eta,r,\Omega) &= \frac{n_i n_j}{r^2} \left(\frac{a(\eta-r)}{a(\eta)}\right)^2 \frac{dL}{d\Omega}(\eta-r, \theta) \\
	\frac{dL}{d\Omega} &\equiv \frac{1}{1024\pi \GN}\left(\frac{5 \GN M}{2}\right)^{5/4}\frac{1}{(t_\rmi{coal}-t(\eta-r))^{5/4}} (1+6 \cos ^2 \theta +\cos ^4 \theta ) \ .\label{lumicosmo}
\end{align}
Note the redshift factors that make this different from the corresponding formula in Minkowski spacetime, Eq. \nr{tijflat}. For later computations it is useful to note the relation between luminosity and total energy of the radiating system:
\begin{equation}\label{dEdt}
	\frac{dL}{d\Omega} = \frac{d E}{dt d\Omega} = \frac{1}{a(\eta)} \frac{d E}{d\eta d\Omega}, \quad \frac{dE}{d\Omega}(t) = \int_{t_0}^t dt' \frac{dL}{d\Omega'}(t') \ ,
\end{equation}
where $dE/d\Omega$ is the total radiated energy per unit solid angle.

%Check the 00 and 0i contributions. These should be exactly vanishing or negligible.

%%%%%%%%%%%%%%%%%%%%%%%%%%%
\section{Nonlinear memory induced by radiation from a black hole binary}\label{radbyrad}
%%%%%%%%%%%%%%%%%%%%%%%%%%%
Equipped with the results above, we may now set out to solve the main problem of our paper, namely, computing the nonlinear memory signal.
The solution for the second order GWs is similar by structure to the first order solution (see Appendix \ref{perturb}):
\begin{equation}\label{gwsol2}
	h\ijd\ttu = 4 \GN \int d^4 x' \frac{\delta(u)}{\abs{{\bf x - x'}}} \frac{a(\eta')}{a(\eta)}t\ijd\ttu(\eta',{\bf x'}) + 4 \GN \int d^4 x' B(x,x')\theta(u) \frac{a(\eta')}{a(\eta)}t\ijd\ttu(\eta',{\bf x'}) \ ,
\end{equation}
with $u = \eta - \eta' - \abs{{\bf x - x'}}$, and the stress-energy $t\ijd$, given in Eq. (\ref{lumicosmo}), of first order GWs as the source. The first term is the nonlinear light cone memory signal, the second one the nonlinear tail memory signal. In a more explicit form, derived later, the two terms are
\begin{align}\label{together}
	h\ijd\ttu &= \frac{4\GN}{a(\eta)} \int d\Omega'  (n_i' n_j')\ttu \int_{\eta_0}^{\eta-r} du' a(u')^2 \frac{dL}{d\Omega}(u',\Omega') \left[
	\frac{a(\emax(\eta,r,u',\Omega'))^{-1}}{\eta-u'-r\cos \theta_{xx'}} \right. \nn \\
	& \left. \hspace{8.5cm}+ \int_{u'}^{\emax(\eta,r,u',\Omega')}\hspace{-6mm} d\eta' \frac{B(\eta,\eta')}{a(\eta')}
	\right] \ .
\end{align}
Here the delta function in the light cone term has made it possible to do one integral, which has to be done numerically for the tail part. Notice that the TT projection is performed in the direction of the observer $\hat{x}^i$.

\subsection{Computation of the light cone memory}
First we compute the light cone memory strain.
The analysis parallels closely the one reviewed in the flat spacetime context, in Sec. \ref{flatspacemem}, with $t$ and $t'$ now simply replaced by $\eta$ and $\eta'$, and with the only complications coming from the cosmological scale factors in the integrand. Using \nr{tijcosmo} to the source, the light cone part is
\begin{equation}
	h^{\rmi{LC,TT}}_{ij} = 4 \GN \int d\Omega' dr' d\eta' \frac{\delta(u)}{\abs{{\bf x - x'}}}  (n'_i n'_j)\ttu \frac{a(\eta'-r')^2}{a(\eta)a(\eta')} \frac{dL}{d\Omega'}(\eta'-r',\Omega') \ .
\end{equation}
The trick was to write the integral in spherical coordinates and introduce a retarded time coordinate $u' = \eta' - r'$ by inserting a delta function inside the integral and then perform the $\eta'$-integral. After that, we are able to get rid of the final delta function by doing the $r'$-integral, thereby restricting $r'$ and $\eta'$ to 
\begin{equation}
	r'_0 =  \frac{(\eta - u')^2 - r^2}{2(\eta - u' - r \cos \theta_{xx'})} \ , \quad \eta' = u' + r'_0 \ .
\end{equation}
Manipulating the integral with delta function identities then yields:
\begin{equation}\label{lconefull}
	h^{\rmi{LC,TT}}_{ij} = \frac{4 \GN}{a(\eta)} \int d\Omega' \int_{\eta_0}^{\eta-r} du' \frac{(n'_i n'_j)\ttu}{\eta - u' - r \cos\theta_{xx'}} \frac{a(u')^2}{a(u' + r'_0)} \frac{dL}{d\Omega'}(u',\Omega') \ ,
\end{equation}
where $\cos \theta_{xx'}$ is the angle between ${\bf x}$ and ${\bf x'}$, and the lower integration limit is the moment $\eta_0$ at which the source is switched on.
%We may then extract the present day physical distance from the denominator and impose the short burst approximation $(\eta-r-u')/r \approx 0$. 
%\begin{equation}
%	\underline{\chi}_{ij} \approx \frac{4 \GN}{a(\eta) r} \int d\Omega' \int_{\eta_0}^{\eta-r} du' \frac{n'_i n'_j}{1- \cos\theta_{xx'}} \frac{a(u')^2}{a(u' + r'_0)} \frac{dL}{d\Omega}(u',\Omega').
%\end{equation}
Now, pulling out a factor of inverse coordinate distance and using instead the luminosity distance to the source via $d_L = a(\eta) r (1+z)$, we get 
\begin{equation}\label{lconebasic}
	h^{\rmi{LC,TT}}_{ij} \approx (1+z)\frac{4 \GN}{d_L} \int d\Omega' \int_{\eta_0}^{\eta-r} du' \frac{(n'_i n'_j)\ttu}{\frac{\eta - u'}{r}+1- \cos\theta_{xx'}} \frac{a(u')^2}{a(u' + r'_0)} \frac{dL}{d\Omega'}(u',\Omega') \ ,
\end{equation}
which is almost the flat spacetime memory effect enhanced by a redshift factor and with coordinate distance replaced by luminosity distance, apart from the scale factors inside the integral. To further evaluate the above integral, it would be convenient to write it in a form where we have a term that is the flat spacetime memory times a redshift factor, plus small corrections parametrized by Hubble constant. The first guess would be to Taylor expand the integrand but doing this, however, yields terms that render the integral divergent. Instead, we make use of the fact
\begin{equation}
	\frac{dL}{d\Omega} = \frac{d E}{dt d\Omega} = \frac{1}{a(u)} \frac{d E}{du d\Omega}(u,\Omega)
\end{equation}
from \nr{dEdt} where we defined $dE/d\Omega$ as the total radiated energy per solid angle. This
allows us to perform a partial integration over the retarded binary frame time $u'$ from $u'=\eta_0$ to $u'=\eta-r$, whereby we have
\begin{align}\label{flatplusHubble}
	h^{\rmi{LC,TT}}\ijd &= (1+z)\frac{4G}{d_L}\int d\Omega' \left[ \frac{(n'_i n'_j)\ttu}{1-\cos \theta_{xx'}} \frac{dE}{d\Omega'}(\eta-r,\Omega') \right. \nn \\ 
	&\left.- \int_{\eta_0}^{\eta-r}du' \frac{dE}{d\Omega'}(u',\Omega')\frac{(n'_i n'_j)\ttu}{\eta-u'-r\cos\theta_{xx'}}\left(\frac{a(u')/a(u'+r'_0)}{\eta-u'-r\cos\theta_{xx'}} + \frac{d}{du'}\left( \frac{a(u')}{a(u'+r_0')} \right)\right)\right] \nn \\
	&\equiv h^\rmi{LC,flat}_{ij} + h^\rmi{LC,novel}_{ij} \ .
\end{align}
Here the first term is the substitution term from integration by parts. It only contains the contribution from the upper limit $\eta-r$. The contribution from the lower limit contains $dE/d\Omega'$ evaluated at $\eta_0$, which vanishes by definition ($E$ is the total radiated energy from a source switched on at $\eta_0$). For retarded times above $\eta_\rmi{coal}$, all the energy has been radiated and $h^\rmi{LC,flat}_{ij}$ is a constant. Redshifted to the observer frame this means that, after the arrival of the merger signal at $\eta_f=\eta_\rmi{coal}+r$, this piece of the metric perturbation is constant in time. The second term $h^\rmi{LC,novel}_{ij}$ has contributions from all values of the retarded time and is found to be a negative contribution monotonically growing in absolute value. The derivative of the scale factor term is, explicitly
\begin{equation}
	\frac{d}{du'}\left( \frac{a(u')}{a(u'+r_0')} \right) = \frac{a(u')}{a(u'+r_0')}\left(
	\mathcal{H}(u') - \left(1+\frac{dr_0'}{du'}\right)\mathcal{H}(u'+r_0') 
	\right) \ ,
\end{equation}
where $\mathcal{H} \equiv a H$ is the conformal Hubble parameter.

We should emphasize that the first term in \nr{flatplusHubble} is exactly the result for memory found in \cite{Bieri:2015jwa,Bieri:2017vni,Kehagias:2016zry,Tolish:2016ggo,Chu:2016qxp} in a cosmological setting, whereas the second one is a novel cosmological correction to the known result. Integration of the first term can be done analytically as in Sec. \ref{flatspacemem}, the new term requires numerical methods. We study these two terms separately in what follows.

\subsubsection{The redshifted flat spacetime memory term}

To write the memory strain in an informative form, we will use the fact that the time and solid angle dependence of radiated energy factorize:
\begin{equation}\label{factorized}
	\frac{dE}{d\Omega}(u,\Omega) = \frac{5 E(u)}{32\pi}\mathcal{F}(\cos\theta),
\end{equation}
where the angular distribution $\mathcal{F}$ was defined in \nr{Fdef}, and the numerical factor comes from the fact that $\int d\Omega \mathcal{F}(\cos\theta) = 32\pi/5$. 

The angular distribution for both the inspiral and the ringdown radiation will be the same in our model, whereas the time profile will be different from its pre-merger counterpart. 
We deal with the angle dependence first.
Again we point the reader to Appendix \ref{angularintegrals} but outline here the main idea. An integral over $S^2$ that includes a function $f({\bf \hat{x}\cdot n})$ with ${\bf \hat{x}}$ a constant unit vector, and tensor product of unit vectors $n_i$, can be written as a linear combination of symmetrized tensor products of flat 3d metrics and $\hat{x}_i$'s. Above we have such integrals with 2, 4, and 6 $n_i$'s coming from different terms of $\mathcal{F}(\cos\theta')$, and the corresponding linear combinations are
derived in the aforementined appendix. Using Eq. \nr{ttgeneral}, the result is
\begin{equation}\label{flatresult}
	h_+^{\rmi{LC,flat}} = (1+z)\frac{G E(\eta-r)}{24 \,d_L} \sqrt{2} (17 + \cos^2 \theta_x ) \sin^2 \theta_x
\end{equation}
similarly with the estimate in \cite{Garfinkle:2016nhe}. Here $\theta_x$ is the angle between the normal vector of the binary rotation plane and the observer position vector. 
Radiated energy in the MWM is, explicitly,
\begin{align}
	&E(u) = 
	\begin{dcases}
		0, & t(u) \leq t_0, \\
		E_\rmi{insp}(u) , \quad & t_0 \leq t(u) \leq t_\rmi{m}, \\
		E_\rmi{ring}(u), \quad & t(u) \geq t_\rmi{m},
	\end{dcases}\label{MWMenergy}\\
	&E_\rmi{insp}(u) = \frac{M}{12}\left( \frac{1}{1+\frac{32}{405 G M}(t_m-t_0)} \right)^{1/4}\left[\left(1+\frac{\frac{32}{405 G M}(t-t_0)}{1+\frac{32}{405 G M}(t_m-t)}\right)^{1/4}-1\right], \\
	&E_\rmi{ring}(u) = \frac{M}{12}\left[1-\left(\frac{1}{1+\frac{32}{405 GM}(t_m-t_0)}\right)^{1/4}\right]+ \nn\\ 
	&\qquad\qquad \frac{M_f}{16\pi} \sum_{n,n'=0}^{n_\rmi{max}}\frac{\mathcal{A}_{22n}\mathcal{A}^*_{22n'}\sigma_{22n}\sigma^*_{22n'}}{\sigma_{22n} + \sigma^*_{22n'}}e^{i(\phi_{22n}-\phi_{22m})}\left[1 - e^{-(\sigma_{22n}+\sigma^*_{22n'})(t(u)- t_\rmi{m})/(G M_f)}\right],
\end{align}
where $t_\rmi{m}$ is the matching time between the inspiral and ringdown phases (see Sec. 2.2). Recall that the matching radius was chosen to be $R_\rmi{m} = (3/2)r_S = 3 G M$, which fixes the matching time in the approximation. Notice that as the radiated energy is proportional to mass of the system, the $1+z$ factor in \nr{flatresult} has the effect of redshifting the mass, just as it has for the oscillating GW signal. Evolution of the strain \nr{flatresult} is essentially the same as in Fig. \ref{flatmemplot}, the only modification being that the timescales are redshifted due to the retarded conformal time dependence in $E(\eta-r)$.

\subsubsection{The novel cosmological term}

\begin{figure}
	\begin{center}
		\includegraphics[width=0.90\textwidth]{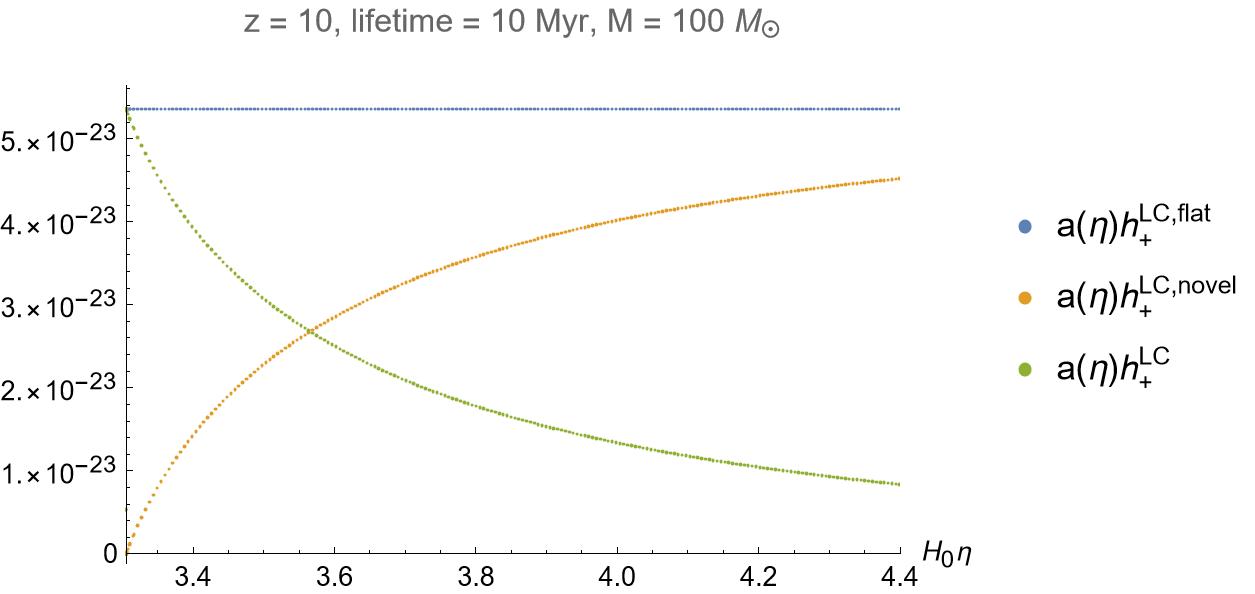}
	\end{center}
	\caption{\small The constant flat spacetime memory term $h_+^{\rmi{LC,flat}}$, the negative novel cosmological term $h_+^\rmi{LC,novel}$ increasing in absolute value (the Figure plots the absolute value), and the decreasing full light cone memory strain $h_+^\rmi{LC}=h_+^\rmi{LC,flat}+h_+^\rmi{LC,novel}$ after the merger. The quantities are computed for a binary coalescence taking place at redshift $z=10$, with lifetime 10 Myr, and
	%lifetime of $10^8$ yrs (right panel), 
	with BH mass = $100$ $M_\odot$, and the results are scaled by $a(\eta)$ (recall that in \nr{flatplusHubble} the memory strain involves a factor of $1/d_L = 1/(a(\eta)r(1+z))$). On the \(x\)-axis we have conformal time in the units of Hubble time with $H_0\eta_f=H_0\eta_\rmi{today}=3.3051$.
	}\label{lconefig}
\end{figure}

%In the case of \nr{lcfinal}, the $f$-function reads
%%%\end{equation}
%\ack{renew this equation}
%The subscript "LC" means that the function is associated with the light cone part of the GW, to distinguish later from the corresponding function for the tail part. Note that the $C_{i,j}$'s are constants with respect to angles but of course depend on the parameters of the setup, namely, redshift $z$ of the coalescence and the binary lifetime, as we can see from the explicit expression for the $f$-function. 
We were able to compute the flat spacetime memory strain analytically due to the fact that the variables in the integrand were separable. In the novel cosmological term this is not the case due to the mixed time and solid angle dependence in a scale factor argument. We can still use the same angular integration techniques but the integrals need to be computed numerically.
The result, again, only has a plus polarized mode:
\begin{equation}
	h^{\rmi{LC,novel}}_+ = -(1+z)\frac{4G}{d_L} \frac{\sqrt{2}}{45}(60 C_{4,0}-12 C_{6,0}-C_{6,2}(1-3 \cos ^2 \theta_x  )) \sin ^2\theta_x.
\end{equation}
The $C$-coefficients above are 
\begin{align}
	&C_{4,0} =\frac{3\pi}{4} \int_{-1}^{1} (1 - \cos^2\theta)^2 f(\cos\theta)d\cos\theta \\
	&C_{6,0} =\frac{5\pi}{8} \int_{-1}^{1} (1 - \cos^2\theta)^3 f(\cos\theta)d\cos\theta \\
	&C_{6,2} =\frac{15\pi}{8} \int_{-1}^{1} (1 - \cos^2\theta)^2(7 \cos^2 \theta - 1) f(\cos\theta)d\cos\theta \ ,
\end{align}
where
\begin{align}
	&f(\cos\theta) = \int_{\eta_0}^{\eta-r}du' \frac{5 E(u')}{32\pi}\frac{\mathcal{F}(\cos\theta)}{\eta-u'-r\cos\theta}\frac{a(u')}{a(u'+r_0')}\left(
	\frac{1}{\eta-u'-r\cos\theta} \right. \nn \\
	&\left. \qquad \qquad \qquad \qquad\qquad\qquad\qquad\qquad\qquad+\mathcal{H}(u') - \left(1+\frac{dr_0'}{du'}\right)\mathcal{H}(u'+r_0')
	\right) \ ,
\end{align}
and
\begin{equation}
	r_0' = \frac{1}{2}\frac{(\eta-u')^2-r^2}{\eta-u'-r\cos\theta}\ . \nn
\end{equation}
We evaluate the $C$-coefficients numerically. The novel term is plotted for a few representative parameter values in Fig. \ref{lconefig} where it is contrasted to the flat spacetime term computed in the previous subsection, and where also the full light cone memory strain is shown. During the inspiral, the novel term is insignificant compared to the flat spacetime term, and it only grows to a comparable magnitude over cosmological timescales after the merger.

\begin{figure}[!t]
	\begin{center}
		\includegraphics[height=0.5\textheight]{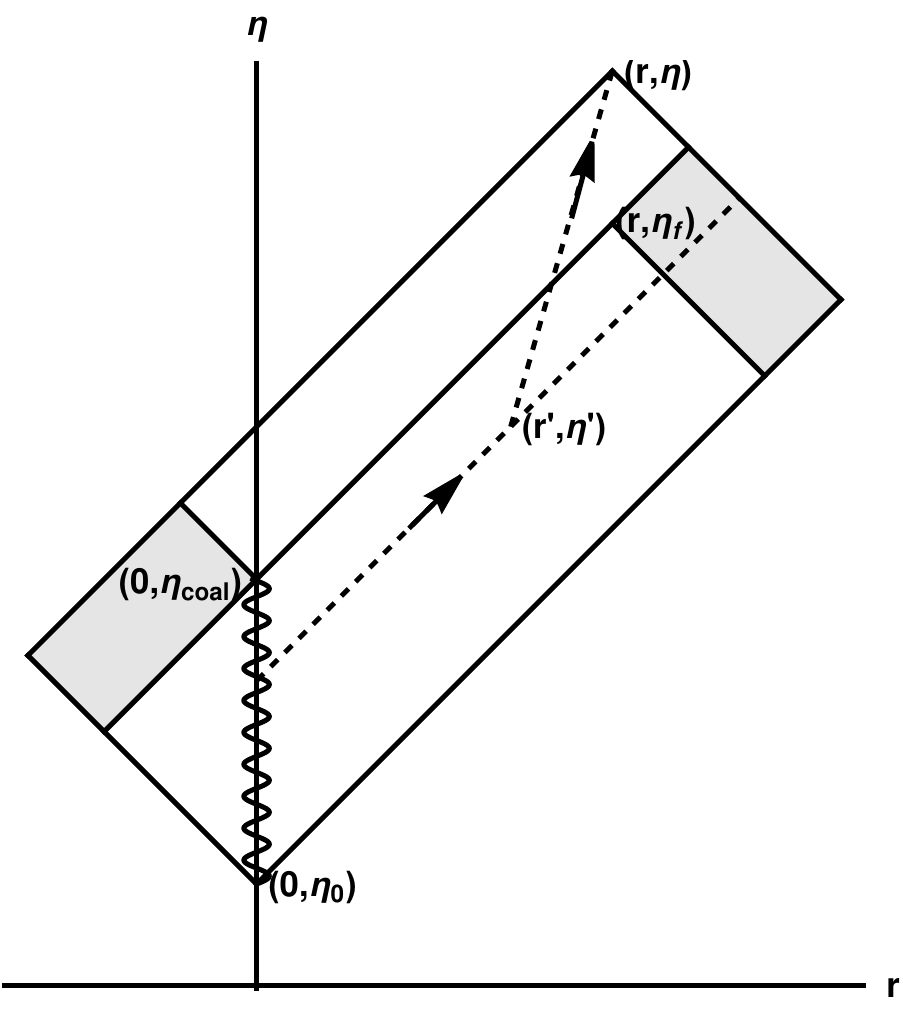}
	\end{center}
	\caption{\small A 2D diagram of the integration region. The compact binary is the wiggly curve between $(0,\eta_0)$ and $(0,\eta_\rmi{coal})$, the merger takes place at $(0,\eta_\rmi{coal})$, the merger pulse arrives at the detector at  $(r,\eta_f)$, GW emitted by the binary moves along the light cone and emits at $(r',\eta')$ a subluminal tail pulse which arrives at the detector at $r$ at the time $\eta>\eta_f$ after the merger pulse. GW in the two shaded regions can also source a tail pulse to $(r,\eta)$. Similar post-merger first order tail radiation (emission of dashed line from the wiggly binary curve) would also be possible but is canceled by the dynamics of the process (subsection \ref{binrad}). %For the leading linear radiation in Section 4 the compact source region cannot send a subluminal tail signal to $\eta>\eta_f$, only to $\eta<\eta_f.$
	}\label{kuvaintregion}
\end{figure}

The full light cone (LC) contribution after the merger, for $\eta>\eta_f$, thus consists of a constant part $h_+^\rmi{LC,flat}$ and a negative part $h_+^\rmi{LC,novel}$. The constancy of the former is understood simply on the basis of the expression in Eq. (4.8), the only $\eta$ dependence there is in $dE(\eta-r,d\Omega')/d\Omega'$ and after the merger this is constant, the total radiated energy. To understand the $\eta$ dependence, it is also useful to consider the null lines followed by LC radiation from the binary to the observer using the causal diagram in Fig. \ref{kuvaintregion} (see also Fig. \ref{intersectingcones} with one more dimension), in which the observer is at the top point $\eta>\eta_f=\text{merger signal arrival}$. All the second order LC signals arrive at the observer along the top sides of the rectangle (backward light cone). The part $h_+^\rmi{LC,novel}$ is sourced by all the dashed LC lines emanating from the binary source and there are lots of reasons for $\eta$ dependence. Why the effect is a monotonically growing negative one is not obvious, it just follows from numerics.

%\begin{figure}
%	\begin{center}
%		\includegraphics[width=0.90\textwidth]{lcone_ltime_10^7_100Msun}
		%\includegraphics[width=0.49\textwidth]{future_ltime_10^8_100Msun}
%	\end{center}
%	\caption{\small Evolution of the light cone memory signal over the binary inspiral. The signal is computed for binaries at redshifts $z=1,2,5,10,20$, with lifetime of $10^7$ yrs,
	%lifetime of $10^8$ yrs (right panel), 
%	with BH mass = $100$ $M_\odot$. The GW strain is scaled by $r(1+z)^{-1/4}$ where $r$ is the coordinate distance to the source and $z$ is the redshift corresponding to the source time. With this scaling the final values of signals from different redshifts match perfectly. On the \(x\)-axis we have conformal time in the units of Hubble time, and on the \(y\)-axis we have the strain $h_+$ of an edge-on binary.
%	}\label{lconefig}
%\end{figure}

\subsection{Computation of the tail memory integral}\label{nonlintail}

We now focus on the tail term of Eq.\nr{gwsol2}, which we denote by $h^\rmi{tail}_{ij}$. To work it out, we have to do a 4d integral over $x'=(\eta',r',\cos\theta',\phi')$ at fixed $x=(\eta,{\bf x})$ so that within the range of integration
\begin{equation}
	\eta - \eta' > \abs{{\bf x} - {\bf x'}}, \quad \eta' - \eta_0 > \abs{{\bf x'}}.
\end{equation}
This domain is the union of the interiors of two null cones whose symmetry axes are aligned but do not coincide. To do this integral, we first convert the $r'$ integral to an integral over the retarded time $u'=\eta'-r'$, then do the integral over $\eta'$ so that the above inequalities are satisfied, then do the integral over $u'$ so that this retarted time covers the lifetime of the binary and finally do the integral over the angles $\Omega'=(\cos\theta',\phi')$. For example, the integration region for a binary with lifetime 10 Myr at $z=10$ and observer at the present time is parametrized by $\eta = 3.305, \tau = 0.008, r = 2.204, \eta_0 = \eta - r - \tau = 1.093$ in the units of Hubble time.

\begin{figure}[!t]
	\begin{center}
		\includegraphics[width=0.49\textwidth]{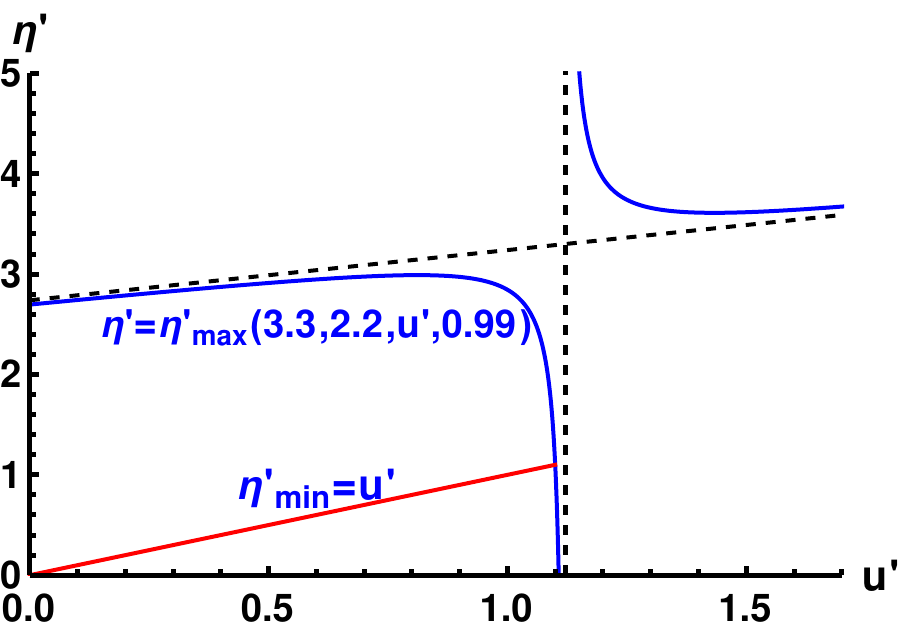}
		\includegraphics[width=0.49\textwidth]{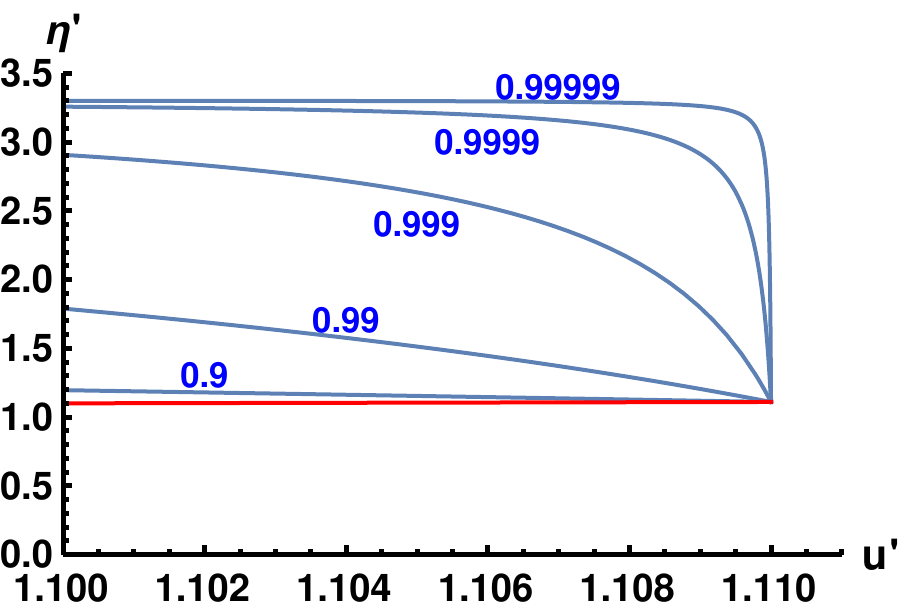}
	\end{center}
	\caption{\small Left: The dependence of $\eta'_\rmi{max}$ on $u'$ for fixed integration angle $\cos \theta' = 0.99$, and for an observer at $(\eta,r)=(3.31,2.2)$. Both branches of the hyperbola (blue curves) and the asymptotes $\eta'=\fra12 (u'+\eta+r \cos\theta')$, $u'=\eta-r \cos\theta'$ are shown (dashed lines). The max and min (red line) curves intersect at the point $\eta'=u'=\eta-r$. Right: The physical region for various $\cos\theta'$ (values increase bottom-up) when the binary radiation starts at conformal time $\eta_0=1.1$ and lasts for a conformal time 0.01. Thus $u'_\rmi{max}=1.11$ which has to be $=\eta-r$ so that $r=2.2$ leads to $\eta=3.31$. For $-1<\cos\theta'<1$ the region is bounded by the lines $\eta'=\fra12(\eta\pm r+u')$ and $u'=\eta_0$, $u'=\eta_0+\tau$.}\label{phregs}
\end{figure}

We now turn to evaluate the integral. Towards this purpose, we write the tail part of (\ref{gwsol2}) more explicitly:
\begin{align}\label{tailint}
	h^{\rmi{tail,TT}}\ijd(\eta,{\bf x}) = & \,4G\int_{\mathbb{R_+}\times \mathbb{R}^3} d^4 x' B(x,x')\theta(\eta - \eta' - \abs{{{\bf x} - {\bf x'}}})\frac{a(\eta')}{a(\eta)} t\ijd\ttu(x') \nn \\
	= & \,4G\int_\mathcal{V} d\Omega' d\eta' dr' B(\eta,\eta') \frac{a(\eta'-r')^2}{a(\eta)a(\eta')} (n'_i n'_j)\ttu \frac{dL}{d\Omega}(\eta'-r',\Omega') \ ,
\end{align}
where $dL/d\Omega$ is the luminosity of the binary source in the source frame, evaluated at retarded conformal time $\eta'-r'$, as given in Eq. \nr{lumicosmo}, and the integration volume $\mathcal{V}$ is the region constrained between the past light cone of the observer and the future lightcone of the first moment of the binary. Note that $B$ is in general also $r'$-dependent but based on our numerical results in Sec. \ref{perttheory} we estimate that in our setup this dependence can be neglected and we can use the analytic interpolation (\ref{interpol}). Again, in analogy with the flat spacetime memory computation, we insert a delta function by $1 = \int du'\delta(u' - \eta' + r')$ but then we leave the $\eta'$-integral intact and perform the $r'$-integral instead, which creates a factor $\theta(\eta' - u')$ since the $r'$-integral is defined on the interval $(0,\infty)$. The upper limit for the $u'$ integral is $\eta - r$ and the lower limit is the moment of time $\eta_0$ at which the source is switched on. Due to the step function $\theta(\eta' - u')$, the lower limit for $\eta'$-integral is set to $u'$.  On the other hand, the other step function $\theta(\eta - \eta' - \abs{{{\bf x} - {\bf x'}}})$ along with the substitution $r'\rightarrow \eta'-u'$ imposes an upper limit for the $\eta'$ integral:
\begin{equation}\label{etaupper}
	\emax(\eta,r,u',\Omega') = u' + \frac{(\eta - u')^2 - r^2}{2(\eta - u' - r \cos \theta_{xx'})} = \frac{1}{2}\frac{\eta^2 - u'^2 - r^2 - 2 r u' \cos \theta_{xx'} }{\eta - u' - r \cos \theta_{xx'}} \ ,
\end{equation}
where $\theta_{xx'}$ is the angle between ${\bf x}$ and ${\bf x'}$. This is where the integral meets the past light cone of the observer. Notice that $\emax$ is actually the same function as $u'+r_0'$ above; here we named it differently because of its role as the maximum value of the $\eta'$-integration range. For fixed $- 1 <\cos \theta_{xx'}< 1$, $\eta$, and $r$, $\emax$ traces a segment of a hyperbola as $u'$ runs from $\eta_0$ to $\eta - r$, and with $\cos\theta_{xx'} =\pm 1$, $\emax$ gives two straight lines that bound the hyperbolic segments from below and above, see Fig. \ref{phregs}.

If the GW burst is very short, so short that the scale factor $a(u')$ stays very close to constant during the process, we can pull it out of the integral. In the same way as with the light cone solution, we may use the fact that luminosity is given by the time derivative of the radiated energy to integrate by parts, which yields, at fixed $\cos\theta',\phi'$,
\begin{align}
	 &\int_{\eta_0}^{\eta-r} du' a(u')^2 \frac{dL}{d\Omega}(u') \int_{u'}^{\emax} d\eta' \frac{B(\eta,\eta')}{a(\eta')} \nn\\ \approx
	 &a(\etco) \left[\Bigg\rvert^{u'=\eta-r}_{u'=\eta_0} \frac{dE}{d\Omega'}(u') \int_{u'}^{\emax} d\eta' \frac{B(\eta,\eta')}{a(\eta')} - \int_{\eta_0}^{\eta-r}du' \frac{dE}{d\Omega'}(u') \left(\frac{B(\eta,\eta'_\rmi{max})}{a(\eta'_\rmi{max})}\frac{d\emax}{du'} - \frac{B(\eta,u')}{a(u')}\right)\right] \ ,
\end{align}
where we did not write $u'$- and angle-dependence in $\emax$ explicitly, to make the expression \nr{tailint} more compact.
Now, notice that the first term is actually zero. First off, $\emax(u'=\eta-r) = \eta-r$, which is why substitution to the upper limit vanishes identically if $dE/d\Omega$ is regular at that moment (of course, the total radiated energy always stays finite). Secondly, $dE/d\Omega = 0$ at $\eta_0$ as in the light cone part computation. Hence, the whole first term vanishes and we are left with a single integral over the burst duration. We get that the tail is given by
\begin{equation}\label{tailmain}
	h^{\rmi{tail,TT}}\ijd = \frac{4G}{a(\eta)(1+z)} \int d\Omega' (n_i' n_j')\ttu \int_{\eta_0}^{\eta-r}du' \frac{dE}{d\Omega'}(u',\Omega') \left[ \frac{B(\eta,u')}{a(u')} - \frac{B(\eta,\emax)}{a(\emax)}\frac{d\emax}{du'} \right] \ ,
\end{equation}
where redshift $z$ corresponds to the source event.
Here one should note that we cannot apply the short burst approximation to the cosmological factors inside the integral anymore, unless we restrict the retarded time $\eta-r$ to be very close to $\etco$. After the burst, $dE/d\Omega$ quickly asymptotes to a constant value so the integrand has support all the way to $\eta-r$, even if $\eta$ runs up to the de Sitter singularity in the infinitely distant future. Therefore, the post-merger signal evolution is basically determined by the tail two-point function.

Eq. \nr{tailmain} is our main result. We still need a numerical evaluation thereof, which requires specification of energy production from the binary. For this we can use the Minimal Waveform Model or a simple step function approximation.

\subsubsection{MWM approximation}
\begin{figure}
	\begin{center}
		\includegraphics[width=0.90\textwidth]{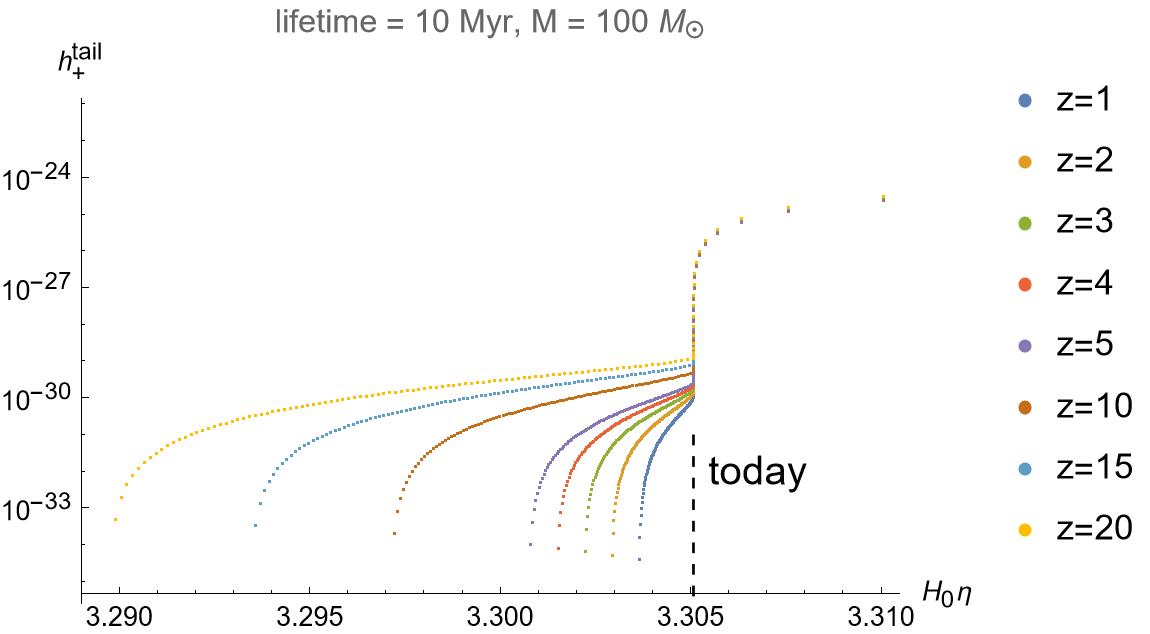}
	\end{center}
	\caption{\small Log-linear plot of the tail strain for binary at redshifts $z=1,2,3,4,5,10,15,20$ %with lifetime of $10^4$ yrs (left panel), 
	with lifetime of $100$ Myr, with BH mass = $100$ $M_\odot$, and observer living at the present time. On the \(x\)-axis we have conformal time in the units of Hubble time, and on the \(y\)-axis we have the strain $h_+$ of an edge-on binary (maximal signal strength). The black dashed line marks the location of the present time on the time axis.
	}\label{hplusinsp}
\end{figure}

\begin{figure}
	\begin{center}
		\includegraphics[width=0.90\textwidth]{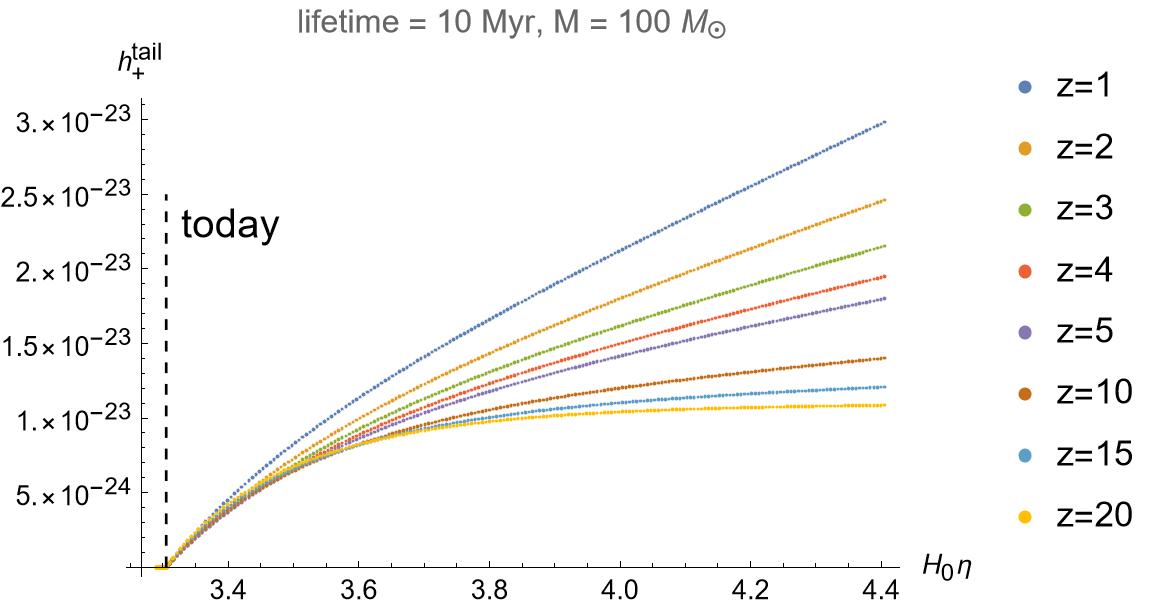}
	\end{center}
	\caption{\small Evolution of the tail strain after the present time, recall that now the light cone memory is absent. Similarly to Fig. \ref{hplusinsp}, the strain is computed for binaries at redshifts $z=1,2,3,4,5,10,15,20$, with lifetime of $10^7$ yrs,
	%lifetime of $10^8$ yrs (right panel), 
	with BH mass = $100$ $M_\odot$. On the \(x\)-axis we have conformal time in the units of Hubble time, and on the \(y\)-axis we have the strain $h_+$ of an edge-on binary. The black dashed line marks the location of the present time on the time axis.
	}\label{hplusfuture}
\end{figure}

\begin{figure}
	\begin{center}
		\includegraphics[width=0.90\textwidth]{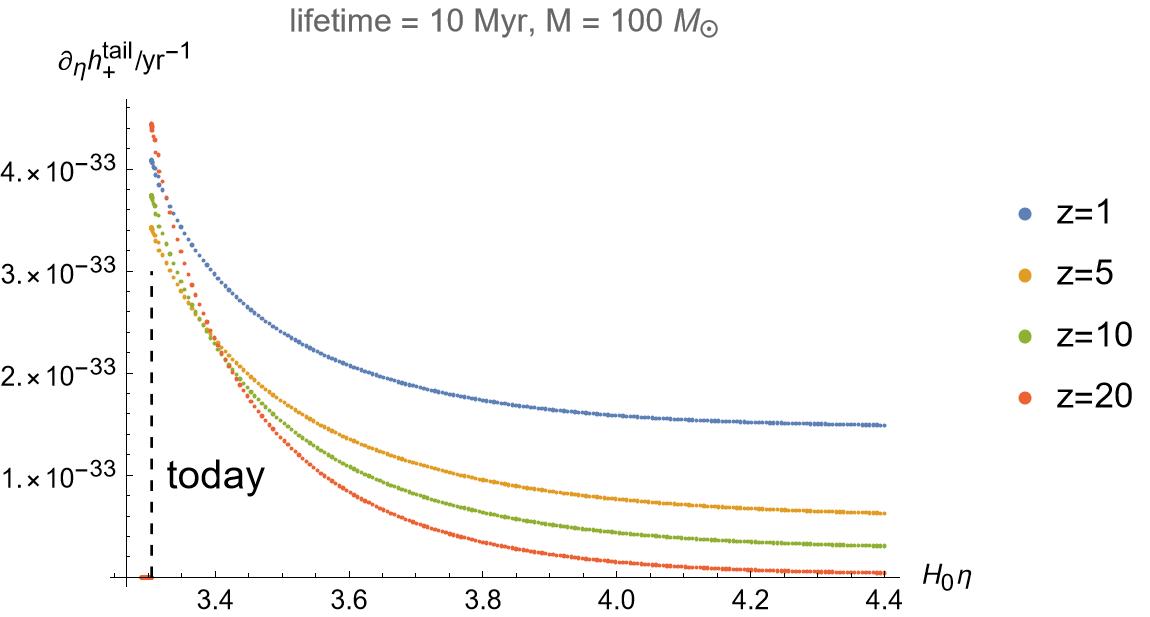}
	\end{center}
	\caption{\small Conformal time derivative of the tail strain. The quantity is computed for binaries at redshifts $z=1,5,10,20$, with lifetime of $10^7$ yrs,
		%lifetime of $10^8$ yrs (right panel), 
		with BH mass = $100$ $M_\odot$. On the \(x\)-axis we have conformal time in the units of Hubble time, and on the \(y\)-axis we have the time derivative $\pt_\eta h_+$ of an edge-on binary in the units of $\text{yr}^{-1}$. The black dashed line marks the location of the present time on the time axis.
}\label{tailderiv}
\end{figure}

We use Eq. \nr{MWMenergy} for radiated energy in MWM and plug this into \nr{tailmain}. Notice that in this model the total radiated energy is $E_\rmi{tot} \approx 0.149 M$. Again applying the techniques of Appendix \ref{angularintegrals}, with TT projection in the direction $\hat{\bf x}$, the plus polarized mode is
\begin{equation}
	h\tailup_+ = \frac{G}{a(\eta)(1+z)}\frac{\sqrt{2}}{72\pi}(60 C_{4,0}-12 C_{6,0}-C_{6,2}(1-3 \cos ^2 \theta_x  )) \sin ^2\theta_x  \ ,
\end{equation}
where $\theta_x$ is the inclination angle of the observer position vector, and
\begin{align}
	&C_{4,0} =\frac{3\pi}{4} \int_{-1}^{1} (1 - \cos^2\theta)^2 f(\cos\theta)d\cos\theta \label{C1}\\
	&C_{6,0} =\frac{5\pi}{8} \int_{-1}^{1} (1 - \cos^2\theta)^3 f(\cos\theta)d\cos\theta \label{C2}\\
	&C_{6,2} =\frac{15\pi}{8} \int_{-1}^{1} (1 - \cos^2\theta)^2(7 \cos^2 \theta - 1) f(\cos\theta)d\cos\theta \label{C3}\\
 	&f(\cos\theta) = \frac{5}{8\pi} \int_{\eta_0}^{\eta-r}du' E(u') \left(\frac{B(\eta,u')}{a(u')}-\frac{B(\eta,\emax)}{a(\emax)}\frac{d\emax}{du'}\right)\ . \label{fcos}
\end{align}
We remind that the $\cos\theta$-dependence of the rhs in \nr{fcos} is entirely in $\eta'_\rmi{max}$, see Eq. \nr{etaupper}.
We evaluated the tail strain with various large redshifts for binary lifetimes $1000$ yr to 10 Myr and BH masses $1,10,100,1000 M_\odot$. As lifetime can be basically anything greater than zero \cite{vanSon:2021zpk}, we have run our numerics with a wide spectrum of different lifetime scales, however, with a restriction to scales much smaller than the Hubble time to simplify the computation. The pre-merger and early post-merger evolution of the tail strain is shown in Fig. \ref{hplusinsp}, and the result for the long time tail is shown in Fig. \ref{hplusfuture}. Notice that the latter plot continues the former one; the range is just larger while the $y$-axis scale changes from logarithmic to linear. We also plot the time derivative of the tail strain in Fig. \ref{tailderiv}.
%We see that the magnitude of the tail is strongly dependent on the binary lifetime, although the functional profile is less so. 

One observes that the magnitude of the pre-merger tail is very small, of the order of $10^{-33}$, but that the post-merger tail increases very rapidly and essentially independently of the redshift of the binary. Note that the change is still relatively slow compared to, e.g., that of Fig. \ref{flatmemplot}, as time is now expressed in the units of Hubble time. After a conformal time of about $\Delta\eta = 0.1/H_0\approx 10^9$ yr, one reaches the level of about $10^{-24}$. For any merger observed today, the tail would be totally negligible, but the tails of merger signals, which have passed through our region earlier, could grow to sizeable values today. There thus would be a stochastic background of these tails, the total effect of which remains to be evaluated.

\subsubsection{Step function approximation}
\begin{figure}
	\begin{center}
		\includegraphics[width=0.90\textwidth]{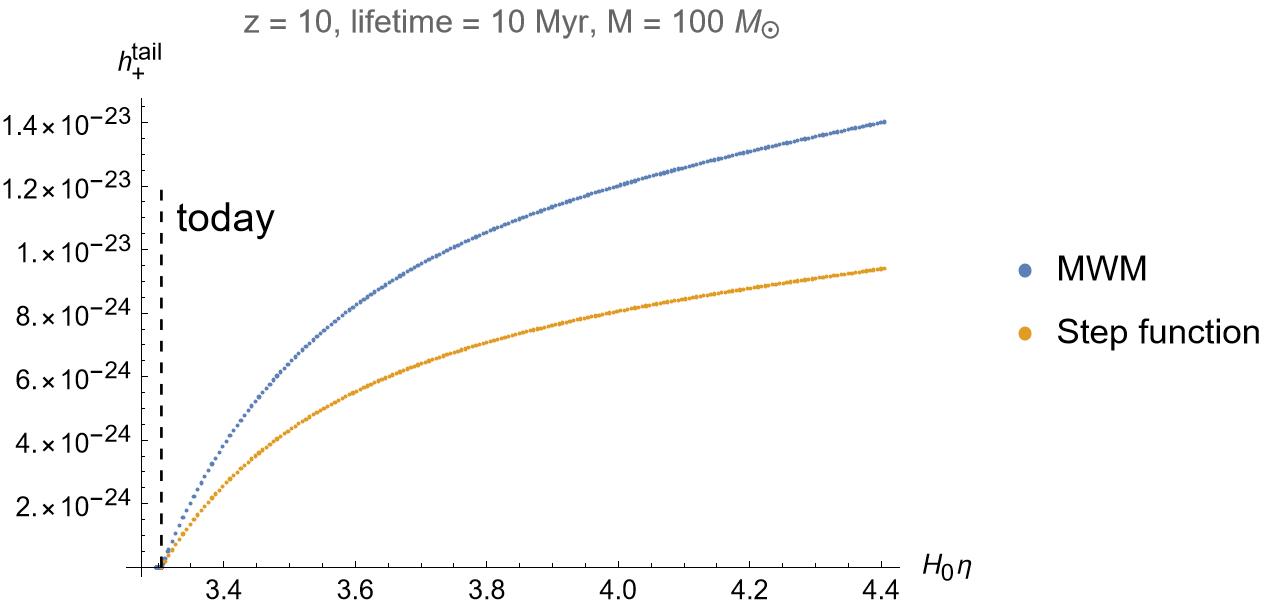}
	\end{center}
	\caption{\small The tail strain from MWM and the step function approximation compared. The quantity is computed for a binary at redshift $z=10$, with lifetime of $10$ Myr,	with BH mass = $100$ $M_\odot$. On the \(x\)-axis we have conformal time in the units of Hubble time, and on the \(y\)-axis we have the tail strain $h_+\tailup$ of an edge-on binary. The ratio between $h_+\tailup$ computed from the step function model and $h_+\tailup$ computed from MWM asymptotes quickly after the merger to roughly the value $0.67$. This is apparently due to the fact that the total radiated energy in the step function model divided by the total radiated energy in MWM is estimately $0.67$. The black dashed line marks the location of the present time on the time axis.
	}\label{mwm-vs-stepf}
\end{figure}

A great deal of the radiation is emitted during the merger so from a large scale perspective, all the energy is released instantaneously at the moment of coalescence, approximately.
To capture the relevant physics, we might simply consider total radiated energy given by a step function singularity at the moment of coalescence. The tail then is
\begin{equation}
	h\ijd^{\rmi{tail,TT}} = \frac{5}{8\pi}\frac{G E_\rmi{tot}}{a(\eta)(1+z)} \int d\Omega' (n_i' n_j')\ttu \mathcal{F}(\cos\theta')\int_{\etco}^{\eta-r}du' \left[ \frac{B(\eta,u')}{a(u')} - \frac{B(\eta,\emax)}{a(\emax)}\frac{d\emax}{du'} \right] \ ,
\end{equation}
where $E_\rmi{tot}$ is the total radiated energy and $\mathcal{F}(\cos\theta')$ is the angular distribution of radiation given by \nr{Fdef}. The second term inside the integral yields, via integration by substitution in reverse order, an integral over $\emax$ from $\emax(u'=\etco)$ to $\eta - r$ with the same integrand as in the first term. Combining the two integrals, we just get
\begin{equation}
	h\ijd^{\rmi{tail,TT}} = \frac{5}{8\pi} \frac{G E_\rmi{tot}}{a(\eta)(1+z)} \int d\Omega' (n_i' n_j')\ttu \mathcal{F}(\cos\theta')\int_{\etco}^{\emax}d\eta' \frac{B(\eta,\eta')}{a(\eta')} \ ,
\end{equation}
where
\begin{equation}
	\emax=\emax(\eta,r,u'=\etco,\cos\theta_{xx'})=\etco +\frac{1}{2} \frac{(\eta-\etco)^2 - r^2}{\eta-\etco-r \cos\theta_{xx'}}, \quad \cos\theta_{xx'}={\bf \hat{x}\cdot n'} \ .
\end{equation}
The integration now only takes place over the $\eta'$-coordinate of the null line emanating from the coalescence point in the direction $n'_i$. We can use the same Eqs. \nr{C1}--\nr{C3} for the $C$-coefficients depending on the function $f$ but now the function just is
\begin{equation}
	f(\cos\theta) = \int_{\etco}^{\emax(\eta,r,u'=\etco,\cos\theta)} d\eta' \frac{B(\eta,\eta')}{a(\eta')} \ .
\end{equation}
The corresponding plus polarized mode then is
\begin{equation}
	h\tailup_+ = \frac{G E_\rmi{tot}}{a(\eta)(1+z)}\frac{\sqrt{2}}{72\pi}(60 C_{4,0}-12 C_{6,0}-C_{6,2}(1-3 \cos ^2 \theta_x  )) \sin ^2\theta_x \ .
\end{equation}
We approximate that $E_\rmi{tot} \approx 0.1 M$, based on $M_f \approx 1.9 M$ for the equal-mass binary  (recall the estimate in Sec. 2.2).
We show the result from numerical integration for $h\tailup_+$ in Fig. \ref{mwm-vs-stepf}, where it is compared to the tail strain computed from MWM.

Note first the this approximation is valid only for $\eta>r+\eta_\rmi{coal}=\eta_f$, for the post-merger tail. The numerical evaluation shows that this simple approximation agrees  well with the more complicated MWM model, the observed difference is even undertood in terms of total radiation energies. There is some similarity with the simple estimate of \cite{Garfinkle:2016nhe}, where the magnitude of the main light cone signal is basically the Coulomb-like term $G E_\rmi{tot}/r$ with angular effects, here one has $G E_\rmi{tot}\times$ a twice TT-projected kinematically constrained integral of $B(\eta,\eta')/a(\eta')\sim 1/\eta\,\,{\rm terms}$. It would be nice to have a physical interpretation of this integral. Could it be related to the conservation law \nr{conserv} which also couples $B$ and $a$?

\subsection{Memory effect as an observable}

Recall from Sec. \ref{flatspacemem} that the memory effect manifests itself as a distance change between two freely falling test masses. This we were able to see from the geodesic deviation equation \nr{deviation}, which describes the effect of spacetime curvature on geodesic motion. In FRW background, there is also an effect due to background spacetime curvature, which makes it initially nonstraightforward to read off effects due to curvature of GWs from the equation. In \cite{Chu:2019ssw} it was shown that adopting the locally flat Fermi Normal Coordinates centered around the observer makes it possible to extract away the effect of background curvature. The full geodesic deviation equation is
\begin{equation}\label{jacobi}
	\frac{d^2 \xi^i}{dt^2} = \tensor{R}{^i_0_0_j} \xi^j = \left(\tensor{\bar{R}}{^i_0_0_j} + \delta \tensor{R}{^i_0_0_j} \right) \xi^j \ ,
\end{equation}
where $\tensor{\bar{R}}{^i_0_0_j}$ is the background Riemann tensor, and $\delta \tensor{R}{^i_0_0_j}$ is a perturbation thereof.
Integrating this twice over time, we get
\begin{equation}
	\Delta \xi^i = \int dt \int dt' \tensor{\bar{R}}{^i_0_0_j} \xi^j + \int dt \int dt' \delta \tensor{R}{^i_0_0_j} \xi^j \ ,
\end{equation}
where the first term is deviation due to background curvature and the second one deviation due to GW. The former potentially represents a large effect due to background curvature so the standard approximation that $\xi^i$ stays constant to leading order is in general inapplicable. However, measurement takes place over small timescales, during which the effect of background curvature is negligible. Moreover, in practice the detectors are positioned inside a gravitationally bound region, such as a galaxy, where Hubble expansion does not take place so the deviation due to FRW background can be anyway neglected. Hence, as a locally measurable observable the memory effect is, as in \nr{memoryeq}, given by
\begin{equation}
	\Delta \xi^i \approx \frac{1}{2}\Delta h\ttu\ijd \xi^j \ ,
\end{equation}
i.e., $\xi^i$ experiences a small shift due to gravitational radiation. This simple corollary of the geodesic deviation equation is applicable for measurement processes with short duration. However, e.g. Figures \ref{hplusinsp}, \ref{hplusfuture} exhibit a long-time effect where this approximation clearly breaks down. For these long-time effects, the memory observable is still in principle determined from \nr{jacobi}, from which $\xi^i$ can be solved since the background is known, although this probably requires numerics. But how to measure such a long-time observable is an open question that we do not try to answer in this paper.

%%%%%%%%%%%%%%%%%%%%%%%%%%%
\section{Discussion}\label{disc}
%%%%%%%%%%%%%%%%%%%%%%%%%%%
In this paper, we studied gravitational radiation sourced by a quasicircular binary  in the $\Lambda$CDM model, especially in view of computing in some detail the tail of the radiation spectrum. The tail is due to the curvature of the FRW spacetime in which the radiation is propagating, in flat spacetime the radiation moves along the light cone. Earlier significant work on tail memory has, for example,  been carried out by Chu \cite{Chu:2015yua,Chu:2016ngc}.

The  $\Lambda$CDM model, also called the concordance model, contains cold dark matter (with $\Omega_m=0.3$) and the cosmological constant ($\Omega_\Lambda=0.7)$, all other components are neglected. We have seen that in this unified treatment the potential $V(\eta)=a''(\eta)/a(\eta)$ has a surprisingly symmetric two-peak appearance with poles at $\eta=0$ (only matter) and at $\eta_\rmi{max}=4.4457H_0^{-1}$ (future comoving visibility limit, only cosmological constant). The leading terms of $V$ are $2/\eta^2,\, 2/(\eta_\rmi{max}-\eta)^2$ and coefficients of expansions around $\eta=0$ and $\eta=\eta_\rmi{max}$ are exactly the same until a difference appears to order $(\eta_\rmi{max}-\eta)^{13}$. 

For early history of the binary we used a simple equal-mass time-dependent Keplerian model tuned to given values of the redshift of the merger, age of the binary, and mass of the constituents. These lead to fixed initial  values, formation time, and initial size of the binary, but leave the details of complicated astrophysical formation phenomena \cite{vanSon:2021zpk} aside. For the late inspiral and merger phases of the binary a more detailed Minimal Waveform Model was used. %These details of the history of the binary should not affect the essential properties of the tail.
Actually, it appeared that the dominant source of the second order tail is simply the main first order light cone merger signal.

For first order oscillating GW the light cone part is, of course, very well known, this is the part that has been observed. The computation of the tail involves a time integration over the binary from formation up to some source time. The integrand, however, is (approximately) an exact differential. The integration can thus be carried out and leaves only contributions from the upper limit (on the past light cone of the observation point) and the lower limit. The upper limit corresponds to tail  radiation effectively propagating along the light cone, the lower limit is time independent and represents the radiation from the formation of the binary. It also would represent the tail memory, there is no memory from the light cone contribution. This contribution to the first order tail memory does not violate the theorem in \cite{Tolish:2016ggo} since the method for extracting the memory there is different.

The computation of the 2nd order tail, sourced by first order radiation, is more complicated, but also leads to a much more interesting result. One finds that there is a growing contribution to the tail {\em after} the arrival of the light cone signal from the merger. This is concretely due to subluminal propagation of tail radiation, GW in curved spacetime is dispersive, like electromagnetic radiation in matter. The growth takes place over cosmological times, times of the order of $H_0^{-1}$. We talk here of "radiation", but no statement of the physical substance of tail is intended. We have a definite equation for the metric perturbation and "tail" is simply a name for its mathematical solution. It is common to describe the tail as arising from back scattering against curvature of space, but we see no trace of this. For mass effects, see \cite{Kilicarslan:2018bia}.

We note that the light cone gravitational wave memory has been associated with asymptotic symmetries due to massless modes propagating to null infinity \cite{Strominger:2017zoo}, in cosmological context see \cite{Bonga:2020fhx,Enriquez-Rojo:2020miw,Enriquez-Rojo:2022onp}. The GW tail has to do with "massive" modes, but if the relationship between memory and asymptotic symmetries be true (for possible reasons to the contrary, however, see \cite{Garfinkle:2022dnm}), then the tail part of the memory should give rise to novel symmetries. We have identified a conserved quantity \nr{conserv} which may be linked with some new supertranslation invariance property (at future null infinity).

The tail memory of any GW burst observed today is negligible, but the real issue is what is the total strain arising from a stochastic distribution of tails from all the mergers which have taken place earlier over cosmological times. For the light cone GW memory this stochastic GW memory background has been discussed in \cite{Zhao:2021zlr}. Having reliable information on $N(m_1,m_2,z)$, number of binaries of masses $m_1,m_2$ at various redshifts $z$, should permit a reliable estimate of this background also in our case of tail memory.

\vspace{0.8cm}
{\bf Acknowledgements} We thank J.~Kastikainen for useful discussions at different stages of this work.
N.~J. and M.~S. have been supported in part by the Academy of Finland grant no. 1322307. M.~S. is also supported by the Finnish Cultural Foundation.

\appendix
\section{Concordance model}\label{concordance}
In this appendix, we summarize some basic facts about the concordance model, i.e., the FRW model with $\epsilon=\epsilon_\rmi{vac}+\epsilon_m=\Omega_\Lambda \rho_c+\Omega_m\rho_c/a^3$, 
$p=-\epsilon_\rmi{vac}$, $H_0^2=8\pi G\rho_c/3$, $H_0^{-1}=h^{-1} 9.78\,{\rm Ga}$, $\Omega_\Lambda+\Omega_m=1=0.7+0.3$, satisfying the Friedmann equations
\ba
{\dot a^2\over a^2}&=&{8\pi \GN\over 3}\epsilon=H_0^2\left(\Omega_\Lambda+\Omega_m \frac{1}{a^3}\right)\\
{\ddot a\over a}&=&-{4\pi \GN\over 3}(\epsilon +3 p)=-{4\pi \GN\over 3}(\epsilon_m -2\epsilon_\rmi{vac})=
H_0^2\left(\Omega_\Lambda-\Omega_m{1\over 2a^3}\right) \ .
\ea
These equations imply that
\be
{a''(\eta)\over a(\eta)}=\dot a^2+a\,\ddot a=H_0^2\left(2\OL a^2+\fra12 \Om a^{-1}\right) \ .
\label{appa}
\ee

Expansion factor is solved from
\be
{da\over dt}=H_0\sqrt{\OL a^2+\Om a^{-1}} \ .
\ee
Integrating and inverting
\be
a(t)=\left({\Om\over \OL}\right)^{1/3}\sinh^\fra23\left(\fra32\sqrt{\OL}H_0t\right),\qquad 
{\epsilon_\Lambda\over \epsilon_m}={\OL\over\Om}a^3 =\sinh^2\left(\fra32\sqrt{\OL}H_0t\right) \ .
\ee

\begin{figure}[!t]
	\begin{center}
		\includegraphics[width=0.49\textwidth]{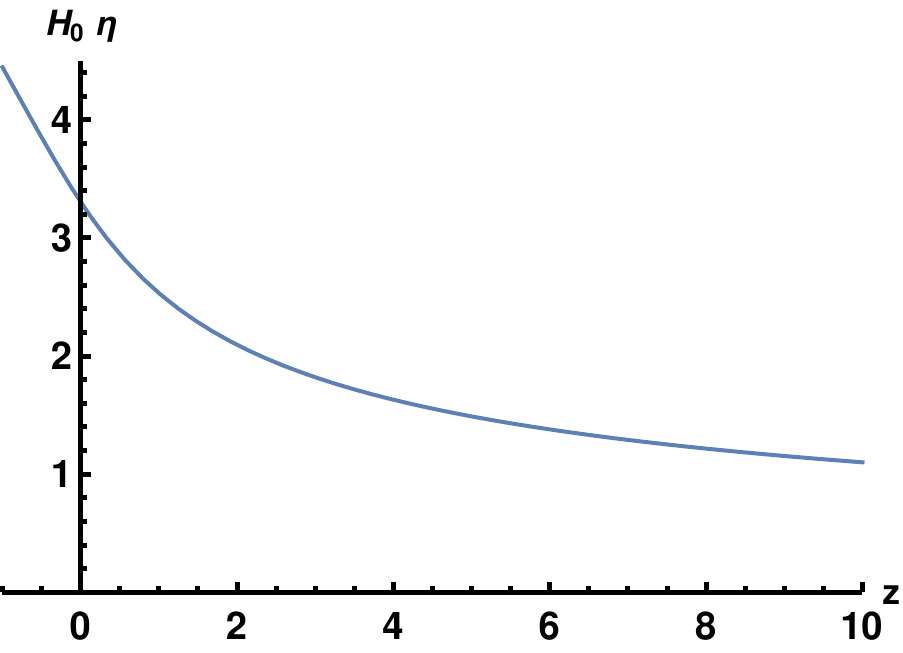}
		\includegraphics[width=0.49\textwidth]{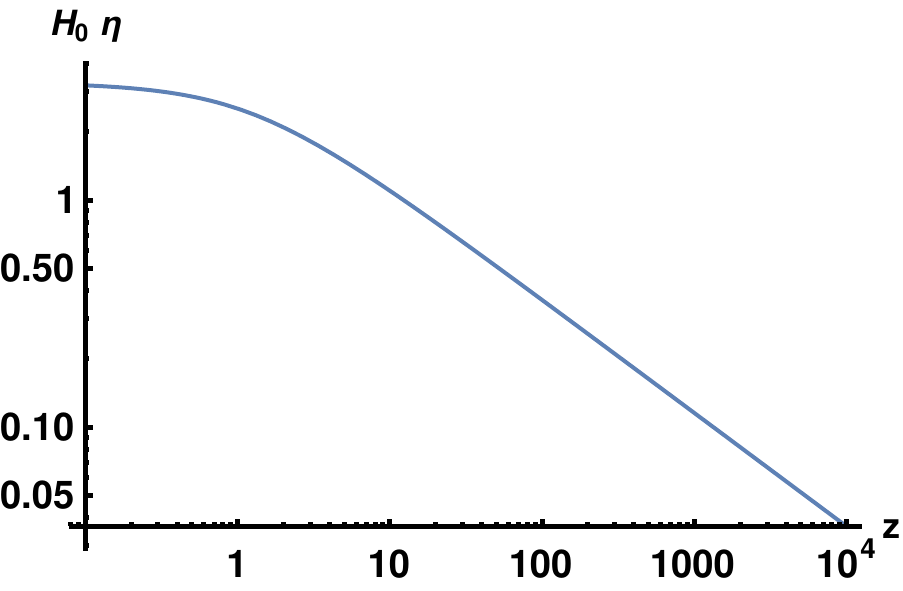}
	\end{center}
	\caption{\small Conformal time $\eta(z)$ times Hubble constant $H_0$, as a function of redshift $z, \,\,a=1/(1+z)$. Left panel is for small $z$ (also for $-1<z<0$),
		right panel for large $z$.
	}\label{etazfig}
\end{figure}

Conformal time $\eta$ and redshift $z$ are defined by 
\begin{equation}
	d\eta = \frac{dt}{a(t)} \ , \quad 1+z = \frac{1}{a} \ .
\end{equation}
Fixing the constant by $\eta(t=0)=0$ and integrating:
\be
H_0\eta(t)=H_0\int_0^t {dt'\over a(t')}={2\over\Om^{1/2}}a(t)^{1/2}\, {}_2F_1(\fra16,\fra12,\fra76,-{\OL\over\Om}a(t)^3) \ .
\label{a(eta)}
\ee 
This is a monotonically increasing curve with small $t$ or $a$ limit
\be
H_0\eta ={2\over\sqrt{\Om}}\sqrt{a}-{\OL\over 7\Om^{3/2}}a^3\sqrt{a}+ \ldots
\label{smalleta}
\ee
and the large $t$  limit
\be
H_0\eta= {2\over (\Om^2\OL)^{1/6}}{\Gamma(\fra13)\Gamma(\fra76)\over\sqrt{\pi}} -{1\over\sqrt{\OL}}\,{1\over a}
+{\Om\over 8\OL^{3/2}}\,{1\over a^4}+ \ldots \ .
\label{largeeta}
\ee
The asymptotic value
\be
H_0\,\eta_\rmi{max}={2\over (\Om^2\OL)^{1/6}}{\Gamma(\fra13)\Gamma(\fra76)\over\sqrt{\pi}}\approx 4.445744
\ee
is the comoving visibility limit, the largest value of $\eta$ ever attains. Today or $a=1$ corresponds to
\be
H_0\,\eta_\rmi{today} \approx 3.3051 \ ,
\ee
the expansion turns from deceleration to acceleration at $\epsilon_m=2\epsilon_\rmi{vac}$ or at 
$a=(\Om/(2\OL))^{1/3}$ or at
\be
H_0\,\eta_\rmi{accel} \approx 2.740 \ ,
\ee
halfway to comoving visibility limit 
\be
H_0\,\eta_\rmi{middle} \approx 2.22287 \ ,
\ee
was reached at redshift $z\approx1.653$ and 
at redshift 10 or for $a=1/11$ finally
\be
H_0\,\eta(z=10)\approx1.1008 \ .
\ee
The full dependence of $H_0\eta(z)$ on the redshift $z$ is plotted in Fig. \ref{etazfig}. Note that in the text conformal time is mostly quoted without the factor $H_0$.

\begin{figure}[!t]
	\begin{center}
		\includegraphics[width=0.49\textwidth]{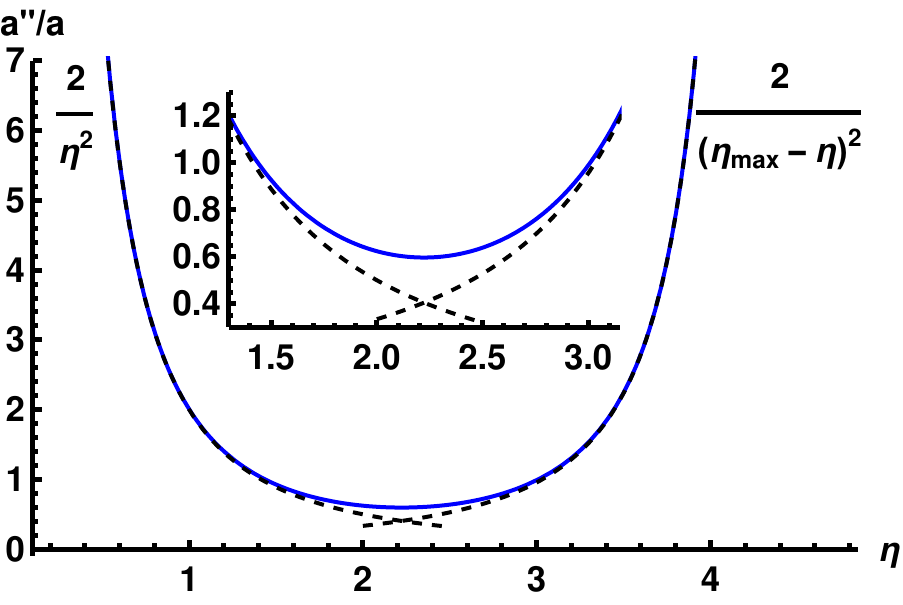}
	\end{center}
	\caption{\small The function $a''/a$ as a function of $\eta$. The curve is very well approximated by the two peaks (black dashed curves)
	corresponding to mass dominated expansion ($\eta\to0$) or de Sitter expansion ($\eta\to\eta_\rmi{max}=4.4457$). The
	inset plot shows a close-up of the transition region in the middle.
	}\label{appovera}
\end{figure}

From the representation
\be\label{potential}
{a''\over a}=H_0^2\left(2\OL a^2+\fra12 \Om a^{-1}\right)
\ee
it is obvious that $a''/a$ will have two singularities corresponding to mass dominated expansion at
$a\to0$ and dS expansion at $t\to\infty$ or $\eta\to\eta_\rmi{max}$. In these limits the curve (\ref{a(eta)})
is easy to invert, for small $\eta$ from (\ref{smalleta}) $a\to \fra14\Om\eta^2$ and
\be
{a''\over a}\to {2\over \eta^2} \ , \quad \eta \ll 1
\ee
and for large $\eta$ from (\ref{largeeta})
\be
\eta=\eta_\rmi{max} -{1\over\sqrt{\OL} a}
\ee
so that 
\be
{a''\over a}\to {2\over (\eta_\rmi{max}-\eta)^2} \ , \quad \eta \lesssim \eta_\rmi{max}
\ee
with exactly the same coefficient as at small $\eta$. A full numerical evaluation of $a''/a$ is plotted in
Fig. \ref{appovera} and one sees that the full curve is very well approximated by the two peaks.

To study the exact form of $a''/a$ we expand at small $\eta$ in powers of $\eta$ and at larger $\eta$ in powers 
of $\eta_\rmi{max}-\eta$. First, inverting the equation (\ref{a(eta)}) one has
\be
a=\fra14\Om\eta^2\left[1+{1\over 7\cdot 2^6}\OL\Om^2\eta^6+{1\over 7^2\cdot13\cdot 2^{10}}(\OL\Om^2\eta^6)^2+
{1\over 7^2\cdot13\cdot19\cdot 2^{17}}(\OL\Om^2\eta^6)^3+...\right]
\ee
Inserting this to $2\OL a^2+\fra12 \Om/a$ gives the result 
\be
{a''\over a}={2\over\eta^2}+{3^3\,\OL\Om^2\over 7\cdot 2^5}\eta^4+{3^6\,\OL^2\Om^4\over7^2\cdot13\cdot 2^{11}}\eta^{10}
+{3^9\,\OL^3\Om^6\over 7^3\cdot13\cdot19\cdot 2^{17}}\eta^{16}
+{3^{13}\,\OL^4\Om^8\over 5\cdot 7^4\cdot13^2\cdot19\cdot 2^{23}}\eta^{22}+
{\cal O}(\eta^{28})
\label{smalletaexp}
\ee
Near the other peak one expands (\ref{a(eta)}) for large $a$, moves the asymptotic limit
$\eta_\rmi{max}$ to the left hand side of the equation and writes it in the form
\be
\Delta\eta=\eta_\rmi{max}-\eta ={1\over\sqrt{\OL}a}-{\Om\over 8\OL^{3/2}a^4}+
{3\Om^2\over 56\OL^{5/2}a^7}-{\Om^3\over 32\OL^{7/2}a^{10}}+{\cal O}\left({1\over a^{13}}\right) \ .
\ee
This is then inverted perturbatively to give $a=a(\Delta\eta)$:
\be
a(\Delta\eta)={1\over\sqrt{\OL}\Delta\eta}-{\Om\over 8}\Delta\eta^2+{3\sqrt{\OL}\Om^2\over 448}\Delta\eta^5
-{\OL\Om^3\over 3584}\Delta\eta^8+{\cal O}(\Delta\eta^{11}) \ .
\ee
When this is inserted to $2\OL a^2+\fra12 \Om/a$ one finds near the dS peak the expansion
\be
{a''\over a}={2\over\Delta\eta^2}+{3^3\,\OL\Om^2\over 7\cdot 2^5}\Delta\eta^4+
{3^6\,\OL^2\Om^4\over7^2\cdot13\cdot 2^{11}}\Delta\eta^{10}
+{3\,\OL^{5/2}\Om^5\over 7^2\cdot13\cdot 2^{11}}\Delta\eta^{13}
+{3^2\cdot71\,\OL^3\Om^6\over 7^3\cdot13\cdot 2^{17}}\Delta\eta^{16}+{\cal O}(\eta^{22}) \ .
\label{dSexp}
\ee
Perhaps strikingly, when measured by the distance from the singularity, the three leading terms are exactly
the same! The first time a difference appears is that near the dS peak there is an order $\Delta\eta^{13}$
term (with a small numerical coefficient) with no $\eta^{13}$ counterpart near the mass peak.

\begin{figure}[!t]
	\begin{center}
		\includegraphics[width=0.49\textwidth]{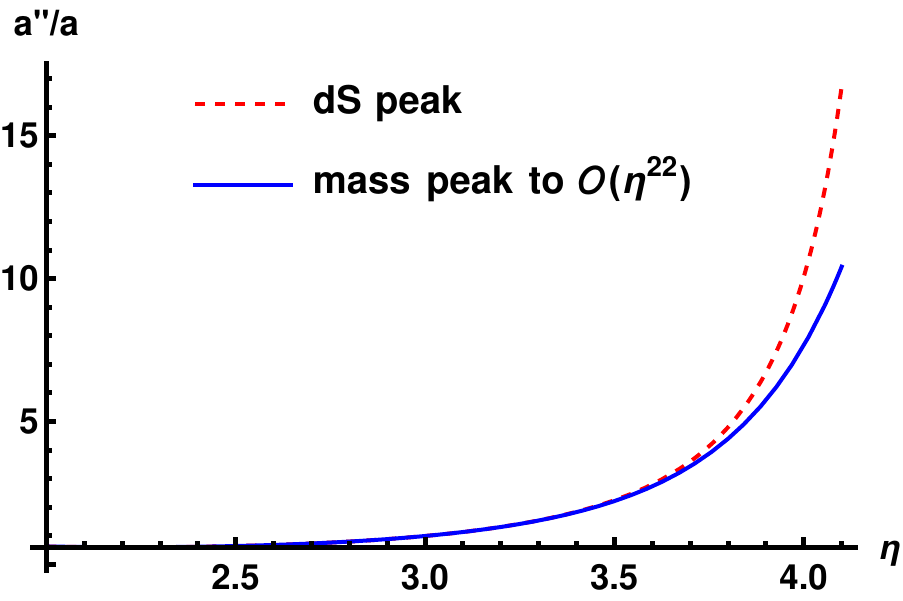}
		\includegraphics[width=0.49\textwidth]{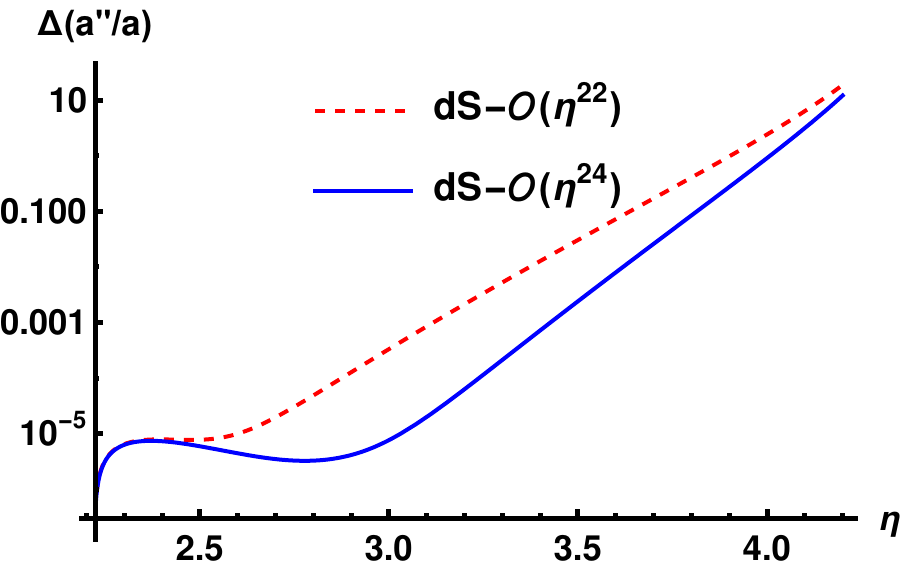}
	\end{center}
	\caption{\small Left panel: The small $\eta$ expansion of $a''/a$ to order $\eta^{22}$ extended to the region close to the
		dS peak to order $(\eta-\eta_\rmi{max})^{25}$ at $\eta_\rmi{max}=4.4457$. 
		The right panel shows how the small $\eta$ expansions near $\eta_\rmi{max}$
		deviate from the dS peak expansions. In more detail, for $\eta_\rmi{today}=3.3$, the dS peak expansion 
		(\protect\ref{dSexp}) to order
		$(\eta_\rmi{max}-\eta)^{25}$ gives the value $1.53665$, while the mass peak small $\eta$ expansion  
		(\protect\ref{smalletaexp}) to order $\eta^{22}$ gives the value $1.53141$ and to order $\eta^{34}$
		the value $1.53644$.
	}\label{opppeak}
\end{figure}

Of practical interest, for approximating the potential $a''/a$ over a wider range, is that the expansions can
be extended to near the opposite peak. This is illustrated in Fig. \ref{opppeak}, in which the expansion 
(\ref{smalletaexp}) with the $\eta^{22}$ term is extended to the neighborhood of the opposite dS peak.
One sees that the expansion produces the overall structure of the potential very well. Differences start to be significant 
when one approaches $\eta_\rmi{max}$. Deviations can easily be decreased by adding more terms to the $\eta$
expansion.

\section{Equations of motion to second order}\label{perturb}
In this appendix, overbars refer to background quantities (except in the trace-reversed perturbation defined below). Indices $(1),(2)$ refer to the perturbative order of the quantity indexed.

Expand the metric to second order:
\begin{align}
	g\munud = \bar{g}\munud +  h\munud\puf +  h\munud\pus \ .
\end{align}
The perturbed Einstein equation then reads
\begin{equation}
	\bar{G}\munud + G\munud\puf + G\munud\pus = 8\pi G \left(\bar{T}\munud + T\munud\puf + T\munud\pus\right) \ .
\end{equation}
We solve this equation order by order. Here we assume that the linear order matter stress-energy perturbations are restricted to a compact region of spacetime, and that the second order matter stress-energy perturbations are negligible as GW sources. The system we solve is therefore
\begin{align}
	\bar{G}\munud &= 8\pi G \,\bar{T}\munud \label{bgeq} \\
	G\munud\puf &= 8\pi G \, T\munud\puf \label{1stord}\\
	G\munud\pus &= 8\pi G \, T\munud\pus \ . \label{2ndord}
\end{align}
The background equations of motion \nr{bgeq} in our case are of course the two Friedmann equations. We first turn to analyze the equation for linear perturbations \nr{1stord}. We just denote $h\munud \equiv h\munud\puf$ and $T\munud \equiv T\munud\puf$ for now, to keep the expressions cleaner.
In terms of the trace-reversed perturbation
\begin{equation}
	\bar{h}\munud = h\munud - \frac{1}{2}\bar{g}\munud h, \quad h \equiv \Tr h\munud \ ,
\end{equation}
the first order equations of motion are
\begin{equation}\label{linearEinstein}
	\nab\ad\nab\au\bh\munud - 2\nab\au \nab_{(\mu}\bh_{\nu)\alpha} + \bg\munud \nab\au \nab\bu \bh\abd - \bg\munud \bh \abu \br\abd + \bh\munud \br = - 16 \pi G T\munud \ .
\end{equation}
One may then apply the Ricci identity 
\begin{equation}
	\left[\na\mud\,,\na\nud\right]V\ad = R_{\mu\nu\alpha\beta}V\bu
\end{equation}
on the symmetrized term, and the lhs of \nr{linearEinstein} becomes
\begin{equation}
	\nab\ad\nab\au\bh\munud + \bg\munud \nab\au\nab\bu \hb\abd - 2 \nab_{(\mu}\nab\au \bh_{\nu)\alpha} - 2 \br_{\alpha\mu\nu\beta}\bh\abu - 2\tensor{\br}{^\alpha_{(\mu}}\bh_{\nu)\alpha} - \bg\munud \bh \abu \br\abd + \bh\munud \br \ .
\end{equation}
This holds for a generic background and gauge. If we impose the Lorenz gauge condition
\begin{equation}
	\nab\mup \bh\munud = 0,
\end{equation}
then the linearized Einstein equation simplifies to
\begin{equation}\label{lorenzeinstein}
	\nab\ad\nab\au\bh\munud - 2 \br_{\alpha\mu\nu\beta}\bh\abu - 2\tensor{\br}{^\alpha_{(\mu}}\bh_{\nu)\alpha} - \bg\munud \bh\abu \br\abd + \bh\munud \br = -16 \pi G T\munud \ .
\end{equation}
Now take the background geometry to be FRW and write the perturbed metric to first order as
\begin{equation}
	ds^2 = a(\eta)^2 \left(-d\eta^2 + (\delta\ijd + h\ttu\ijd )dx^i dx^j\right) \ ,
\end{equation}
i.e., we only consider perturbations in the tensor sector, $h\ttu\ijd$ being the gauge-invariant TT perturbation. The Einstein equation then becomes, to first order in perturbation theory,
\begin{equation}\label{htteom}
	\Box h\ttu\ijd - 2 \frac{a'}{a} \pt_\eta h\ttu\ijd = - 16\pi G T\ijd\ttu \ .
\end{equation}
Here we needed to be careful with the perturbation of stress-energy, as linearizing yields
\begin{equation}
	T\munud\puf = \left( g_{\mu\rho}T^\rho\nud \right)\puf = h_{\mu\rho}\puf \bar{T}^\rho_\mu + \bg_{\mu\rho} \delta T^\rho\nud \ ,
\end{equation}
where $\delta T^\rho\nud$ is the linear order stress-energy perturbation. The first of these terms, after applying the background Einstein equation, exactly cancels a term proportional to $h_{\mu\rho}$ in the linearized Einstein tensor; we are then left with the expression on the lhs of \nr{htteom}, and the remaining stress-energy term gives the rhs.

We then turn our attention to the second order Einstein equation \nr{2ndord}. Notice that even though we assume that the second order matter stress-energy is negligible, an analysis similar to the previous equation reveals that terms involving coupling between the first order metric perturbations and linearized stress-energy may now serve as a second order source:
\begin{equation}\label{nonlinTmunu}
	T\munud\pus = h_{\mu\rho}\pus \bar{T}^\rho\nud + h_{\mu\rho}\puf \delta T^\rho\nud + \bg_{\mu\rho} \delta^2 T^\rho\nud \ ,
\end{equation}
$\delta^2 T^\rho\nud$ being the second order stress-energy perturbation. As already stated, we shall assume that $\delta^2 T^\rho\nud$ is negligible. As for the GW--source coupling term, it describes scattering of GWs off of the gravitational potential of the first order source restricted to the compact region. Interesting as these scattering phenomena are, we will focus here on a different effect and, consequently, ignore the GW--source coupling term in the second order equation of motion. We thus get the equation
\begin{equation}\label{2ndEinstein}
	G\munud\pus - 8\pi G h_{\mu\rho}\pus \bar{T}^\rho\nud = G\munud\pus - h_{\mu\rho}\pus \bar{G}^\rho\nud = 0 \ .
\end{equation}
Next, decompose the nonlinear Einstein tensor to a piece that is linear in the second order metric perturbation and a piece that is quadratic in the first order perturbation of the metric:
\begin{equation}
	G\munud\pus = G^{(1)}\munud\left[h\abd\pus\right] + G^{(2)}\munud\left[h\abd\puf\right] \ .
\end{equation}
We then move the terms that depend on the first order perturbation $h\puf\munud$ to the rhs of \nr{2ndEinstein}, and interpret the resulting quantity as effective stress-energy that gives rise to the second order metric perturbations on the lhs:
\begin{equation}\label{unavg}
	G^{(1)}\munud\left[h\abd\pus\right] -  h_{\mu\rho}\pus \bar{G}^\rho\nud = - G^{(2)}\munud\left[ h\abd\puf\right] \equiv 8\pi G \tau\munud \ .
\end{equation}
In general, on the rhs, the tensor
\begin{equation}
	\tau\munud = -\frac{1}{8\pi G} G^{(2)}\munud\left[ h\abd\puf\right]
\end{equation}
involves both long and short wavelength modes. We therefore apply the well-known "coarse-graining" scheme where we integrate out the fast oscillating modes to get an equation for GWs induced by stress-energy of first order GWs. Taking an average over several wavelengths of both sides of \nr{unavg}, we have
\begin{equation}\label{avgEinstein}
	\Big \langle G^{(1)}\munud\left[h\abd\pus\right] -  h_{\mu\rho}\pus \bar{G}^\rho\nud \Big\rangle= 8\pi G\, t\munud,
\end{equation}
where now
\begin{equation}
	t\munud = \frac{1}{32\pi G} \Big\langle \pt\mud h\ttu\abd \pt\nud h\ttd\abu \Big\rangle \ ,
\end{equation}
where we again dropped the superscript $(1)$ to get a tidier expression. This is just the famous Isaacson stress-energy tensor for gravitational waves \cite{Isaacson:1968zza}. 
%Note that deriving the Isaacson formula, we used the first order vacuum Einstein equation $\puf G\mup\nud = 0$, which we are able to do outside the compact region containing first order stress-energy sources \ack{check this in detail}. 

On the lhs of \nr{avgEinstein}, we have the linearized Einstein tensor with second order metric perturbation substituted inside minus the second order metric perturbation coupling to the background Einstein tensor. This expression also involves in general both high and low frequency modes, the former of which gets integrated out under the average. We therefore get an equation of motion for the slowly varying piece of the second order metric perturbation.
Concretely, this procedure yields the same equation as \nr{linearEinstein} but now for $h\munud\pus$ instead of $h\munud\puf$ and with the lhs averaged. If we pick the Lorenz gauge for $h\munud\pus$ too, the equation becomes the same as \nr{lorenzeinstein}. Further, going into the TT gauge reduces this to Eq. \nr{htteom}. However, to be precise, we are not allowed to transform into the TT gauge in the presence of sources, which are exactly nonvanishing even at large distances from the first order source. The stress-energy of gravitational waves falls off as inverse distance squared so for an observer far away from the compact source, vacuum approximation is very accurate and hence the TT gauge choice is permissible. Therefore, we get
\begin{equation}
	\Big \langle \Box h^{(2),\rmi{TT}}\ijd - 2 \frac{a'}{a} \pt_\eta h^{(2),\rmi{TT}}\ijd \Big\rangle \approx -16\pi G\, t\ijd\ttu \ .
\end{equation}
Now, the scale factors change over timescales that are very large compared to the wavelength of the high-frequency GW so they can clearly be pulled out of the average operation. Hence, we only need to concern ourselves with what happens to the derivatives of different modes under averaging. We assume here that a quantity that changes over much larger scales than the scale of averaging has derivatives that only change over scales large compared to the scale of averaging. We may therefore approximate that the derivatives commute with averaging operation, which yields
\begin{equation}\label{averageeom}
	\Box h^{\rmi{low,TT}}\ijd - 2 \frac{a'}{a} \pt_\eta h^{\rmi{low,TT}}\ijd \approx\Box \big \langle h\ttu\ijd\big\rangle - 2 \frac{a'}{a} \pt_\eta \big \langle h\ttu\ijd\big\rangle \approx -16\pi G\, t\ijd\ttu \ ,
\end{equation}
where $h^{\rmi{low,TT}}\ijd$ stands for the long wavelength modes.
Here we have an equation for GWs that are generated by first order GW stress-energy. In the main text, we simply denote $h\ttu\ijd \equiv h^{\rmi{low,TT}}\ijd$ even though we only solve for modes that survive after averaging. The differential operator on the lhs is the same as in the first order equation of motion so we may solve \nr{averageeom} in a similar way as in the first order case. It should be noted that we derived our Eq. \nr{averageeom} using several approximations, and the argument might still need refinements.

%\ack{What happens to the memory, how do the integrated-out modes contribute? The long u integral effectively acts as a "coarse-grainer"? So that the memory effect is actually only determined by the low frequency modes?}

\section{Position dependence of the tail two-point function}\label{Bapp}
In a homogeneous and isotropic background, position dependence enters the tail two-point function via modulus of the spatial separation vector: $B(x,x') = B(\eta, \eta', \abs{{\bf x - x'}})$. Here we show that only in very special cases the tail two-point function $B(x,x')$ is fully position-independent (this is of course a coordinate-dependent statement; here we are assuming the standard comoving frame of reference).

Assume that $\vec{\nabla} B = 0$. Then \nr{Beq} simply becomes
\begin{equation}\label{nabla0}
	B'' - V B = B'' - \frac{a''}{a}B =0 \ .
\end{equation}
As an aside, note that this can be written as
\begin{equation}
	(a B' - a' B)' = 0 \ .
\end{equation}
Integrating this from $\eta'$ to $\eta$ yields
\begin{equation}
	a(\eta)B'(\eta)-a'(\eta)B(\eta) = a(\eta')B'(\eta')-a'(\eta')B(\eta') = const.
\end{equation}
This is just the statement that the Wronskian of $a$ and $B$ is a constant. If there is no position dependence in $B$, we can simply extend the boundary solution \nr{bdrysol} into the bulk and plug it into \nr{nabla0}. This yields the following integral equation for $V$:
\begin{equation}
	V(\eta) + V(\eta') = \frac{2}{(\eta-\eta')} \int_{\eta'}^{\eta} V(\zeta)d\zeta + \frac{1}{2}\left( \int_{\eta'}^{\eta} V(\zeta)d\zeta\right)^2 \ .
\end{equation}
Note that this is symmetric with respect to interchanging $\eta$ and $\eta'$. Denoting $f(\eta,\eta') \equiv \int^\eta_{\eta'} V(\zeta)d\zeta$, we can write this as a partial differential equation
\begin{equation}
	\pt_\eta f - \pt_{\eta'} f = \frac{2 f}{\eta - \eta'} + \frac{1}{2}f^2 \ ,
\end{equation}
whose solution is
\begin{equation}
	f(\eta,\eta') = \frac{2(\eta-\eta')}{\eta\, \eta' + F(\eta+\eta')} \ .
\end{equation}
Here $F$ is a function that must satisfy the condition
\begin{equation}\label{Ccond}
	(F(\eta+\eta')+\eta\,\eta')F''(\eta+\eta') -2 F'(\eta+\eta')(\eta+\eta'+F'(\eta+\eta')) + 2 F(\eta+\eta') = 0
\end{equation}
that comes from the fact that $\pt_{\eta'}\pt_\eta f = \pt_\eta \pt_{\eta'} f = \pt_\eta V(\eta') = \pt_{\eta'} V(\eta) = 0$. Clearly $F = 0$ satisfies the above condition, and it actually yields the known solution
\begin{equation}
	B(\eta,\eta') = \frac{1}{\eta \,\eta'} \ ,
\end{equation}
which corresponds to the potential
\begin{equation}
	V = \frac{2}{\eta^2} \ .
\end{equation}
Recalling the definition of $V = a''/a$, we get the solution for the scale factor:
\begin{equation}
	a(\eta) = c_1 \eta^2 + \frac{c_2}{\eta} \ .
\end{equation}
The first of these terms corresponds to matter-dominated expansion, the second one describes de Sitter expansion. For a generic $F$, the scale factor is determined by
\begin{equation}\label{aeq}
	\frac{a''}{a}(\eta) = \frac{2(\eta'^2+ F(\eta+\eta') - (\eta-\eta')F'(\eta+\eta'))}{(\eta \, \eta' + F(\eta+\eta'))^2} \ .
\end{equation}
Note that despite appearances, rhs only depends on $\eta$ given that $F$ solves \nr{Ccond}.
Any scale factor that satisfies \nr{aeq} for some solution $F$ of \nr{Ccond} only develops position-independent tails (as seen by an observer comoving with the cosmic fluid).

%\subsection{Further facts about $B(x,x')$}
%Let's denote $R \equiv \abs{{\bf x - x'}}$. Since $B(x,x')$ is invariant under spatial rotations about $R=0$, $B(\eta,\eta',R)$ must have a local maximum or minimum at $R=0$, or be a constant on a spatial hypersurface. This means that $\vec{\na} B (\eta,\eta',0) = 0$ for each $\eta, \eta'$.

\section{Angular technicalities}\label{angularintegrals}
In this appendix, we will present some additional details of angular integrations performed in various places in the main text, along with some other useful facts pertaining to angular structure of gravitational radiation. Here we will be quite liberal about index placements: computations are done in a 3d flat space so the distinction between up and down indices does not matter too much. Summation is always over repeated indices, whether up or down.

We will evaluate the integral
\begin{equation}\label{avgLambda}
	\Lambda_{ij,pq}({\bf \hat{x}}) \int d\Omega \, n_p n_q \int du' \big\langle
	\Lambda_{kl,mn}({\bf n})\dddot{Q}_{kl} \dddot{Q}_{mn}
	\big\rangle g(u',{\bf \hat{x}\cdot n}) \ ,
\end{equation}
where $Q\ijd$ is the quadrupole moment, $g$ is a function of time and angles, and the angle brackets denote time average over several wavelengths. The lambda tensor was defined in Eq. \nr{ttdef}. The integral can also be written as 
\begin{equation}\label{avgQQ}
	\Lambda_{ij,pq}({\bf \hat{x}}) \int d\Omega \, n_p n_q \int du' \big\langle \dddot{Q}_{kl} \dddot{Q}^{kl} - 2 \dddot{Q}_{kl} \dddot{Q}^{km}n^p n_m + \frac{1}{2}\dddot{Q}_{kl} \dddot{Q}_{mp} n^k n^l n^m n^p\big\rangle g(u',{\bf \hat{x}\cdot n}) \ .
\end{equation}
This is the integral we encountered several times in the main text.
The quantity inside the brackets comes from the TT projection in the direction $n^i$ of the first order GW; the TT projection on the $i,j$ indices is performed with respect to direction $\hat{x}^i$.

The lambda tensor is purely angle-dependent, thus we can pull it out of the time average:
\begin{equation}
	\big\langle
	\Lambda_{kl,mn}({\bf n})\dddot{Q}_{kl} \dddot{Q}_{mn}
	\big\rangle = \Lambda_{kl,mn}({\bf n})\big\langle \dddot{Q}_{kl} \dddot{Q}_{mn}
	\big\rangle \ .
\end{equation}
Assuming the quasicircular inspiral quadrupole moment described by \nr{quad} and \nr{R(t)}, and evaluating the average, we get
\begin{equation}\label{factorized}
	\Lambda_{kl,mn}({\bf n})\big\langle \dddot{Q}_{kl} \dddot{Q}_{mn}
	\big\rangle = \frac{G^3 M^5}{2 R(u')^5}\Lambda_{kl,mn}({\bf n}) I_{kl}I_{mn}^* \ , \quad (I_{kl}) = \begin{pmatrix}
		1 & i & 0 \\
		i & -1 & 0 \\
		0 & 0 & 0
	\end{pmatrix}
\end{equation}
(see also Eq. \nr{Y22}). In terms of angular variables, this gives the structure that appears over and over again in the main text:
\begin{equation}
	\Lambda_{kl,mn}({\bf n}) I_{kl}I_{mn}^* = I\ijd I^*\ijd - 2 I_{ik} I_{jk}^* n_i n_j + \frac{1}{2}I\ijd I_{kl}^*n_i n_j n_k n_l = \frac{1}{2}(1+6\cos^2 \theta + \cos^4 \theta) \ ,
\end{equation}
where the last expression is just $\mathcal{F}(\cos\theta)$ defined in (\ref{Fdef}).
Now we have a time- and angle-independent tensor product ${\bf I}\otimes {\bf I^*}$ that can be pulled out of the both integrals in \nr{avgLambda}, and the rest of the factors are either purely time- or angle-dependent. Hence, we denote
\begin{equation}
	f({\bf \hat{x}\cdot n}) \equiv \int du' \frac{G^3 M^5}{2 R(u')^5} g(u',{\bf \hat{x}\cdot n}) \ ,
\end{equation}
and write
\begin{align}
	&\Lambda_{ij,pq}({\bf \hat{x}}) \int d\Omega \, n_p n_q \int du' \big\langle
	\Lambda_{kl,mn}({\bf n})\dddot{Q}_{kl} \dddot{Q}_{mn}
	\big\rangle g(u',{\bf \hat{x}\cdot n}) \nn \\
	=& \Lambda_{ij,pq}({\bf \hat{x}}) I_{kl}I_{mn}^* \int d\Omega \, n_p n_q \Lambda_{kl,mn}({\bf n}) f({\bf \hat{x}\cdot n}) \ .
\end{align}
The lambda tensor involves terms with $0,2,$ and $4$ $n_i$'s.
Towards evaluating this integral, we therefore need to compute angular integrals of the form
\begin{equation}
	\int d\Omega f({\bf \hat{x} \cdot n}) n_{i_1} ... n_{i_{2k}} \ ,
\end{equation}
where $f$ is a real-valued function and ${\bf \hat{x}}$ is a constant unit vector. Obviously, the result of integration must be symmetric in indices $i_1,...,i_{2k}$, and it can only depend on the metric components and the vector $\hat{x}^i$. This can be seen from rotational covariance of the integral: first defining
\begin{equation}
	C({\bf \hat{x}})_{i_1\cdt i_{2k}} = \int d\Omega f({\bf \hat{x} \cdot n}) n_{i_1}... n_{i_{2k}} \ ,
\end{equation}
we then take an arbitrary orthogonal matrix ${\bf R}$ and transform both sides:
\begin{align}
	\tensor{R}{_{i_1}^{j_1}}...\tensor{R}{_{i_{2k}}^{j_{2k}}} C({\bf \hat{x}})_{j_1\cdt j_{2k}}
	&=\tensor{R}{_{i_1}^{j_1}}...\tensor{R}{_{i_{2k}}^{j_{2k}}} \int d\Omega f({\bf \hat{x} \cdot n}) n_{j_1}... n_{j_{2k}} \\
	&=\int d\Omega f({\bf \hat{x} \cdot n}) \tensor{R}{_{i_1}^{j_1}}...\tensor{R}{_{i_{2k}}^{j_{2k}}} n_{j_1}... n_{j_{2k}} \\
	&=\int d\Omega' f({\bf \hat{x}' \cdot n'}) n'_{i_1}... n'_{i_{2k}} \\
	&=C({\bf \hat{x}'})_{i_1\cdt i_{2k}} \ .
\end{align}
This (and symmetricity) will only be satisfied if the integral can be written as
\begin{equation}\label{ansatz}
	\int d\Omega f({\bf \hat{x} \cdot n}) n_{i_1}...n_{i_{2k}} = \sum_{l = 0}^{k} C_{2k,2l} \, \hat{x}_{(i_1} \hat{x}_{i_2} ... \hat{x}_{i_{2l-1}}\hat{x}_{i_{2l}} \delta_{i_{2l+1} i_{2l+2}} ... \delta_{i_{2k-1} i_{2k})} \ .
\end{equation}
The constants $C_{2k,2l}$ can be determined by contracting both sides with all possible combinations $\hat{x}_{i_1} \hat{x}_{i_2} ... \hat{x}_{i_{2j-1}}\hat{x}_{i_{2j}} \delta_{i_{2j+1} i_{2j+2}} ... \delta_{i_{2k-1} i_{2k}}$ and solving the resulting system of algebraic equations. After a contraction, the lhs of (\ref{ansatz}) looks like
\begin{equation}\label{lhscont}
	\int d\Omega \, ({\bf \hat{x} \cdot n})^{2j} f({\bf \hat{x} \cdot n}) ,
\end{equation}
with $j = 0,1,...,k$. The coordinate system on $S^2$ may be rotated such that ${\bf \hat{x}}$ points to the north pole, i.e., ${\bf \hat{x} \cdot n} = \cos \theta$ in polar coordinates. The azimuthal integral can thus be done trivially, and (\ref{lhscont}) becomes
\begin{equation}
	\int d\Omega \, ({\bf \hat{x} \cdot n})^{2j} f({\bf \hat{x} \cdot n}) = 2\pi \int_{-1}^{1} d\cos\theta \, \cos^{2j}\theta \, f(\cos\theta) \ .
\end{equation}
The constant factors are then given by
\begin{equation}
	C_{2k,2l} = 2\pi \sum_{j = 0}^{k} \int_{-1}^{1} d\cos\theta \,  P_{2k,2l,2j}\cos^{2j} \theta f(\cos\theta) \ ,
\end{equation}
where $P_{2k,2l,2j}$ are numbers that can be determined combinatorially.  

Looking at \nr{avgQQ}, we see that the relevant combinations for us are
\begin{align}
	&\int d\Omega  f({\bf \hat{x}\cdot n}) n_{i_1}n_{i_2} =  C_{2,0}\,\delta_{i_1 i_2} +C_{2,2}\,\hat{x}_{i_1} \hat{x}_{i_2}\label{2n} \\
	&\int d\Omega  f({\bf \hat{x}\cdot n})n_{i_1} n_{i_2}n_{i_3} n_{i_4} = C_{4,0} \,\delta_{(i_1 i_2}\delta_{i_3 i_4)}+C_{4,2} \,\hat{x}_{(i_1}\hat{x}_{i_2}\delta_{i_3 i_4)}+C_{4,4} \hat{x}_{i_1} \hat{x}_{i_2} \hat{x}_{i_3}\hat{x}_{i_4}\label{4n} \\
	&\int d\Omega f({\bf \hat{x}\cdot n}) n_{i_1} n_{i_2}n_{i_3}n_{i_4}n_{i_5} n_{i_6} = C_{6,0} \, \delta_{(i_1 i_2}\delta_{i_3 i_4}\delta_{i_5 i_6)} + C_{6,2}\, \hat{x}_{(i_1} \hat{x}_{i_2} \delta_{i_3 i_4} \delta_{i_5 i_6)} \nn \\
	& \qquad \qquad \qquad \qquad + C_{6,4} \, \hat{x}_{(i_1} \hat{x}_{i_2} \hat{x}_{i_3} \hat{x}_{i_4} \delta_{i_5 i_6)} + C_{6,6}\,\hat{x}_{i_1} \hat{x}_{i_2} \hat{x}_{i_3} \hat{x}_{i_4} \hat{x}_{i_5} \hat{x}_{i_6} \ .\label{6n}
\end{align}
Under TT projection with respect to the observer direction ${\bf x}$, all terms in \nr{2n}--\nr{6n} that involve less than two Kronecker deltas vanish identically, due to the fact that 
\begin{equation}
	\Lambda_{ij,pq}({\bf \hat{x}}) \hat{x}_q = \Lambda_{ij,pq}({\bf \hat{x}}) \delta_{pq} = 0 \ .
\end{equation}
Thus, the coefficients relevant for the final TT projected result are just $C_{4,0}, C_{6,0},$ and $C_{6,2}$.
We only computed the numbers $P_{2k,2l,2j}$ in the special cases corresponding to these $C$-factors, using Mathematica:
\begin{align}
	&P_{4,0,0} = \frac{3}{8}, \quad P_{4,0,2} = -\frac{3}{4},\quad P_{4,0,4} = \frac{3}{8}, \nn \\ 
	&P_{6,0,0} = \frac{5}{16}, \quad P_{6,0,2} = -\frac{15}{16},\quad P_{6,0,4} = \frac{15}{16},\quad P_{6,0,6} = -\frac{5}{16}, \nn \\ 
	&P_{6,2,0} = \frac{15}{16}, \quad P_{6,2,2} = -\frac{225}{16}, \quad P_{6,2,3} = \frac{525}{16}, \quad  P_{6,2,6} = -\frac{315}{16} \ .
\end{align}
The $C$-factors are then straightforward to compute and they are given by
\begin{align}
	&C_{4,0} =\frac{3\pi}{4} \int_{-1}^{1} (1 - \cos^2\theta)^2 f(\cos\theta)d\cos\theta \\
	&C_{6,0} =\frac{5\pi}{8} \int_{-1}^{1} (1 - \cos^2\theta)^3 f(\cos\theta)d\cos\theta \\
	&C_{6,2} =\frac{15\pi}{8} \int_{-1}^{1} (1 - \cos^2\theta)^2(7 \cos^2 \theta - 1) f(\cos\theta)d\cos\theta \ . 
\end{align}
The final TT projected result then is
\begin{align}\label{ttgeneral}
	&\Lambda_{ij,pq}({\bf \hat{x}}) I_{kl}I_{mn}^* \int d\Omega \, n_p n_q \Lambda_{kl,mn}({\bf n}) f({\bf \hat{x}\cdot n}) \nn\\
	=&\frac{1}{45}(60 C_{4,0}-12 C_{6,0}-C_{6,2}(1-3 \cos ^2 \theta_x  )) \sin ^2\theta_x \sqrt{2} e^+\ijd \ ,
\end{align}
where $\theta_x$ is the inclination angle of ${\bf x}$. 
%To verify this relation, it is useful to transform to spherical coordinates so that the plus polarization becomes manifest. 
Note that in the original Cartesian coordinates the lhs is a very complicated 3$\times$3 matrix. When transformed to polar coordinates with $e_{\bf r}={\bf \hat x}$, its plus-polarized structure becomes manifest, as in Eq. \nr{polarization}. The $\cos\theta$ integrals in the constants $C_{2k,2l}$ can be combined so that the rhs becomes
\be
{\pi\over8}\int_{-1}^1 d\cos\theta f(\cos\theta)(1-\cos^2\theta)^2\left(7-\cos^2\theta+(7\cos^2\theta-1)\cos^2\theta_x\right)\sin^2\theta_x \sqrt2 e_{ij}^+.
\ee
We shall anyway use the full form.

For instance, in the calculation of the flat spacetime nonlinear memory effect we take
\begin{equation}
	f({\bf \hat{x} \cdot n})=\int du' \frac{G^3 M^5}{2 R(u')}\frac{1}{1-{\bf \hat{x} \cdot n}} \ .
\end{equation}
In this case the time integral can be done separately so we just drop the time integral along with the time-dependent factor and focus on the angular structure. We then need to evaluate the angular integrals
\begin{equation}
	\int d\Omega \frac{n_i n_j}{1 - {\bf\hat{x}}\cdot {\bf n}}, \quad \int d\Omega \frac{n_i n_j n_k n_l}{1 - {\bf\hat{x}}\cdot {\bf n}},\quad \int d\Omega \frac{n_i n_j n_k n_l n_p n_q}{1 - {\bf\hat{x}}\cdot {\bf n}} \ .
\end{equation}
The integrals involve collinear singularities but the TT projection removes all the diverging terms, indicating that these infinities were unphysical. The relevant $C$-factors are
\begin{equation}
	C_{4,0}= \pi, \quad C_{6,0} = \frac{2\pi}{3}\quad C_{6,2} = \pi \ .
\end{equation}
Plugging these into \nr{ttgeneral} yields the well-known structure
\begin{equation}
	\frac{\pi}{15} (17+\cos^2 \theta_x)\sin^2 \theta_x \sqrt{2}\,e^+\ijd \ .
\end{equation}
For the novel light cone memory term as well as the tail discussed in Sec. \ref{radbyrad}, $f$ is a much more complicated function that cannot be factorized into time- and angle-dependent functions, so the $C$-coefficients must be computed numerically. However, the recipe for solution is still given by the general form \nr{ttgeneral}.

Here we also collect some facts we need to know about symmetric trace-free (STF) tensors and TT tensors. Given a unit vector ${\bf n}$ on the sphere,
\begin{equation}
	{\bf n} = (\sin \theta \cos \phi, \sin\theta\sin\phi, \cos\theta) \ ,
\end{equation}
we may define a basis for STF tensors by
\begin{equation}
	Y_{lm}(\theta,\phi) = \mathcal{Y}_{K_{l}}^{lm} N^{K_{l}} \ , \quad N_{K_{l}} = n_{i_1}\cdot\cdot\cdot n_{i_l} \ ,
\end{equation}
where $Y_{lm}$ are the usual spherical harmonic functions and $K_l$ is a multi-index symbol denoting $l$ indices $i_1,...,i_l$.

In flat spacetime, we can in general write the spin-2 GW in the wave zone as
\begin{equation}\label{tensorexpansion}
	h\ijd \ttu = \frac{G}{r} \sum_{l=2}^{\infty}\sum_{m=-l}^{l} h_{lm} ({\bf T}^{E2}_{lm})\ijd + v_{lm} ({\bf T}^{B2}_{lm})\ijd \ ,
\end{equation}
where $h_{lm}$ and $v_{lm}$ are modes related to mass multipoles and current multipoles of the source, respectively. ${\bf T}^{E2}_{lm}$ and ${\bf T}^{B2}_{lm}$ are TT tensor spherical harmonics, given by \cite{Maggiore:2007ulw}
\begin{align}
	&({\bf T}^{E2}_{lm})\ijd = c_l\, r^2 \Lambda_{ij,i'j'}({\bf n})\pt_{i'}\pt_{j'}Y_{lm} \label{TE2def}\\
	&({\bf T}^{B2}_{lm})\ijd = c_l\, r\, \Lambda_{ij,i'j'}({\bf n}) \frac{i}{2}(\pt_{i'}L_{j'} + \pt_{j'}L_{i'})Y_{lm} \\
	&c_l \equiv \left(2\frac{(l-2)!}{(l+2)!}\right)^{1/2} \ ,
\end{align}
where ${\bf L} = -i {\bf r}\times {\bf \nabla}$ is the orbital angular momentum operator.

Here we are only interested in mass multipole radiation so we set $v_{lm}=0$ in \nr{tensorexpansion}.
Furthermore, gravitational radiation is dominated by the $l=2$ mode in the expansion \nr{tensorexpansion}. Therefore we truncate the series and approximate
\begin{equation}
	\dot{h}\ttu\ijd \dot{h}\ttu\ijd \approx \frac{G^2}{r^2}\sum_{m,m'=-2}^{2} \dot{h}_{2m}\dot{h}_{2m'}^* ({\bf T}^{E2}_{2m})\ijd ({\bf T}^{E2}_{2m'})\ijd ^* = \frac{G^2}{3 r^2} \Lambda_{ij,kl}\sum_{m,m'=-2}^{2} \dot{h}_{2m}\dot{h}_{2m'} ^* \mathcal{Y}^{2m}\ijd \mathcal{Y}^{2m'*}_{kl} \ .
\end{equation}
Here all the angular dependence is in the lambda tensor and the time-dependence in the $h_{2m}$ modes. We used Thorne's \cite{Thorne:1980ru} Eq. (2.39e) in deriving the latter equality. Notice that 
\begin{equation}\label{Y22}
	\mathcal{Y}^{22}\ijd  = \sqrt{\frac{15}{32\pi}} I_{ij} \ ,
\end{equation}
where the constant STF tensor $I\ijd$ appeared above in \nr{factorized}. 
%The contraction
%\begin{equation}
%	\Lambda_{ij,kl}\mathcal{Y}^{2m}\ijd \mathcal{Y}^{2m'*}_{kl} = \frac{15}{64\pi} (1+6\cos^2\theta+\cos^4\theta)
%\end{equation}
%has the same angular structure as

Let us then relate these to spin-2 spherical harmonics.
Spin-weighted spherical harmonics are in general defined in terms of ordinary spherical harmonics by \cite{Newman:1966ub}
\begin{equation}
	{}_{s}Y_{lm} =
	\begin{cases}
		\sqrt{\frac{(l-s)!}{(l+s)!}}\eth ^s Y_{lm} \ ,  &\quad 0 \leq s \leq l \ ,\\
		(-1)^s \sqrt{\frac{(l+s)!}{(l-s)!}} \bar{\eth}^{-s} Y_{lm} \ , &\quad -l \leq s \leq 0 \ ,
	\end{cases}
\end{equation}
where $\eth$ and $\bar{\eth}$ are covariant derivative operators defined on the two-sphere. Explicitly, let $f$ be a function with spin-weight $s$ on the sphere, which means that the function transforms like $f \rightarrow e^{i s \psi} f$ under a rotation with angle $\psi$ about the polar axis. Then the action of the operators on it is defined by
\begin{align}
	&\eth f = - (\sin\theta)^s \left[\pt_\theta + \frac{i}{\sin\theta} \pt_\phi\right]\left[(\sin \theta)^{-s} f\right] \\
	&\bar{\eth} f = -(\sin\theta)^{-s} \left[\pt_\theta -\frac{i}{\sin\theta}\pt_\phi \right]\left[ (\sin\theta)^s f\right] \ .
\end{align}
$\eth$ raises the spin-weight of $f$ by 1, and $\bar{\eth}$ lowers it by 1. With \nr{TE2def}, we get the relation
\begin{equation}
	(e^+\ijd -i e^\times\ijd) ({\bf T}^{E2}_{lm})\ijd = {}_{-2} Y_{lm} \ ,
\end{equation}
where ${}_{-2} Y_{lm}$ is a spin-2 spherical harmonic and the polarization tensors were defined in \nr{polar}. For the dominant part of gravitational radiation, the relevant harmonics are
\begin{equation}
	{}_{-2}Y_{22} = \frac{1}{8}\sqrt{\frac{5}{\pi}}(1+\cos\theta)^2 e^{2i\phi} \ , \quad {}_{-2}Y_{2-2} = \frac{1}{8}\sqrt{\frac{5}{\pi}}(1-\cos\theta)^2 e^{-2i\phi} \ .
\end{equation}
Then we find the relation
\begin{equation}
	\abs{{}_{-2}Y_{22}}^2+\abs{{}_{-2}Y_{2-2}}^2 = \frac{2}{3}\Lambda_{ij,kl} \mathcal{Y}^{22}\ijd \mathcal{Y}^{22*}_{kl} \ .
\end{equation}
This relation can be used when expressing the angular dependency of the leading quasinormal modes in Eq. (\ref{complexstrain}) in terms of unit radial vectors $n^i$. This also accounts for the angular distribution in \nr{Lring}, which is the same as in \nr{factorized}. The angular integration in the ringdown phase of MWM is therefore identical to what was done above for the inspiral phase.

\bibliographystyle{JHEP}
\bibliography{refs}

\end{document}